\documentclass{jnmp}

\usepackage{amsmath}

\JNMPnumberwithin{equation}{section}

\newtheorem{proposition}{Proposition}[section]
\newtheorem{corollary}{Corollary}[section]

\theoremstyle{definition}

\newtheorem{remark}{Remark}[section]
\newtheorem{problem}{Problem}[section]
\newtheorem{example}{Example}[section] 

\begin{document}

%
\renewcommand{\evenhead}{B~A~Kupershmidt}
\renewcommand{\oddhead}{Dark Equations}

%
\thispagestyle{empty}

\FirstPageHead{8}{3}{2001}{\pageref{kupersh-firstpage}--\pageref{kupersh-lastpage}}{Article}

\copyrightnote{2001}{B~A~Kupershmidt}

\Name{Dark Equations}
\label{kupersh-firstpage}

\Author{B~A~KUPERSHMIDT}

\Address{The University of Tennessee Space Institute, Tullahoma, TN  37388, USA\\
E-mail: bkupersh@utsi.edu}

\Date{Submitted March 28, 2001; Accepted April 29, 2001}

\vspace{5mm}

\rightline{\bfseries\itshape To the memory of Auberon Waugh}

\rightline{\bfseries\itshape (1939--2001), a light in the darkness.}

\vspace{5mm}

\begin{abstract}
\noindent
Observing the Universe, astronomers have 
concluded that the motion of stars can not be accounted for unless 
one assumes that most of the mass in the Universe is carried on by 
a ``dark matter", so far impervious to all attempts at being detected.  
There is now a similar concept of ``dark energy".  
I shall discuss a different subject, ``dark equations".  These have 
never indicated that they influence anything or even exist, 
but if one supposes that they do exist, one can systematically discover 
them and study their properties, some of which turn out to be 
strange and others mysterious.  These equations are similar in spirit 
to what one gets when linearizing a given system, or studies how an 
external linear wave interacts with a particular solution of a given 
system.  We define and study linear extensions 
of dynamical systems in
general, and integrable and Hamiltonian systems in particular.  
Systems discussed include the KdV and mKdV equations and the 
associated Miura maps, the Burgers hierarchy and the associated 
Hopf--Cole transformations, long wave equations, the Benney hierarchy, 
and the KP hierarchy.
\end{abstract}

\centerline{\large \bf Contents}
\begin{enumerate}
\itemsep0mm
\item[1.] Introduction
\item[2.] Tangent and cotangent bundles, the functional case
\item[3.] Duality involution
\item[4.] Morphisms of extensions
\item[5.] Extended Lax equations
\item[6.] The first two KdV flows
\item[7.] Solutions of the extended KdV equations
\item[8.] Miura maps for the extended KdV equations
\item[9.] KdV and mKdV extensions of order $\leq 1$
\item[10.] The Burgers hierarchy
\item[11.] The equation $u_t + uu_x = 0$ and its hierarchy
\item[12.] Long wave equations
\item[13.] The Benney hierarchy
\item[14.] The KP hierarchy
\item[15.] One-dimensional linear extensions of linear Hamiltonian matrices
\item[16.] Nonlinear derivations of Lie algebras
\item[17.] Linear extensions as Symbols of nonlinear ones
\item[18.] Scalar extensions associated with scalar Lax operators of order $\geq 4$
\item[19.] Composition of Hamiltonian extensions
\end{enumerate}

\section{Introduction}
\begin{flushright}
\begin{minipage}{45mm}
\it Politics is for the moment.\\
An equation is for eternity. \\
\null \hfill  Einstein
\end{minipage}
\end{flushright}

Given a system (of equations, say) deserving our attention, we may well
inquire whether this system is a part of a larger whole still interesting
in one way or another.

In other words, we are talking about extensions, a fruitful mathematical
concept.  Two typical examples are: 1)~given a field ${\mathcal F}$, consider
all possible field extensions ${\mathcal F}^\prime \supset {\mathcal F}$; and 2)~given a vector
field $X$ on a manifold $M$, consider (say, vector) bundles $\pi: E \rightarrow
M$ and lifts $X \mapsto X^{\rm ext}$ of $X$ from $M$ into $E$ such that
$\pi^* X = X^{\rm ext} \pi^*$.  If we want to lift up not just an individual
vector field $X$ but the whole Lie algebra ${\mathcal D} (M)$ of vector fields on $M$,
we are looking at connections on bundles with the base $M$.

The 2$^{nd}$ example is close to what we are after.  We start with the 
following concrete problem:

\begin{problem}\label{problem:1.1}
Consider the 2$^{nd}$ and 3$^{rd}$ KdV 
equations: 
\setcounter{equation}{1}
\begin{gather}
(X) \quad  u_t = \left(3u^2 + u_{xx}\right)_x, \label{1.2}\\
(Y) \quad u_t = \left(10u^3 + 5u_x^2 + 10uu_{xx} + u_{xxxx}\right)_x. \label{1.3}
\end{gather}
The flows $X$ and $Y$ commute.  We seek {\it linear extensions}
of these flows, $X^{\rm ext}$ and $Y^{\rm ext}$, of the form
\begin{subequations} \label{1.4}
\renewcommand{\theparentequation}{\arabic{section}.\arabic{parentequation}}
\begin{gather}
u_t = \left(3u^2 + u_{xx}\right)_x,  \label{1.4a}\\
\varphi_t = {\mathcal A}_1 (\varphi), \quad \left(X^{\rm ext}\right), \label{1.4b}\\
\refstepcounter{parentequation}\setcounter{equation}{0}
u_t = \left(10u^3 + 5u^2_x + 10uu_{xx} + u_{xxxx}\right)_x, \label{1.5a}\\
\varphi_t = {\mathcal A}_2 (\varphi),  \quad \left(Y^{\rm ext}\right) \label{1.5b}
\end{gather}
\end{subequations}
which {\it still commute}; here $\varphi$ is a vector, and ${\mathcal A}_1$ and 
${\mathcal A}_2$ are matrix linear differential operators whose coefficients depend only 
upon $u$, {\it not} upon $\varphi$.
\end{problem}

We shall discuss the above Problem \ref{problem:1.1} later on, in Section~6.  Let us now take 
a general view on this problem.

It can be looked at from at least four different points of view:

({\it A}) Given two {\it commuting} dynamical systems $X$ and $Y$:
\begin{gather}
(X) \quad u_t = X^\sim, \label{1.6}\\
(Y) \quad u_t = Y^\sim, \label{1.7}
\end{gather}
find/classify all their linear extensions $X^{\rm ext}$ and $Y^{\rm ext}$ 
that continue to commute:
\begin{subequations} \label{1.8}
\begin{gather} 
\left(X^{\rm ext}\right) \quad u_t = X^\sim, \quad \varphi_t = {\mathcal A}_1 (\varphi), \label{1.8a}\\
\left(Y^{\rm ext}\right) \quad u_t = Y^\sim, \quad \varphi_t = {\mathcal A}_2 (\varphi); \label{1.8b}
\end{gather}
\end{subequations}

({\it B}) Given an infinite commuting hierarchy of dynamical systems $\{X_n\}$, 
find its linear extensions $\left\{X_n^{\rm ext}\right\}$ that are still commuting.
\begin{equation}
\left(X_n^{\rm ext}\right) \quad u_t = X^\sim_n, \quad \varphi_t = {\mathcal A}_n (\varphi); 
\label{1.9}
\end{equation}

({\it C}) Given the whole Lie algebra of dynamical systems (evolution fields) 
$D^{\rm ev} = \{ X \}$, find its linear extensions each of which is isomorphic 
to $D^{\rm ev}$ as a Lie algebra.  In other words, if $Z = [X, Y]$, for any 
$X, Y, Z \in D^{\rm ev}$, then the linear extensions $X^{\rm ext}, Y^{\rm ext}, 
Z^{\rm ext}$ satisfy
\begin{equation}
Z^{\rm ext} = \left[X^{\rm ext}, Y^{\rm ext}\right]. \label{1.10}
\end{equation}
This equality can be further rewritten as
\begin{equation}
{\mathcal A}^{[X, Y]} = X\left({\mathcal A}^Y\right) - Y \left({\mathcal A}^X\right) - 
\left[{\mathcal A}^X, {\mathcal A}^Y\right], \label{1.11}
\end{equation}
where
\begin{equation}
\left(X^{\rm ext}\right) \quad u_t = X^\sim, \quad \varphi_t = {\mathcal A}^X (\varphi), \label{1.12}
\end{equation}
is the general form of a particular linear extension determined by a map 
$X \mapsto {\mathcal A}^X$.

When $X$ and $Y$ commute, the characteristic equation (\ref{1.11}) becomes
\begin{equation}
X\left({\mathcal A}^Y\right) - Y \left({\mathcal A}^X\right) = 
\left[{\mathcal A}^X, {\mathcal A}^Y\right], \label{1.13}
\end{equation}
which can be thought of as a generalized zero-curvature equation; 

{\samepage
({\it D}) Given a Lie algebra of {\it Hamiltonian} vector fields
\begin{equation}
(X_H) \quad u_t = B \left(\frac{\delta H}{\delta u} \right), \label{1.14}
\end{equation}
find its linear {\it Hamiltonian} extensions.  This amounts to finding 
{\it Hamiltonian} matrices $B^{\rm ext}$, of the form
\begin{equation}
B^{\rm ext} = \begin{array}{c} u \\ \varphi\end{array}\!\!\!
\left(\begin{array}{c|c}
 \mbox{\raisebox{5mm}[0pt][0pt]{$u$}}\hspace{-2mm}B & 
 \mbox{\raisebox{5mm}[0pt][0pt]{\hspace{2mm}$\varphi$}}\hspace{-5mm}-b^\dagger \\
\hline
&\\[-4mm]
 b & (*) \end{array}\right) \label{1.15}
\end{equation}
where $b$ is {\it linear} in $\varphi$.}

For example, our original Problem \ref{problem:1.1}, with the two KdV flows (\ref{1.2}) and 
(\ref{1.3}), can be looked at as the particular case
\begin{equation}
B = \partial \label{1.16}
\end{equation}
or
\begin{equation}
B = u \partial + \partial u + \frac 12 \partial^3 \label{1.17}
\end{equation}
of the alternative ({\it D}).

We shall meet every one of the above mentioned 4 cases ({\it A})--({\it D}) later on.  
At the moment let us notice that all these 4 cases are related but neither 
one covers apriori any other.  Also, in cases ({\it A})--({\it C}), the solutions 
$\{{\mathcal A} = {\mathcal A}^X\}$ can be {\it direct-summed}, like (vector) bundles over a 
fixed base $M$; it's not clear if this observation applies to the Hamiltonian 
case ({\it D}), and it's very likely that in general it doesn't.

The plan of the paper is as follows.  In the next three Sections 
we discuss the general case ({\it C}).  Section~5 addresses the case ({\it B}) 
for Lax equations, both continuous and discrete.  In Section~6 we 
investigate scalar extensions of the first two nontrivial KdV 
flows (\ref{1.2}) and (\ref{1.3}) and find all such extensions, 8 in total.  In 
Section~7 we discuss solutions of the extended KdV equations, and in Section~8 
we look at Miura maps for these equations.  
In Section~9 we examine extensions given by 
differential operators of order $\leq 1$, for the KdV and mKdV 
hierarchies, and find the associated extensions of the classical Miura 
map.  In Section~10 we analyze the Burgers equation and the Hopf--Cole 
transformation, and then generalize all the extensions obtained for the 
whole Burgers hierarchy.  In Section~11 we consider the 
zero-dispersion/zero-dissipation limit of the Burgers hierarchy, and find some of 
its extensions which do not have a dispersive/dissipative origin.  
Section~12 is devoted to the Dispersive Water Waves hierarchy and its 
quasiclassical limit, while in Section~13 we study the Benney hierarchy.  
The Benney hierarchy is the quasiclassical limit of the KP 
hierarchy, and we examine the latter's extensions, two of 
them, in Section~14.  The second of these extensions is of a Lie-algebraic 
nature, and we develop the relevant mathematical notion~--- 
generalized 
derivations of Lie algebras~--- in Section~15.  In Section~16 we analyze 
Lie-algebraic  
meaning of scalar extensions provided by operators of order zero, 
and arrive at nonlinear/relative generalizations of the notion of 
a derivation of a Lie algebra.  In Section~17 we arrive at the conclusion 
that the linear extensions of integrable systems are 1$^{st}$-order 
approximations to nonlinear extensoins.  In Section~18 we find all scalar 
extensions associated with Lax equations with scalar Lax operators 
of order $\geq 4$:  it turns out that there exist no such extensions 
in adition to the Lax-type ones and their duals.  In Section~19 we 
prove that (nonlinear) Hamiltonian extensions of some general type 
can be composed.

\section{Tangent and cotangent bundles, the functional case}

Let $E$ be a manifold and $q = (q_i)$ be a local coordinate system on $E$.  
Let $TE$ be the tangent bundle of $E$, with the induced local coordinates 
$(q, v) = (q_i, v_i)$.  Every vector field $X$ on $E$ generates a (local) 
1-parameter group of diffeomorphisms $\{A_t\}$ of $E$; this group lifts 
uniquely into a group $\{\hat A_t\}$ of diffeomorphisms of $TE$ and this 
latter group's generating vector field $\hat X$ on $TE$ covers the vector 
field $X$ on $E$.  In local coordinates $q$, $v$, the motion equations for 
$X$ and $\hat X$ are, respecively:
\begin{gather}
(X) \quad \dot q_i = X_i (q), \label{2.1}
\\
(\hat X) \quad  \dot q_i = X_i (q), \quad  
\dot v_i = \sum_j \frac{\partial X_i}{\partial q_j} v_j. \label{2.2}
\end{gather}
By construction, the map $X \mapsto \hat X$ is a homomorphism of the Lie 
algebra ${\mathcal D}(E)$ into the Lie algebra ${\mathcal D} (TE)$.  This is our first example 
of the case ({\it C}), (\ref{1.10})--(\ref{1.12}).

Turning from Mechanics to Field Theory, we should consider instead of a manifold 
$E -{}$ a bundle $\pi: E \rightarrow M$, instead of vector fields on $E$~---
 evolution vector fields on the infinite jet bundle $\pi_\infty:  J^\infty \pi 
\rightarrow M$, etc.  Rather than rush into the geometric wilderness, let us 
restrict ourselves to the purely algebraic setting; this way we lose topological 
aspects but gain generality by allowing in discrete degrees of freedom and even 
noncommutativity.  (Basic facts about differential-difference algebras and 
related notions can be found in [11] and [14].)

Thus, now the components $X_i = X_i \{q\} = X (q_i)$ of an evolution vector field 
$X$ are functions of $\{q$, $x$, various $x$-derivatives and shifts of $q\}$.  The 
lifted vector field $\hat X$ (\ref{2.2}) becomes 
\begin{equation}
(\hat X) \quad \dot q_i = X_i \{q\} , \quad \dot v_i = \sum_j
 \frac{DX_i}{Dq_j} (v_j), \label{2.3}
\end{equation}
where $D = \frac{D}{Dq} = \left(\frac{D}{Dq_i}\right)$ is the Fr\`echet 
derivative operator:
\begin{equation}
\sum_j \frac{Df}{Dq_j} (v_j) =\left. \frac{d}{d \epsilon}\right|_{\epsilon=0} 
(f (q + \epsilon v)),  \quad \forall \; f. \label{2.4}
\end{equation}
 
The map $X \mapsto \hat X$ (\ref{2.3}) is an example of (generalized) 
{\it linearizations}~[13].  That this map is a Lie algebra homomorphism is 
proven in [13] and [14, Ch.~12] for commutative systems and 
in [16, App.~2] for noncommutative ones.  (These 
references also cover the Hamiltonian case ({\it D}), but {\it not} in the form 
(\ref{1.15}): a different form appears there.)  This is our 2$^{nd}$ example of the 
general case ({\it C}), and it obviously subsumes the form (\ref{2.2}). 

To get more examples, we of course could consider any {\it natural bundle} over 
$E$, not just the tangent bundle $TE$; we won't learn much that way, except proving 
that the case ({\it C}) has an infinite number of solutions.  But the case of 
cotangent bundle $T^*E$ will prove to be instructive in the next Section, so 
let's look at this bundle now.
 
Let $p, q = (p^i, q_i)$ be local coordinates on $T^* E$, and let 
$\rho = pdq = \sum\limits_i p^i dq_i$ be the universal 1-form on $T^*E$.  For any 
vector field $X = \sum X_i \frac{\partial}{\partial q_i}$ 
on $E$, the 
expression $X \rfloor \rho = \sum\limits_i p^i X_i$ is well-defined, and the 
Hamiltonian 
vector field $X_H$ on $T^*E$ with the Hamiltonian $H = X \rfloor 
\rho$ has the following motion equations:
\begin{equation}
(X_{X \rfloor \rho}) \quad \dot q_i = \frac{\partial H}{\partial p^i}
 = X_i, \quad \dot p^i =- \frac{\partial H}{\partial q_i} = 
- \sum_j \frac{\partial X_j}{\partial q_i} p^j. \label{2.5}
\end{equation}
The map $X \mapsto X_{X \rfloor \rho}$ is well-known to be a Lie algebra homomorphism; 
this is our 3$^{rd}$ example of the case ({\it C}). 

Let's now turn to the PDE version of the above 
cotangent picture.  Again, 
we shall stick here~--- as elsewhere in this paper~--- to algebraic 
language.  
(Geometric version of what follows in this Section 
can be found in [7, Ch.~3].)  So, if $X$ is an evolution vector field with the motion equations
\begin{equation}
(X) \quad \dot q_i = X_i \{q\}, \label{2.6}
\end{equation}
then its cotangent lift is a Hamiltonian vector field $X_H$ with the 
Hamiltonian $H= X \rfloor \rho = \sum\limits_j X_j p^j$:
\begin{equation}
(X_{X \rfloor \rho}) \quad \dot q_i = \frac{\delta H}{\delta p^i}= 
X_i, \quad \dot p^i = - \frac{\delta H}{\delta q_i} = - \frac{\delta}
{\delta q_i} \left(\sum_j X_j p^j\right). \label{2.7}
\end{equation}
The map $X \mapsto X_{X \rfloor \rho}$ is known 
to be a Lie algebra homomorphism.  
We thus get a 4$^{th}$ example of the case ({\it C}).

\section{Duality involution}

Suppose we have an extension $X \mapsto {\mathcal A}^X$ satisfying the equation 
(\ref{1.11}) --- for all $X$ in the case ({\it C}), or for some $X$ in 
the cases ({\it A}) or ({\it B}).

Set
\begin{equation}
\widetilde{\mathcal A}^X= - \left({\mathcal A}^X\right)^\dagger, \label{3.1}
\end{equation}
where $\dagger$ stands for ``adjoint''.

\setcounter{proposition}{1}
\setcounter{equation}{2}
\begin{proposition}\label{proposition:3.2}
If the family $\left\{{\mathcal A}^X\right\}$ satisfies 
the equation (\ref{1.11}) then so does the family
 $\left\{\widetilde{\mathcal A}^X\right\}$.
\end{proposition}

\begin{proof}
Applying the operator $-\dagger$ to the equality 
(\ref{1.11}) we get
\begin{equation*}
\widetilde{\mathcal A}^{[X,Y]} = X \left(\widetilde{\mathcal A}^Y\right) - Y  
\left(\widetilde{\mathcal A}^X\right) - 
\left[\widetilde{\mathcal A}^X,  \widetilde{ \mathcal A}^Y \right]. \label{3.3}
\tag*{\qed}
\end{equation*}
\renewcommand{\qed}{}
\end{proof}

Thus, extensions come in pairs --- if we agree to double-count self-dual 
extensions.  An example of the latter is the trivial extension:
\begin{equation}
{\mathcal A}^X = 0, \quad \forall \; X. \label{3.4}
\end{equation}

We have seen in Section~2 two classes of general extensions:  the tangent 
(\ref{2.3}) and the cotangent one (\ref{2.7}).

\setcounter{proposition}{3}
\setcounter{equation}{4}
\begin{proposition}\label{proposition:3.5}
The tangent extension (\ref{2.3}) and the 
cotangent extension (\ref{2.7}) are dual to each other (in the sense of 
Definition (\ref{3.1})).
\end{proposition}

\begin{proof}
 For the tangent map $X \mapsto {\mathcal A}^X$ (\ref{2.3}), 
the matrix ${\mathcal A}^X = \left({\mathcal A}^X\right)_{ij}$ has the matrix elements
\begin{equation}
\left({\mathcal A}^X\right)_{ij} = \frac{DX_i}{Dq_j}. \label{3.6}
\end{equation}
Hence, by formula (\ref{3.1}), 
\begin{equation}
\left( \widetilde{\mathcal A}^X\right)_{ij} = \left(- \left( {\mathcal A}^X\right)^\dagger\right)_{ij} = -  
\left(\left({\mathcal A}^X\right)_{ji}\right)^\dagger = - 
\left(\frac{DX_j}{Dq_i}\right)^\dagger. \label{3.7}
\end{equation}
Therefore, the corresponding extension has the form
\begin{equation}
\dot p^i = \sum_j \left( \widetilde{\mathcal A}^X\right)_{ij} (p^j) = - \sum_j 
\left(\frac{DX_j}{Dq_i}\right)^\dagger \left(p^j\right). \label{3.8}
\end{equation}
Comparing this formula and formula (\ref{2.7}), we see that we need to check 
the identity
\begin{equation}
\frac{\delta}{\delta q_i} \left(X_j p^j\right) = \left(\frac{D X_j }{Dq_i} \right)^\dagger \left(p^j\right),
 \quad X_j = X_j \{q\}, \label{3.9}
\end{equation}
which follows directly from the definition of the variational derivatives.
\end{proof}

\section{Morphisms of extensions}

Suppose we are given {\it two different} evolution vector fields $X$ 
and $Y$ over two different rings $C_u$ and $C_v$.  For example, the field 
$X$ defines the KdV equation (\ref{1.2}) and the field $Y$ defines the modified 
mKdV equation
\begin{equation}
(Y) \quad v_t = \left(-2v^3 + v_{xx}\right)_x. \label{4.1}
\end{equation}
Let $\Phi: C_u \rightarrow C_v$ be a homomorphism compatible with the 
vector field $X$ and $Y$:
\begin{equation}
\Phi X = Y \Phi. \label{4.2}
\end{equation}
For example, 
\begin{equation}
\Phi (u) = - v^2 + \epsilon v_x, \quad \epsilon = \pm 1, \label{4.3}
\end{equation}
in one such (the original Miura map) for the KdV (\ref{1.2}) and mKdV (\ref{4.1}) fields. 

The question is, how do such homomorphisms interact with extensions?  Clearly, 
unless the map $\Phi$ is invertible --- and thus an isomorphism ---
 only 
{\it some} vector fields on $C_u$ can be pushed  forward by $\Phi$; hence, 
we are in the realm of alternatives ({\it A}) and ({\it B}) ---
 and possibly ({\it D}) --- but not ({\it C}).

Now, if ${\mathcal A}^X$ provides an extension of $X$, $\Phi \left({\mathcal A}^X\right)$ provides 
an extension of $Y$:
\begin{gather}
\left(X^{\rm ext}\right) \quad u_t = X^\sim, \quad  \varphi_t = {\mathcal A}^X (\varphi), \label{4.4}\\
\left(Y^{\rm ext}\right) \quad v_t = Y^\sim, \quad \psi_t = \Phi 
\left({\mathcal A}^X\right) (\psi), \label{4.5}
\end{gather}
with the extended homomorphism $\Phi^{\rm ext}: C_{u, \varphi} 
\rightarrow C_{v, \psi}$ acting as 
\begin{equation}
\Phi^{\rm ext} (u) = \Phi (u), \quad \Phi^{\rm ext} (\varphi) = \psi. \label{4.6}
\end{equation}
This mechanism, while general, is however not the {\it only} one possible.  
An interesting situation appears when $\Phi$-related fields $X$ and $Y$ 
are 
extended tangently, by linearization (see Section~2): 
\begin{gather}
\left(X^{\rm ext}\right) \quad u_t = X^\sim, \quad  \varphi_t = \frac{DX^\sim}{Du} (\varphi), 
\label{4.7}\\
\left(Y^{\rm ext}\right) \quad v_t = Y^\sim, \quad  \psi_t = \frac{DY^\sim}{Dv} (\psi). 
\label{4.8}
\end{gather}
The general theory [13, 14] then guarantees that the extended fields 
$X^{\rm ext}$ (\ref{4.7}) and $Y^{\rm ext}$ (\ref{4.8}) are related by the extended 
homomorhpisms $\Phi^{\rm ext}: C_{u, \varphi} \rightarrow C_{v, \psi}$, 
acting as 
\begin{equation}
\Phi^{\rm ext} (u) = \Phi(u), \quad \Phi^{\rm ext} (\varphi) = \frac{D \Phi(u)}{Dv} (\psi). \label{4.9}
\end{equation}
(The general theory also guarantees that the map $\Phi^{\rm ext}$ (\ref{4.9}) is 
Hamiltonian provided the map $\Phi: C_u \rightarrow C_v$ is.  However, this 
is not applicable to the alternative ({\it D}) because our form (\ref{1.15}) is not of 
the type furnished by the general linearization theory.)

For example, the KdV-mKdV case (\ref{1.2}), (\ref{4.1}), (\ref{4.2}) yields:
\begin{subequations}\label{4.10}
\begin{gather}
\left(X^{\rm ext}\right) \quad u_t = \left(3u^2 + u_{xx}\right)_x , \quad \varphi_t = (6 u \varphi + 
\varphi_{xx})_x, \label{4.10a}\\
\left(Y^{\rm ext}\right) \quad v_t = \left(-2v^3 + v_{xx}\right)_x, \quad
 \psi_t = \left(- 6v^2 \psi + \psi_{xx}\right)_x, 
\label{4.10b}\\
\Phi^{\rm ext} (u) = - v^2 + \epsilon v_x, \quad  \Phi^{\rm ext} (\varphi) = - 
2v \psi + \epsilon \psi_x, \quad \epsilon = \pm 1. \label{4.10c}
\end{gather}
\end{subequations}

Of course, (\ref{4.5}) and (\ref{4.9}) are not the only kinds of extended homomorphisms 
possible.

It is interesting to observe that the duality involution in general 
destroys 
extended homomorphisms --- unless they are of timid type (\ref{4.5}). 
 The tangent 
point of view, in this sense, is vastly richer than the 
cotangent one --- an inversion of the natural order of Physics.

The importance of extended homomorphisms can not be overestimated, for they 
are the most effective organizing principle in studying and classifying 
extensions.  (See an example of how this works in Section~8.)

\section{Extended Lax equations}

Lax equations are equations of the form
\begin{equation}
L_t = [A, L], \label{5.1}
\end{equation}
where $L$ and $A$ are operators --- in one sense or another.  In the 
differential case [26], $L$ is a matrix ordinary differential 
(or pseudodifferential) operator of the form
\begin{equation}
L = \sum^N u_i \partial^i, \quad \partial = \partial/\partial x, \label{5.2}
\end{equation}
satisfying two conditions:

(i) $u_N$ is a constant invertible diagonalizable matrix; 

(ii)  $u_{N-1} \in \mbox{Im}\; \mbox{ad}(u_N)$.

(For example, when $L$ is scalar, $u_N = 1 $ and $u_{N-1} = 0.)$  The 
general theory [26] then furnishes all possible $A$'s  entering formula 
(\ref{5.1}) as
\begin{equation}
A = P_+, \quad P \in Z^c (L), \label{5.3}
\end{equation}
when $Z^c (L)$ is the ({\it abelian} center of the) centralizer of $L$, 
\begin{equation}
P = \sum^k p_j \partial^j, \label{5.4}
\end{equation}
with $p_k$ being a constant matrix belonging to the (center of the)
 centralizer 
$Z^c(u_N)$ of $u_N$; the $+$ and $-$ notation employed in formula (\ref{5.3}) 
means:
\begin{equation}
\left(\sum_s a_s \partial^s\right)_+ = \sum_{s \geq 0} a_s 
\partial^s; \quad \left(\sum_s a_s \partial^s\right)_- 
= \sum_{s < 0} a_s \partial^s. \label{5.5}
\end{equation}
The main result from the general theory [26] we shall make use of 
below is:

Denote by $X_P$ the evolution derivation (\ref{5.1}) with $A=P_+$: 
\begin{equation}
X_P (L) = [P_+, L], \quad P \in Z^c (L). \label{5.6}
\end{equation}
If $Q$ is another element of $Z^c (L)$, 
\begin{equation}
X_Q (L) = [Q_+, L], \label{5.7}
\end{equation}
then
\begin{equation}
X_P (Q) = [P_+, Q] \label{5.8}
\end{equation}
and
\begin{equation}
[X_P, X_Q] = 0. \label{5.9}
\end{equation}

{\samepage
With minor changes, similar results apply to discrete 
Lax equations [8, 11]. 
 Here $L$ is a scalar operator
\begin{equation}
L = \sum^N_{- \infty} u_i \triangle^i, \quad u_N =1; \label{5.10}
\end{equation}
$\triangle$ is the (dual) shift acting on functions of $n \in {\mathbb Z}$:
\begin{gather}
\left(\triangle^s (f)\right) (n) = f(n+s), \quad n, s \in {\mathbb Z}; \label{5.11}\\
P = L^m, \quad m \in {\mathbb N}; \nonumber\\
\left(\sum_s a_s \triangle^s\right)_+ = \sum_{s \geq 0} a_s \triangle^s, \quad
\left(\sum_s a_s \triangle^s\right)_- = \sum_{s <0} a_s \triangle^s.
 \label{5.12}
\end{gather}
All other formulae (\ref{5.1})--(\ref{5.9}) remain unchanged.}

Now, let $P$ run over the generators of $Z^c(L)$, and let $X_P$ be, 
for each $P$, the corresponding evolution derivation (\ref{5.6}).  We are, 
therefore, in the alternative ({\it B}) of Section~1.  We define the following 
linear extension of the derivations $X_P$:
\begin{subequations}\label{5.13}
\begin{gather}
L_t = [P_+, L], \label{5.13a}\\
\varphi_t = P_+ (\varphi).  \label{5.13b}
\end{gather}
\end{subequations}

\setcounter{proposition}{13}
\setcounter{equation}{14}
\begin{proposition} \label{proposition:5.14}
All the extended flows $X^{\rm ext}_P$ 
(\ref{5.13}) commute between themselves.
\end{proposition}

\begin{proof} Let $P, Q \in Z^c (L)$.  By formula (\ref{1.13}), we 
need to check that
\begin{equation}
X_P (Q_+) - X_Q (P_+) - [P_+, Q_+] = 0. \label{5.15}
\end{equation}
This is one of the Wilson's formulae [26], and it follows from formula 
(\ref{5.8}) and the following calculation:
\begin{gather*}
X_P (Q_+) - X_Q (P_+) - [P_+, Q_+] = [P_+, Q]_+ - [Q_+, P]_+ - 
[P_+, Q_+] \\
\qquad {}= [P_+, Q_+ + Q_-]_+ + [P_+ + P_-, Q_+]_+ - [P_+, Q_+]\\
\qquad {}= [P_+, Q_-]_+ + [P_+, Q_+]_+ + [P_-, Q_+]_+ \\
\qquad {}= [P_+ + P_-, Q_+ +  Q_-]_+ =  [P, Q]_+ =0, 
\end{gather*}
since $P$ and $Q$ commute, being elements of an {\it{abelian}} center 
of the centralizer $Z^c (L)$.
\end{proof}

\setcounter{remark}{15}
\setcounter{equation}{16}
\begin{remark} \label{remark:5.16}
Let $K = 1 + \cdots$ be the dressing operator 
for $L$, conjugating $L$ into its highest term:
\begin{subequations}\label{5.17}
\begin{gather}
L = K u_N \partial^N K^{-1}, \label{5.17a}\\
L = K \triangle^N K^{-1}, \label{5.17b}
\end{gather}
\end{subequations}
for the differential and difference cases respectively.  The Lax motion 
equations (\ref{5.6}) result from the Wilson motion equations [27] in the 
dressing language:
\begin{equation}
\partial_P (K) = - P_- K. \label{5.18} 
\end{equation}
Then the extended system (\ref{5.13}) results from the extended system
\begin{subequations}\label{5.19}
\begin{gather}
K_t =  - P_-K, \label{5.19a}\\
\varphi_t = P_+ (\varphi). \label{5.19b}
\end{gather}
\end{subequations}
Obviously, all the extended systems (\ref{5.19}) still commute in the larger 
dressing space.
\end{remark}

\setcounter{remark}{19}
\setcounter{equation}{20}
\begin{remark}\label{remark:5.20}
 Set
\begin{equation}
\widetilde{L} = - L^\dagger, \quad \widetilde{ P} = - P^\dagger, 
\label{5.21}
\end{equation}
and apply the duality involution to the system (\ref{5.13}).  We get:
\begin{subequations}\label{5.22}
\begin{gather}
\widetilde{L}_t  = [ \widetilde{P}_+, \widetilde{L}], \label{5.22a}\\
\varphi_t = \widetilde{P}_+ (\varphi),  \label{5.22b}
\end{gather}
\end{subequations}
which is a system of the {\it same type} as (\ref{5.13}), though  not necessarily 
the same system: (\ref{5.13}) and (\ref{5.22}) are dual to each other.
\end{remark}

\setcounter{example}{22}
\setcounter{equation}{23}

\begin{example}\label{example:5.23}
Let 
\begin{equation}
L = \partial^2 + u \label{5.24}
\end{equation}
be the KdV Lax operator.  Then $P = \left(L^{1/2}\right)^{2n+1}$, 
$n \in {\mathbb Z}_+$, and therefore
\begin{equation}
L^\dagger = L, \quad  P^\dagger = - P \ \Rightarrow \ \widetilde{L} = 
- L, \quad  \widetilde{P} = P, \label{5.25}
\end{equation}
and all the extended KdV flows (\ref{5.13}) are {\it self-dual}. 
\end{example}

\setcounter{remark}{25}
\setcounter{equation}{26}
\begin{remark}\label{remark:5.26}
What does the duality involution do to the 
extended dressing system (\ref{5.19})?  Denote by $L_0$ the highest term of $L$
entering the equations (\ref{5.17}), so that
\begin{equation}
L = K L_0 K^{-1}. \label{5.27}
\end{equation}
Taking adjoint, we get
\begin{equation}
\widetilde{L}  = \widetilde{K} \; \widetilde{L}_0 \; \widetilde{K}^{-1}, \label{5.28}
\end{equation}
where
\begin{equation}
\widetilde{K}  = K^{-1\dagger}. \label{5.29}
\end{equation}
Therefore, by formula (\ref{5.19a}), 
\begin{equation}
\widetilde{K}_t = - K^{\dagger-1} \left(K^\dagger\right)_t K^{\dagger-1}
 = - K^{\dagger-1} \left(-K^\dagger P^\dagger_-\right) K^{\dagger-1} = P_-^\dagger 
\widetilde{K} = - \widetilde{P}_- \widetilde{K}. 
\label{5.30}
\end{equation}
Thus, the dual form of (\ref{5.19}) is:
\begin{subequations}\label{5.31}
\begin{gather}
\widetilde{K}_t =  -  \widetilde{P}_- \widetilde{K}, \label{5.31a}\\
\varphi_t = \widetilde{P}_+ (\varphi).  \label{5.31b}
\end{gather}
\end{subequations}
\end{remark}

\setcounter{remark}{31}
\setcounter{equation}{32}

\begin{remark}\label{remark:5.32} The Lax equation (\ref{5.13a}), 
\begin{equation}
L_t = [P_+, L], \label{5.33}
\end{equation}
results from the compatibility analysis of the Lax eigenvalue 
problem
\begin{subequations}\label{5.34}
\begin{equation}
L (\varphi) = \lambda \varphi, \label{5.34a}
\end{equation}
supplemented by the time evolution of the eigenfunction $\varphi$:
\begin{equation}
\varphi_t = P_+ (\varphi). \label{5.34b}
\end{equation}
\end{subequations}
The Lax system (\ref{5.34}):
\begin{equation}
L (\varphi) = \lambda \varphi, \quad \varphi_t = P_+ (\varphi) 
\label{5.35}
\end{equation}
is {\it not the same} as our ``extended Lax system'' (\ref{5.13}):
\begin{equation}
L_t = [P_+, L], \quad \varphi_t = P_+ (\varphi). \label{5.36}
\end{equation}
Our system (\ref{5.36}) is more general, for it does {\it not} assume 
that $\varphi$ is an eigenfunction of anything; in other words, the 
constraint 
\begin{equation}
\left\{ \varphi^{-1} L (\varphi) = \mbox{const}\right \} \label{5.37}
\end{equation}
is compatible with our extended system (\ref{5.36}) but it is not implied 
by that system.
\end{remark}

\section{The first two KdV flows}

Let us now tackle the Problem \ref{problem:1.1}.  We need to make some assumptions 
to make this problem manageable.  We make three such assumptions:
\begin{equation}\label{6.1}\arraycolsep=0em
\begin{array}{ll}
{\rm (a)}  & \mbox{\ \rm $\varphi$ is {\it scalar};}\vspace{1mm}\\
{\rm (b)} & \mbox{\ \rm The extension operators ${\mathcal A}_1$ (\ref{1.4b}) and ${\mathcal A}_2$ (\ref{1.5b}) 
are {\it polynomial} in $u$}\\
& \mbox{\ \rm and $x$-derivatives of $u$;}\vspace{1mm}\\
{\rm (c)}  & \mbox{\ \rm The extended systems preserve homogeneuity of the original 
ones.}
\end{array}
\end{equation}

The homogeneuity count is based on the ranks
\begin{gather}
rk \left(u^{(i)}\right) = i+2, \quad  rk (\partial) =1, \label{6.2}\\
rk (X) = 3, \quad rk (Y) = 5, \label{6.3}
\end{gather}
where $u^{(i)} = \frac{\partial^i u}{\partial x^i}$. 
 Thus, we look for the KdV extensions of the form:
\begin{subequations}\label{6.4}
\begin{gather}
u_t = \partial \left(3 u^2 + u_{xx}\right) \label{6.4a}\\
\varphi_t = \left(\alpha u_x + \beta u \partial + \gamma \partial^3\right) (\varphi), \label{6.4b}
\end{gather}
\end{subequations}
\vspace{-9mm}
\begin{subequations}\label{6.5}
\begin{gather}
u_t = \partial \left(10u^3 + 5u_x^2 + 10uu_{xx} + u^{(4)}\right), \label{6.5a}\\
\varphi_t = \langle\left(au_{xxx} + buu_x\right) + \left(cu_{xx} + du^2\right) \partial + 
eu_x \partial^2 + fu \partial^3 + g \partial^5 \rangle(\varphi), \label{6.5b}
\end{gather}
\end{subequations}
where $\alpha$, $\beta$, $\gamma$, $a$, $b$, $c$, $d$, $e$, $f$, $g$ are unknown
 constants.

We assume, {\it in this Section}, that ${\mathcal A}_1$ and ${\mathcal A}_2$ are 
of maximal order so that $\gamma \not= 0$ and hence $g \not= 0$.  
The case $\{ \gamma = 0 \Rightarrow g = f = e = 0 \}$ is treated 
in Section~9. 

Equating the like-terms in the equation
\begin{equation}
X ({\mathcal A}_2) - Y ({\mathcal A}_1) = [{\mathcal A}_1, {\mathcal A}_2], \label{6.6}
\end{equation}
we arrive at the following relations:
\begin{subequations}\label{6.7}
\begin{gather}
f = \frac{5}{3} \frac{g}{\gamma} \beta, \label{6.7f}\\
e = \frac{5}{3} \frac{g}{\gamma} (\alpha + \beta), \label{6.7e}\\
d = \frac{5}{3} \frac{g}{\gamma^2} \beta \left(1 + \frac{\beta}{3}\right), \label{6.7d}\\
c = \frac{5}{3} \frac{g}{\gamma} \left(\alpha + \beta \frac{2 + \gamma^{-1}}{3}\right), \label{6.7c}\\
b = \frac{10}{3} \frac{g}{\gamma^2} \alpha \left(1 + \frac{\beta}{3}\right), \label{6.7b}\\
a = \frac{5}{3} \frac{g}{\gamma} \alpha \frac{2 + \gamma^{-1}}{3}, \label{6.7a}
\end{gather}
\end{subequations}

\newpage

\vspace*{-8mm}

\begin{subequations}\label{6.8}
\begin{gather}
d = \frac{30 \beta}{12 - \beta}, \label{6.8a}\\
b = \frac{60 \alpha}{12 - \beta}, \label{6.8b}
\end{gather}
\end{subequations}
\begin{equation}
\beta = \left[ \frac{5}{9} g \left(\gamma^{-1} - 1\right) \left(\gamma^{-1} + 2\right) + 
(g - 1) \right] = 0, \label{6.9}
\end{equation}

\vspace{-5mm}

\begin{subequations}\label{6.10}
\begin{gather}
(g - 1) \alpha = (\gamma - 1) a, \label{6.10a}\\
3 (f \alpha - a \beta + b) = \alpha (c + 10), \label{6.10b}\\
\alpha (f - 10) + a(6 - \beta) + b (1 - \gamma) = 0, \label{6.10c}\\
\beta (f - 10) + c (6 - \beta) + 2d (1- \gamma) - 3 \gamma b + 
3 \alpha f = 0, \label{6.10d}\\
18c - 20\beta - 6d \gamma - 9 \gamma \beta + (e + c) \beta + 2 
e \alpha = 0, \label{6.10e}\\
(e - c - f) \alpha + a \beta = 0. \label{6.10f}
\end{gather}
\end{subequations}

We break the analysis of the system (\ref{6.7})--(\ref{6.10}) into four steps, 
as follows.

{\bfseries\itshape Step 1:} $\alpha = \beta = 0$.  Then $a = b = c 
= d = e = f = 0$, and we get the {\it decomposed} systems
\begin{gather}
u_t = \partial \left(3 u^2 + u_{xx}\right), \quad  \varphi_t = \gamma \partial^3 
(\varphi), \label{6.11}\\
u_t = \partial \left(10u^3 + 5u_x^2 + 10uu_{xx} + u^{(4)}\right), \quad 
\partial_t = g \partial^5 (\varphi). \label{6.12}
\end{gather}
Obviously, such decomposed extensions are available for {\it any} 
system of vector fields anywhere.  For the KdV hierarchy, we then 
have
\begin{equation}
\left(X_n^{\rm ext}\right) \quad u_t = X^\sim_n, \quad \varphi_t = \gamma_n \partial
^{2n-1} (\varphi), \quad n \in {\mathbb N}, \quad \gamma_n = {\rm const}, \label{6.13}
\end{equation}
and all these are self-dual extensions.

{\bfseries \itshape Step 2:} $\beta = 0$, $\alpha \not= 0$:  no 
solutions.

{\bfseries \itshape Step 3:} $\alpha = 0$, $\beta \not= 0$:  there are 
three possible solutions:
\begin{equation}
\alpha = 0, \quad \beta= 6, \quad  \gamma = 1: \quad
\varphi_t = \left(6 u \partial + \partial^3\right) (\varphi), \label{6.14}
\end{equation}
and this is dual to the linearized case (\ref{4.10a}); 
\begin{gather}
\alpha = 0, \quad  \beta = 3, \quad \gamma = 1; \quad f = e = c = 5, \quad  d = 10, 
\quad g = 1, \quad a = b = 0: \label{6.15}\\
u_t = \partial \left(3 u^2 + u_{xx}\right), \quad \varphi_t = \left(3 u \partial + 
\partial^3\right) ( \varphi), \label{6.16}
\end{gather}
\vspace{-8mm}
\begin{subequations}\label{6.17}
\begin{gather}
u_t = \partial  \left(10 u^3 + 5u_x^2 + 10 uu_{xx} + u^{(4)}\right), 
\label{6.17a}\\
\varphi_t = \langle\left(5u_{xx} + 10 u^2\right) \partial + 5u_x \partial^2 + 
5u \partial^3 + \partial^5 \rangle (\varphi); \label{6.17b}
\end{gather}
\end{subequations}
\vspace{-9mm}
\begin{gather}
\alpha = 0, \quad  \beta = - 6, \quad \gamma = -2; \quad f = e = 
- 20, \quad d = c = - 10, \nonumber\\
 g = -4, \quad  a = b = 0: \label{6.18}\\
u_t = \partial \left(3 u^2 + u_{xx}\right), \quad \varphi_t = - \left(6u \partial + 2 
\partial^3\right)(\varphi), \label{6.19}
\end{gather}

\vspace*{-10mm}

\begin{subequations}\label{6.20}
\begin{gather}
u_t = \partial \left(10 u^3 + 5u_x^2 + 10 uu_{xx} + u^{(4)}\right), \label{6.20a}\\
\varphi_t = - \langle 10 \left(u_{xx} +u^2\right) \partial + 20 u_x \partial^2 
+ 20 u \partial^3 + 4 \partial^5 \rangle(\varphi). \label{6.20b}
\end{gather}
\end{subequations}

{\bfseries \itshape Step 4:} $\alpha \beta \not= 0$.  There are four 
possible solutions:
\begin{gather}
\alpha = 3, \quad \beta = 6, \quad \gamma = 4; \quad a = 15, \quad
 b = d = 30, \quad c = 50, \nonumber\\ 
e = 60, \quad f = 40, \quad g = 16: \label{6.21}\\
u_t = \partial \left(3u^2 + u_{xx}\right), \quad \varphi_t = \left(3u \partial + 3 
\partial u+ 4 \partial^3\right) (\varphi), \label{6.22}
\end{gather}

\vspace{-8mm}

\begin{subequations}\label{6.23}
\begin{gather}
u_t = \partial \left(10 u^3 + 5 u^2_x + 10uu_{xx} + u^{(4)}\right), \label{6.23a}\\
\varphi_t = \langle (15 u_{xxx} + 30 uu_x) + \left(50u_{xx} + 
 30 u^2\right) \partial + 60 u_x \partial^2 + 40 u \partial^3 + 16 \partial^5\rangle(\varphi), 
\label{6.23b} 
\end{gather}
\end{subequations}
and this is the self-dual Lax form (\ref{5.24}), (\ref{5.25}); 
\begin{gather}
\alpha = \beta = 6, \quad \gamma =1; \quad a = f = 10, \quad b = 60, \nonumber\\ 
c = e = 20, \quad d = 30, \quad g = 1: \label{6.24}
\end{gather}
this is the linearized case (\ref{4.10a}), dual to (\ref{6.14})--(\ref{6.17}):
\begin{gather}
u_t = \partial \left(3u^2 + u_{xx}\right), \quad \varphi_t = \langle\partial \left(6u + 
\partial^2\right) \rangle (\varphi), \label{6.25}\\
u_t = \partial \left(10 u^3 + 5 u_x^2 + 10 uu_{xx} + u^{(4)}\right), 
\label{6.26}\\
\varphi_t = \langle(10 u_{xxx} + 60 uu_x) + \left(20 u_x + 30 u^2\right) 
\partial+ 20 u_x \partial^2 + 10 u \partial^3 + \partial^5 \rangle (\varphi) 
\label{6.27}\nonumber\\
\phantom{\varphi_t }= \langle \partial\left \{ \left(3 0u^2 + 10 u_{xx}\right) + 10 u_x \partial + 10 u
 \partial^2 + \partial^4 \right\} \rangle (\varphi); \label{6.28}\\
\alpha = \beta = 3, \quad \gamma = 1; \quad a = 5, \quad b = 20, \quad 
c = d = e = 10, \quad f = 5, \quad  g = 1: \label{6.29}\\
u_t = \partial \left(3u^2 + u_{xx}\right), \quad  \varphi_t = \langle \partial \left(3u + 
\partial^2\right)\rangle (\varphi), \label{6.30}
\end{gather}
\vspace{-8mm}
\begin{subequations}\label{6.31}
\begin{gather}
u_t = \partial \left(10u^3 + 5u^2_x + 10 uu_{xx} + u^{(4)}\right), \label{6.31a}\\
\varphi_t = \langle (5u_{xxx}  + 20 uu_x) + \left(10 u_{xx} + 10u^2\right) 
\partial + 10 u_x \partial^2 + 5u \partial^3 + \partial^5 \rangle(\varphi) \label{6.31b}\nonumber\\
\phantom{\varphi_t} = \langle  \partial \left\{\left(5 u_{xx} + 10u^2\right) + 5u_x \partial + 5u \partial^2 + 
\partial^4 \right\} \rangle(\varphi), \label{6.31b'}
\end{gather}
\end{subequations}
and this is dual to the case (\ref{6.15})--(\ref{6.17}); 
\begin{gather}
\alpha = \beta = - 6, \quad  \gamma = -2; \quad a = d = - 10, \quad
 b = f = -20, \nonumber\\
c = -30, \quad e = - 40, \quad  g = - 4: \label{6.32}\\
u_t = \partial \left(3u^2 + u_{xx}\right), \quad \varphi_t = \langle- \partial \left(6 u + 
2 \partial^2\right)\rangle(\varphi), \label{6.33}
\end{gather}
\vspace{-8mm}
\begin{subequations}\label{6.34}
\begin{gather}
u_t = \partial \left(10u^3 + 5u_x^2 + 10 uu_{xx} + u^{(4)}\right), \label{6.34a}\\
\varphi_t = - \langle(10 u_{xxx} + 20 uu_x) + \left(30 u_{xx} + 10u^2\right) 
\partial + 40u_x \partial^2 + 20 u \partial^3+ 4 \partial^5 \rangle (\varphi) \label{6.34b} \nonumber\\
\qquad {}  =- \langle\partial \left\{ 10 \left(u_{xx} + u^2\right) + 20 u_x \partial + 20 
u \partial^2 + 4 \partial^4 \right\} \rangle(\varphi), \label{6.34b'}
\end{gather}
\end{subequations}
and this is dual to the case (\ref{6.18})--(\ref{6.20}).

Thus, we have found 8 solutions: the decomposed one (\ref{6.11}), (\ref{6.12}) and 
the Lax one (\ref{6.21})--(\ref{6.23}), both self-dual; the linearized one (\ref{6.24})--(\ref{6.26}) 
and its dual (\ref{6.14}); the {\it strange} one (\ref{6.29})--(\ref{6.31}) and its dual 
(\ref{6.15})--(\ref{6.17}); and the {\it mysterious} one (\ref{6.32})--(\ref{6.34}) 
and its dual (\ref{6.18})--(\ref{6.20}). 

Which ones among these 8 extensions are applicable to the whole 
KdV hierarchy and not just to the two nontrivial KdV flows?

The decomposed extension is certainly applicable, and is given by 
formula (\ref{6.13}).  The Lax extension likewise applies to the whole 
hierarchy, by formulae (\ref{5.13}).  Ditto the linearized extension, by 
the results of Section~2, and its dual, by the results of Section~3.

This leaves us with two mutually dual pairs.  The mysterious extension 
(\ref{6.32})--(\ref{6.34}) appears at the moment
 just that, mysterious.  (It will be explained in Section~17.)  
Let's look closely at the strange 
extension (\ref{6.29})--(\ref{6.31}).  Let us compare the strange and the linearized 
extensions:
\begin{subequations}\label{6.35}
\begin{gather}
u_t = \partial\left(3u^2 + u_{xx} \right) \quad (\mbox{KdV}{}_2), \label{6.35a}\\
\varphi_t = \left(6 \partial u + \partial^3\right) (\varphi) \quad {\rm (linearized)},  \label{6.35b}\\
\varphi_t = \left(3 \partial u + \partial^3\right) (\varphi) \quad {\rm (strange)}, \label{6.35c}
\end{gather}
\end{subequations}
\vspace{-9mm}
\begin{subequations}\label{6.36}
\begin{gather}
u_t = \partial \left(10 u^3 + 5u_x^2 + 10 uu_{xx} + u^{(4)}\right) \quad 
(\mbox{KdV}{}_3), \label{6.36a}\\
\varphi_t = \langle \partial \left\{ \left(30u^2 + 10 u_{xx} \right) + 10 u_{x} \partial 
+ 10 u \partial^2 + \partial^4 \right\} \langle (\varphi) \quad {\rm (linearized)}, 
\label{6.36b}\\
\varphi_t = \langle\partial \left\{\left(10u^2 + 5 u_{xx} \right) + 5u_x \partial + 
5u \partial^2 + \partial^4\right\} \rangle (\varphi) \quad {\rm (strange)}.
\label{6.36c} 
\end{gather}
\end{subequations}
We see that in each of the two cases, the strange operator
 ${\mathcal A}_{\rm strange}$
 can be gotten by the formula 
\begin{equation}
{\mathcal A}_{\rm strange} = \frac{D^I X^\sim}{Du} \label{6.37}
\end{equation}
from the vector field $u_t = X^\sim$, where the {\it integrated}
operator $D^I = \frac{D^I}{Du}$ acts on 
differential (-difference) 
polynomials in $u$ (without constant terms) via the rule 
\begin{equation}
\frac{D^I f}{Du} = \frac{1}{r} \frac{Df}{Du} = \int^1_0 dt \frac{Df}{Du} 
\{tu\}, \quad f \{ \lambda u\} = \lambda^r f \{u\} \quad \forall\; \lambda = \mbox{const}, 
\label{6.38}
\end{equation}
where $r > 0$ is the total $u$-degree of a homogeneous polynomial 
$f$; for $f$ non-homogeneous, the definition (\ref{6.38}) is extended by 
additivity.  It is very likely that formulae
\begin{equation}
u_t = X^\sim_n, \quad  \varphi_t = \frac{D^f X^\sim_n}{Du} (\varphi), \quad
 n \in {\mathbb N}, \label{6.39}
\end{equation}
provide strange extensions for the whole KdV hierarchy.
 
In the next two Sections, we shall approach the KdV case from  
different perspectives.

\section{Solutions of the extended KdV equations}

We start with general linearized systems:

\setcounter{equation}{1}
\begin{proposition} \label{proposition:7.1}
Suppose $X$ is an evolution vector field invariant with respect to translations in $x^\alpha$: 
if
\begin{equation}
u_t = X^\sim\label{7.2}
\end{equation}
is the corresponding motion equation then its every solution 
$u = u (\pbf{x}, t)$ generates solutions $u(\pbf{x} + \tau e^\alpha, t)$ $\forall\; \tau$, where 
\begin{equation}
e^\alpha = (0, \ldots, 1, \ldots, 0) \quad (1 \ {\rm at the} \ \alpha^{th} \ {\rm place}). \label{7.3}
\end{equation}
Let 
\begin{equation}
u_t = X^\sim, \quad  \varphi_t = \frac{DX^\sim}{Du} (\varphi) 
\label{7.4}
\end{equation}
be the linearization of $X$.  If $u$ is a solution of (\ref{7.2}) then
\begin{equation}
\left\{u; \ \varphi = \frac{\partial u}{\partial x^\alpha} \right\} \label{7.5}
\end{equation}
is a solution of (\ref{7.4}).
\end{proposition}

\begin{proof}
Translation invariance assures that if 
$u_t = X^\sim \{u\}$ then 
\begin{equation}
\frac{\partial}{\partial t} (u(\pbf{x} + \tau e^\alpha, t))
 = X^\sim  \{u (\pbf{x} + \tau e^\alpha, t) \}, 
\quad \forall \; \tau. \label{7.6}
\end{equation}
Differentiating this with respect to $\tau$ at $\tau = 0$ we get:
\begin{equation}
\frac{\partial}{\partial t} \left(\frac{\partial u}{\partial x^\alpha} 
\right) = \frac{DX^\sim}{Du} \left(\frac{\partial u}{\partial x^\alpha} \right). \label{7.7}
\end{equation}
Thus, (\ref{7.5}) solves (\ref{7.4}).
\end{proof}
 
This is as far as general linearization results go.  
Let's look now at the particular 
case of the KdV equations.

{\samepage
We normalize the KdV flows by requiring the highest $u$-derivative 
in $X_n$,  $u^{(2n-1)}$, enter with coefficient 1:
\begin{equation}
(X_n) \quad u_t = \partial (h_n), \quad h_n = u^{(2n-2)} + \cdots, \quad n \in {\mathbb N}. 
\label{7.8}
\end{equation}
The differential polynomials $h_n$ are in fact variational derivatives 
of some Hamiltonians~$H_n$:
\begin{equation}
h_n = \frac{\delta H_n}{\delta u}, \label{7.9}
\end{equation}
and it is known that
\begin{gather}
\left(\partial^3 + 2u \partial + 2 \partial u\right) (h_n) = \partial 
(h_{n+1}), \quad n \in {\mathbb Z}_+, \label{7.10}\\
h_0 = 1/2, \quad  h_1 = u, \ldots \label{7.11}
\end{gather}
because the KdV equations are bi-Hamiltonian.
}

If we write
\begin{equation}
h_n = \theta_n u^n + \cdots + u^{(2n-2)}, \quad n > 1, \label{7.12}
\end{equation}
then formula (\ref{7.10}) yields
\begin{gather}
4 u \theta_n n u^{n-1} u_x + 2 u_x \theta_n u^n = \theta_{n+1} (n+1) 
u^n u_x \nonumber\\
{} \Rightarrow  \ \theta_{n+1} = \frac{2(2n+1)}{n+1} \theta_n \ \label{7.13}\\
{} \Rightarrow \ \theta_n = \frac{2^{n-1} (2n-1)!!}{n!} , \quad n \in {\mathbb N}.
\label{7.14}
\end{gather}
If we look at stationary solutions of the $n^{th}$ KdV equation, of the 
form
\begin{equation}
u = cx^M, \quad c = \mbox{const}, \label{7.15}
\end{equation}
formula (\ref{7.12}) tells us that (for $n > 1)$
\begin{gather}
Mn = M - (2n-2) \ \label{7.16}\\
{} \Rightarrow \ M = -2. \label{7.17}
\end{gather}
Thus, 
\begin{equation}
u = c/x^2. \label{7.18}
\end{equation}
To find all possible values of $c$, set
\begin{gather}
h_n \left\{ \frac{c}{x^2}\right\} = h_n \{u\} \Big|_{u=c/x^2} = \frac{p_n}{x^{2n}},
 \quad p_n = p_n (c), \label{7.19}\\
p_1 = c. \label{7.20}
\end{gather}
Using formula (\ref{7.10}), we find
\begin{gather}
\partial \left(\frac{p_{n+1}}{x^{2n+2}} \right) = - (2n+2) 
\frac{p_{n+1}}{x^{2n+3}} = \left(\partial^3 + 4 \frac{c}{x^2} 
\partial +2 \left(\frac{c}{x^2} \right)_x \right) \left(\frac{p_n}{x^{2n}} \right)\nonumber\\
\qquad {} =  \frac{p_n}{x^{2n+3}} \left\{ - 2n (2n+1) (2n+2) - 4c\cdot2n - 4c \right\} 
\nonumber\\
\qquad {}= - \frac{(2n+1) p_n}{x^{2n+3}} 4 \{ c + n (n+1) \} 
\nonumber\\
 {} \Rightarrow \ p_{n+1} = \frac{2 (2n+1)}{n+2} \{c + n (n+1)\} p_n \label{7.21}\\
 {} \Rightarrow \ p_n (c) = \frac{2^{n-1} (2n-1)!!}{n!} \prod^{n-1}_{\ell=0} 
\{c + \ell (\ell +1) \}. \label{7.22}
\end{gather}
Thus, all stationary solutions of the $n^{th}$ KdV equation of the 
form $u = c/x^2$ have the $c$-values 
\begin{equation}
c=- \ell (\ell+1), \quad \ell = 0, \ldots, n - 1. 
\label{7.23}
\end{equation}

By Proposition \ref{proposition:7.1}, the corresponding linearized flows are
satisfied by 
\begin{equation}
\varphi = \mbox{const}/x^3, \quad n > 1. \label{7.24}
\end{equation}
By Corollary \ref{corollary:7.35} below, the dual to linearized flows are all 
satisfied by
\begin{equation}
\varphi = \mbox{const}/x^2, \quad n > 1. \label{7.25}
\end{equation}

\setcounter{proposition}{25}
\setcounter{equation}{26}

\begin{proposition}\label{proposition:7.26}
Suppose a Hamiltonian system
\begin{equation}
u_t = \partial (h), \quad h = \frac{\delta H}{\delta u}, 
\label{7.27}
\end{equation}
has a stationary solution $u = f(x)$ such that
\begin{equation}
\left.\frac{\delta H}{\delta u} \right|_{u = f(x)} = 0, \label{7.28}
\end{equation}
and the corresponding linearized system
\begin{equation}
u_t = \partial (h), \quad \varphi_t = \partial \frac{Dh}{Du} (\varphi), \label{7.29}
\end{equation}
has a stationary solution $\{u = f(x)$, $\varphi = \omega (x)\}$ such 
that 
\begin{equation}
\left.\frac{Dh}{Du}\right|_{u = f(x)} (\omega(x)) = 0. \label{7.30}
\end{equation}
Then the dual to linearized system (\ref{7.29}): 
\begin{equation}
u_t = \partial (h), \quad \varphi_t = \left(\frac{D h}{Du}\right)
^\dagger \partial (\varphi) \label{7.31}
\end{equation}
has a stationary solution $\{u = f(x), \ \varphi = \Omega (x) \}$, 
where
\begin{equation}
\Omega (x) = \int^x \omega(\zeta) d \zeta. \label{7.32}
\end{equation}
\end{proposition}

\begin{proof} By the basic property of variational 
derivatives [20],
\begin{equation}
D \left(\frac{\delta H}{\delta u} \right)^\dagger = D \left( 
\frac{\delta H}{\delta u} \right), \label{7.33}
\end{equation}
so that
\begin{equation}
\left(\frac{Dh}{Du} \right)^\dagger = \frac{Dh}{Du}. 
\label{7.34}
\end{equation}
Hence, for the RHS of the $\varphi$-equation (\ref{7.31}) we get
\begin{equation*}
\left.\left(\frac{D h}{Du}\right)^\dagger \right|_{u = f(x)} \partial 
(\Omega (x)) = \left.\frac{Dh}{Du}\right|_{u = f (x)} (\omega (x)) \ 
\overset{\rm [by \ (\ref{7.30})]}{=} \ 0. \tag*{\qed}
\end{equation*}
\renewcommand{\qed}{}
\end{proof}

\setcounter{corollary}{34}
\setcounter{equation}{35}

\begin{corollary} \label{corollary:7.35}
 If $u = f(x)$ is a stationary solution of (\ref{7.27}) and $H$ is $x$-independent then $\{u = f (x), 
\ \varphi = {\rm const}\;  f (x)\}$ solves the dual to linearized system (\ref{7.31}). 
\end{corollary}

\begin{proof}
By Proposition \ref{proposition:7.1}, we can take $\omega (x) = 
f^\prime (x)$.
\end{proof}

By formula (\ref{7.23}) with $\ell =1$, 
\begin{equation}
u = - 2/x^2 \label{7.36}
\end{equation}
is a stationary solution for each KdV flow $\#n$ with $n>1$.  Let 
us see what stationary solutions of the form $\varphi = \mbox{const} \; x^N$ 
look like for the 7~extended KdV$_2$ flows of the preceding Section 
(from the total of~8, with the decomposed extension deleted).

The linearized extension (\ref{4.10a})
\begin{equation}
\partial \left(6 u + \partial^2\right) (\varphi) = 0: \quad N = - 3,4; 
\label{7.37} 
\end{equation}

The dual to the linearized extension (\ref{6.14}) 
\begin{equation}
\left(6 u + \partial^2\right) \partial (\varphi) = 0: \quad N = - 2,5, 
\label{7.38}
\end{equation}
in agreement with Corollary \ref{corollary:7.35}; 

The self-dual Lax extension (\ref{6.22}):
\begin{equation}
\left(3 u \partial + 3 \partial u + 4 \partial^3\right) (\varphi) = 0: \quad
N = -1, 1, 3; \label{7.39}
\end{equation}

The strange extension (\ref{6.30}):
\begin{equation}
\partial \left(3u + \partial^2\right) (\varphi) = 0: \quad N = -2, 3; 
\label{7.40}
\end{equation}

The dual to the strange extension (\ref{6.16}):
\begin{equation}
\left(3 u + \partial^2\right) \partial (\varphi) = 0: \quad N = -1, 4; 
\label{7.41}
\end{equation}

The mysterious extension (\ref{6.33}):
\begin{equation}
- \partial \left(6u + 2 \partial^2\right) (\varphi) = 0: \quad N = -2, 3; 
\label{7.42}
\end{equation}

The dual to the mysterious extension (\ref{6.19}): 
\begin{equation}
- \left(6 u + 2 \partial^2\right) \partial (\varphi) = 0: \quad N = -1, 4. 
\label{7.43}
\end{equation}

We see that the sets of strange and mysterious exponents are 
identical, an artifact of the 2$^{nd}$ extended KdV flow where 
\begin{equation}
X^{\rm ext}_{\rm myst} (\varphi) = - 2 X^{\rm ext}_{\rm strange} (\varphi). 
\label{7.44}
\end{equation}
No similar simple relation exists for the 3$^{rd}$ extended 
KdV flow; but a more complex one may.

\setcounter{remark}{44}
\setcounter{equation}{45}

\begin{remark}\label{remark:7.45}
 Proposition \ref{proposition:7.26} and Corollary \ref{corollary:7.35} 
can be considerably generalized, as follows.

Let
\begin{equation}
{\pbf{q}}_t = B \left(\frac{\delta H}{\delta \pbf{q}} \right) \label{7.46}
\end{equation}
be a Hamiltonian system whose Hamiltonian matrix $B$ is $q$-independent.  
Let
\begin{equation}
{\pbf{q}}_t = B \left(\frac{\delta H}{\delta \pbf{q}} \right), \quad 
{\pbf{v}}_t = B \left[\frac{D}{D \pbf{q}} \left(\frac{\delta H}{\delta \pbf{q}} 
\right) \right] (\pbf{v}) \label{7.47}
\end{equation}
and
\begin{equation}
{\pbf{q}}_t = B \left(\frac{\delta H}{\delta \pbf{q}} \right), \quad 
{\pbf{p}}_t = \left[\frac{D}{D \pbf{q}} \left(\frac{\delta H}{\delta \pbf{q}}
\right) \right] B (\pbf{p}) \label{7.48}
\end{equation}
be the corresponding (tangent and cotangent) linearized extension 
and its dual.
\end{remark}

\setcounter{proposition}{48}
\setcounter{equation}{49}

\begin{proposition}\label{proposition:7.49}
 (i) If 
\begin{equation}
\{ {\pbf{q}} = {\pbf{f}}, \ {\pbf{p}} = {\pbf{g}} \} \label{7.50}
\end{equation}
is a solution of the cotangent system (\ref{7.48}) then
\begin{equation}
\{ {\pbf{q}} = {\pbf{f}}, \ {\pbf{v}} = B ({\pbf{g}}) \} \label{7.51}
\end{equation}
is a solution of the tangent system (\ref{7.47}); 

(ii) The map $\Phi$: 
\begin{equation}
{\pbf{q}} = {\pbf{q}}, \quad  {\pbf{v}} = B ({\pbf{p}}) \label{7.52}
\end{equation}
is a {\it Hamiltonian} map between the canonical Hamiltonian structure 
\begin{equation}
J = \left(\begin{array}{cc}
{\bf{0}} & {\bf{1}} \\
- {\bf{1}} & {\bf{0}} \end{array} \right) \label{7.53}
\end{equation}
of the cotangent system (\ref{7.48}) and the linearized Hamiltonian 
structure [14, p.~192] 
\begin{equation}
{\mathcal B} = \left(\begin{array}{cc} 0 & -B \\
-B & 0 \end{array} \right) \label{7.54}
\end{equation}
of the tangent system (\ref{7.47}). 
\end{proposition}

\begin{proof} (i) Apply the operator $B$ to the 2$^{nd}$ 
equation in (\ref{7.48}).  This results in the 2$^{nd}$ equation in (\ref{7.47}); 

(ii) The Jacobian of the map $\Phi$ (\ref{7.52}) is
\begin{equation}
Jac = \left(\begin{array}{cc}
{\bf{1}} & {\bf{0}} \\
{\bf{0}} & B\end{array}  \right) . \label{7.55}
\end{equation}
Therefore, 
\begin{gather}
Jac \cdot J \cdot (Jac)^\dagger = \left(\begin{array}{cc}
{\bf{1}} & {\bf{0}} \\
{\bf{0}} & B \end{array} \right) 
\left(\begin{array}{cc} {\bf{0}} & {\bf{1}} \\
-{\bf{1}} & {\bf{0}} \end{array} \right) 
\left(\begin{array}{cc} {\bf{1}} & {\bf{0}} \\
{\bf{0}} & -B \end{array} \right) \nonumber\\
\phantom{Jac \cdot J \cdot (Jac)^\dagger}=
\left(\begin{array}{cc} {\bf{0}} & -B \\
-B & {\bf{0}} \end{array} \right) = {\mathcal B}. \label{7.56}
\end{gather}
On the other hand, the corresponding Hamiltonians are also related 
by the map $\Phi$ (\ref{7.52}):  the linearized Hamiltonian [14, p.~182] is
\begin{equation}
-\frac{D H}{D {\pbf{q}}} ({\pbf{v}}) \sim - {\pbf{v}} \cdot \frac{\delta H}{\delta {\pbf{q}}} 
= - B ({\pbf{p}}) \cdot \frac{\delta H}{\delta {\pbf{q}}} \sim - 
{\pbf{p}} \cdot B^\dagger \left(\frac{\delta H}{\delta {\pbf{q}}} \right) = 
{\pbf{p}} \cdot \pbf{X}, \label{7.57}
\end{equation}
where
\begin{equation}
\pbf{X} = B \left(\frac{\delta H}{\delta {\pbf{q}}} \right) \quad (= {\pbf{q}}_t)
\label{7.58}
\end{equation}
is the original evolution vector field (\ref{7.46}).
\end{proof}

\setcounter{proposition}{58}
\setcounter{equation}{59}

\begin{proposition} \label{proposition:7.59}
Let
\begin{equation}
u_t = \partial (h), \quad  h = \frac{\delta H}{\delta u}, \label{7.60}
\end{equation}
be a Hamiltonian system with the Hamiltonian $H$ that is 
$x$-independent.  If 
\begin{equation}
u = f (x, t) \label{7.61}
\end{equation}
is a solution of (\ref{7.60}) then 
\begin{equation}
\{ u = f, \ p = {\rm const} \; f\} \label{7.62}
\end{equation}
is a solution of the cotangent system
\begin{equation}
u_t = \partial (h),  \quad p_t = \frac{Dh}{Du} \partial (p). 
\label{7.63}
\end{equation}
\end{proposition}

\begin{proof}
Since $h$ is $x$-independent, 
\begin{equation}
u_t = \partial (h) = \frac{Dh}{Du} (u_x) = \frac{Dh}{Du} \partial (u), \label{7.64}
\end{equation}
so that
\begin{equation}
f_t = \left.\frac{Dh}{Du}\right|_{u=f} \partial (f), \label{7.65}
\end{equation}
and this is the form of the $p$-equation in (\ref{7.63}).
\end{proof}

The conclusion (\ref{7.62}) applies also to some non-Hamiltonian systems:

\setcounter{proposition}{65}
\setcounter{equation}{66}

\begin{proposition}\label{proposition:7.66}
 Let
\begin{equation}
u_t = X^\sim: \quad  u_{i,t} = X^\sim_i , \quad  i=1, \ldots \label{7.67}
\end{equation}
be a dynamical system where each $X_i^\sim$ is a 
{\it quasipolynomial}, i.e., a sum of quasimonomials, the latter 
being products of the terms of the form
\begin{equation}
\left(u_s^{(\sigma)} \right)^{\pm 1}; \label{7.68}
\end{equation}
in other words, of terms of {\it positive and negative} degrees.  
Assume that no $X^\sim_i$ contains terms of total degree zero, and 
consider the following strange extension of (\ref{7.67}):
\begin{subequations}\label{7.69}
\begin{gather}
u_t = X^\sim, \quad \varphi_t = \frac{D^I X^\sim}{Du} (\varphi): \label{7.69a}\\
u_{i,t} = X^\sim_i, \quad  \varphi_{i,t} = \sum_j \frac{D^I X^\sim_i}{Du_j} (\varphi_j) , \label{7.69b}
\end{gather}
\end{subequations}
where
\begin{equation}
\frac{D^I f}{Du_j} = \frac{1}{{\rm deg}\, (f) } \frac{Df}{Du_j} 
\label{7.70}
\end{equation}
for any quasihomogeneous $f$ of total degree ${\rm deg}\, (f) \not= 0$. 

If $u = f (\pbf{x}, t)$ solves (\ref{7.67}) then
\begin{equation}
\{u = f (\pbf{x},t), \ \varphi = \mbox{\rm const}\; f(\pbf{x}, t) \} \label{7.71}
\end{equation}
solves (\ref{7.69}).
\end{proposition}

\begin{proof} The statement (\ref{7.71}) amounts to the identify
\begin{equation}
\sum_j \frac{D^I f}{Du_j} (u_j) = f \label{7.72}
\end{equation}
for any quasihomogeneous $f$ of ${\rm deg}\, (f) \not= 0$.

Now,
\begin{subequations}\label{7.73}
\begin{gather}
f = \left(u_s^{(\sigma)}\right)^\ell, \quad 0 \not= \ell \in {\mathbb Z}\label{7.73a}\\
 {} \Rightarrow  \ \frac{D^I f}{Du_j} = \delta^s_j \left(u_s^{(\sigma)}\right)^{\ell-1} 
\partial^\sigma \ \Rightarrow \ \sum_j \frac{D^I f}{Du_j} 
(u_j) = f. \label{7.73b}
\end{gather}
\end{subequations}
Finally, if $f$ and $g$ are quasihomogeneous, with $0 \not= {\rm deg} \, (f)$, 
${\rm deg}\, (g)$, ${\rm deg}\, (f) + {\rm deg}\, (g)$, and formula (\ref{7.72}) holds true 
for both $f$ and $g$, then for $h = fg$ we find
\begin{gather*}
\frac{D^I h}{Du_j} = \frac{1}{{\rm deg}\,  (h)} \frac{Dh}{Du_j} = 
\frac{1}{{\rm deg}\, (h)} \frac{D(fg)}{Du_j} = \frac{1}{{\rm deg}\, (h)} 
\left(f \frac{Dg}{Du_j} + g \frac{Df}{Du_j} \right)\\
\qquad {} = \frac{1}{{\rm deg}\, (h)} \left({\rm deg}\, (g) f \frac{D^I g}{Du_j} 
+ {\rm deg}\,(f) g \frac{D^I f}{Du_j} \right)\\
{} \Rightarrow \ \sum_j \frac{D^I h}{Du_j} (u_j) =
\frac{1}{{\rm deg}\, (h)}  [{\rm deg}\, (g) f g+ {\rm deg}\,(f) g f] = h.\tag*{\qed}
  \end{gather*}
  \renewcommand{\qed}{}
\end{proof}

\setcounter{corollary}{73}
\setcounter{equation}{74}

\begin{corollary} \label{corollary:7.74} If $f$ is quasihomogeneous then 
\begin{equation}
\sum_j \frac{Df}{Du_j} (u_j) = {\rm deg}\, (f) f, \quad {\rm deg}\, (f) \in {\mathbb Z}.\label{7.75}
\end{equation}
\end{corollary}

\begin{proof} If ${\rm deg}\, (f) \not= 0$, this is just a 
reformulation of formula (\ref{7.72}).  If ${\rm deg}\,(f) = 0$, we 
decompose $f$ as $f = gh$, with ${\rm deg}\, (g) \,{\rm deg}\, (h) \not= 0$, 
and then use the derivation property of the Fr\'echet 
derivative operator $\frac{D}{Du_j}$.
\end{proof}

It seems likely that, for every fixed scalar differential Lax 
operator $L$, 
\begin{equation}
L = \partial^n + \sum^{n-1}_{i=1} u_j \partial^{n-1-i}, 
\label{7.76}
\end{equation}
all the commuting Lax flows
\begin{equation}
L_t = [P_+, L], \quad P \in Z (L), \label{7.77}
\end{equation}
still commute after being strangely extended.

\section{Miura maps for extended KdV equations}

The KdV equation
\begin{equation}
u_t = \left(3u^2 + u_{xx}\right)_x \label{8.1}
\end{equation}
and the mKdV equation
\begin{equation}
v_t = \left(-2v^3 + v_{xx}\right)_x \label{8.2}
\end{equation}
are related by the classical Miura map
\begin{equation}
u = -v^2 + \epsilon v_x, \quad \epsilon = \pm1. \label{8.3}
\end{equation}
We saw in Section~4 that the linearized equations
\begin{gather}
u_t = \partial \left(3u^2 + u_{xx}\right), \quad  \varphi_t = \partial \left(6u + 
\partial^2\right) (\varphi), \label{8.4}\\
v_t = \partial \left(-2v^3 + v_{xx}\right), \quad \psi_t = \partial \left(-6v^2 + 
\partial^2\right) (\psi) \label{8.5}
\end{gather}
are related by the linearized Miura map (\ref{4.10c}):
\begin{equation}
u = - v^2 + \epsilon v_x, \quad  \varphi = - 2v \psi + \epsilon \psi_x, 
\quad \epsilon = \pm 1. \label{8.6}
\end{equation}

Let us see which general extensions of the KdV 
\begin{equation}
u_t = \partial \left(3u^2 + u_{xx}\right), \quad
\varphi_t = \left(\alpha u_x + \beta u \partial + \gamma \partial^3\right) (\varphi) \label{8.7}
\end{equation}
and of the mKdV
\begin{subequations}\label{8.8}
\begin{gather}
v_t = \partial \left(-2v^3 + v_{xx}\right), \label{8.8a}\\
\psi_t = \langle \left(av_{xx} + b vv_x + cv^3\right) + \left(d v_x + ev^2\right) \partial + 
f v \partial^2 + \gamma \partial^3 \rangle (\psi) \label{8.8b}
\end{gather}
\end{subequations}
are related by a generalized Miura map which, from dimensional 
considerations (assuming it's polynomial), must have the form
\begin{equation}
u = - v^2 + \epsilon v_x, \quad \varphi = \epsilon \psi_x + 
\theta v \psi = (\theta v + \epsilon \partial) (\psi). \label{8.9}
\end{equation}
Here $\alpha$, $\beta$, $\gamma$, $a$, $b$, $c$, $d$, $e$, $f$, $\theta$ are unknown 
constants, with $\gamma \not= 0$.

Calculating $\varphi_t$ in two different ways, from (\ref{8.7}) as 
\begin{subequations}\label{8.10}
\begin{equation}
\varphi_t = \left\{\alpha (-2v v_x + \epsilon v_{xx}) + \beta \left(- v^2 + 
\epsilon v_x\right) \partial + \gamma \partial^3 \right\} (\theta v + \epsilon 
\partial) (\psi), \label{8.10l} 
\end{equation}
and from (\ref{8.9}) as 
\begin{gather}
\varphi_t = \left\{\theta \left(-2v^3 + v_{xx}\right)_x+ (\theta v + \epsilon \partial) 
\langle \left(av_{xx} + bvv_{x} + cv^3\right) \right.\nonumber\\
\phantom{\varphi_t={}} \left.{} + \left(dv_x + ev^2\right) \partial + 
 fv \partial^2 + \gamma \partial^3 \rangle\right\} (\psi), \label{8.10r}
\end{gather}
\end{subequations}
and equating the expressions (\ref{8.10}), we find the following relations:
\begin{subequations}\label{8.11}
\begin{gather}
f=0, \label{8.11f}\\ 
e = - \beta, \label{8.11e}\\
d = \epsilon (\beta + 3 \gamma \theta), \label{8.11d}\\
c = 0, \label{8.11c}\\
b=2 (\beta - \alpha) - 3 \gamma \theta^2, \label{8.11b}\\
a = \epsilon (\alpha - \beta); \label{8.11a}
\end{gather}
\end{subequations}
\vspace{-10mm}
\begin{subequations}\label{8.12}
\begin{gather}
\theta (\gamma \theta^2 - \beta + 2) = 0, \label{8.12a}\\
\theta \beta = 2 (\beta - \alpha) - 3 \gamma \theta^2, \label{8.12b}\\
\theta (\gamma - 1) = \alpha - \beta, \label{8.12c}
\end{gather}
\end{subequations}
If $\theta = 0$, the relations (\ref{8.12}) collapse into 
\begin{subequations}\label{8.13}
\begin{equation}
\alpha = \beta = {\rm arbitrary}, \label{8.13a}
\end{equation}
and this case is not interesting, for all it says is that 
$\varphi$ is a conserved density and one can introduce the 
potential $\psi$ such that
\begin{equation}
\varphi = \epsilon \psi_x. \label{8.13b}
\end{equation}
\end{subequations}

So, suppose
\begin{equation}
\theta \not= 0. \label{8.14}
\end{equation}
Then (\ref{8.12a}) and (\ref{8.12c}) can be considered as providing $\beta$ 
and $\alpha$ in terms of $\gamma$, $\theta$, while (\ref{8.12b}) is the 
desired relation on $\gamma$ and $\theta$:
\begin{gather}
\beta = 2 + \gamma \theta^2, \quad \alpha = 2 + \gamma \theta^2 + 
\theta (\gamma - 1), \label{8.15}\\
0 = - \theta \beta + 2 (\beta - \alpha) - 3 \gamma \theta^2 = 
\theta \{ -2 - \gamma \theta^2 + 2 (1- \gamma) - 3 \gamma \theta\}\nonumber\\ 
\qquad {} = - \theta \gamma (\theta^2 + 3 \theta +2) = 
- \theta \gamma (\theta + 1) (\theta + 2). \label{8.16}
\end{gather}
Since $\theta, \gamma \not= 0$, we obtain
\begin{equation}
\theta = - 1, -2. \label{8.17}
\end{equation}
Thus, 
\begin{equation}
\varphi = (\epsilon \partial + \theta v) (\psi) = 
\left(\begin{array}{c} D^Iu / Dv \\ Du / Dv\end{array} \right)  (\psi) \quad  
{\rm for} \quad  \theta = 
\left(\begin{array}{c} -1 \\ -2\end{array} \right), \label{8.18}
\end{equation}
an interesting and unexpected result.

Summarizing, we get 
\begin{gather}
\theta = \left( \begin{array}{c} -1 \\ -2\end{array}\right), \quad \alpha = \left( \begin{array}{c} 
3 \\ 
4 + 2 \gamma\end{array} \right), \quad \beta = \left(\begin{array}{c} 2 + \gamma \\ 2 + 4 
\gamma\end{array}\right), \label{8.19}\\
e = - \left(\begin{array}{c} 2 + \gamma \\ 2 + 4 \gamma\end{array} \right), \quad d = 2 
\epsilon (1- \gamma), \quad b = \left(\begin{array}{c}-2 -\gamma \\ -4 - 8 \gamma\end{array} 
\right), \nonumber\\
 a = \epsilon \left(\begin{array}{c} 1 - 2 \\ 2 - 2 \gamma\end{array} 
\right), \label{8.20}
\end{gather}
so that formulae (\ref{8.7})--(\ref{8.9}) become
\begin{subequations}\label{8.21}
\begin{gather}
\varphi_t = \langle 3 u_x + (2 + \gamma) u \partial + \gamma \partial^3 \rangle 
(\varphi), \quad \varphi = (\epsilon \partial - v) (\psi), \label{8.21a}\\
\psi_t = \langle (\epsilon (1 - \gamma) v_{xx} - (2 + \gamma) vv_x ) 
+ \left(2 \epsilon (1-\gamma) v_x - (2 + \gamma) v^2\right) \partial + \gamma 
\partial^3 \rangle (\psi), \label{8.21b}
\end{gather}
\end{subequations}
\vspace{-9mm}
\begin{subequations}\label{8.22}
\begin{gather}
\varphi_t = \langle  (4 + 2 \gamma) u_x + (2 + 4 \gamma) u \partial + 
\gamma \partial^3 \rangle (\varphi), \quad \varphi = (\epsilon \partial - 
2 v) (\psi), \label{8.22a}\\
\psi_t = \langle (2 \epsilon (1 - \gamma) v_{xx} - (4 + 8 \gamma) vv_x)\nonumber\\
\phantom{psi_t=}+ \left(2 \epsilon (1- \gamma) v_x - (2 + 4 \gamma) v^2\right) \partial + \gamma 
\partial^3\rangle (\psi). \label{8.22b}
\end{gather}
\end{subequations}

Let us determine which ones of the 7 KdV extensions of Section~6 come out 
of the corresponding mKdV ones.

From formulae (\ref{8.19}) we see that:
\begin{equation}
\alpha = 0 \ \Leftrightarrow  \ \theta = - 2, \ \gamma = -2 \
\Rightarrow \ \beta = - 6, \label{8.23}
\end{equation}
and this is the dual to mysterious extension (\ref{6.18})--(\ref{6.20}); 
\begin{equation}
\theta = -1, \quad \gamma = 4, \quad \alpha = 3, \quad \beta = 6, 
\label{8.24}
\end{equation}
and this is the self-dual Lax extension (\ref{6.21})--(\ref{6.23}); 
\begin{equation}
\theta = -2, \quad \gamma = 1, \quad  \alpha = \beta = 6, \label{8.25}
\end{equation}
and this is the expected linearized case (\ref{8.4})--(\ref{8.6}); 
\begin{equation}
\theta = -1, \quad  \gamma = 1, \quad \alpha = \beta = 3, \label{8.26}
\end{equation}
and this is the strange extension (\ref{6.29})--(\ref{6.31}), with
\begin{subequations}\label{8.27}
\begin{gather}
\varphi_t = \frac{D^I u_t}{Du } (\varphi), \quad \varphi = \frac{D^I u}{Dv}(\psi), \label{8.27a}\\
\psi_t = \langle - 3 v v_x - 3 v^2 \partial + \partial^3 \rangle (\psi) 
\not=  \frac{D^I \left(- 6 v^2 v_x + v_{xxx}\right)}{Dv} (\psi). \label{8.27b}
\end{gather}
\end{subequations}
The latter formula shows that the strange extensions of the 
higher mKdV flows, if they exist, are not given in terms of the 
intergrated operator $\frac{D^I}{Dv}$.

The Lax case (\ref{8.24}) deserves some comment. The formula
\begin{equation}
\varphi = (\epsilon \partial - v)(\psi) \label{8.28}
\end{equation}
is certainly unexpected, for Lax equations and modified Lax 
equations have essentially the same eigenfunctions (see [29]).
Now, with the KdV Lax operator
\begin{equation}
L = \partial^2 + u, \label{8.29}
\end{equation}
$\varphi$ satisfies the eigenvalue problem
\begin{equation}
L (\varphi) = \lambda \varphi, \label{8.30}
\end{equation}
which becomes, in terms of $\psi$ (\ref{8.28}):
\begin{equation}
L (\epsilon \partial - v) (\psi) = \lambda (\epsilon \partial - v) (\psi). \label{8.31}
\end{equation}
Recall [19] that the Miura map (\ref{8.3}) 
\begin{equation}
u = - v^2 + \epsilon v_x \label{8.32}
\end{equation}
comes out of factorization of the Lax operator $\partial^2 + u$:
\begin{equation}
\partial^2 + u = (\partial - \epsilon v) (\partial + \epsilon v). 
\label{8.33}
\end{equation}
Hence, the equation (\ref{8.31}) becomes
\begin{equation}
(\partial - \epsilon v) (\partial + \epsilon v) \epsilon (\partial 
- \epsilon v) ( \psi) = \lambda \epsilon (\partial - e v) (\psi), 
\label{8.34}
\end{equation}
or
\begin{equation}
(\partial + \epsilon v) (\partial - \epsilon v) (\psi) = \lambda 
\psi, \label{8.35}
\end{equation}
which can be rewritten as
\begin{equation}
\widetilde{L}(\psi) = \lambda \psi, \label{8.36}
\end{equation}
where $\widetilde{L}$ is the B\"acklund 
transform of $L$:
\begin{equation}
\widetilde{L} = \partial^2 + \widetilde{u}, \quad \widetilde{u} = - v^2 - \epsilon v_x. \label{8.37}
\end{equation}
The time evolution of $\psi$ (\ref{8.21}) is 
\begin{gather}
\psi_t = \langle 3\left(- \epsilon v_x - v^2\right)_x + 6 \left(- \epsilon v_x - 
v^2\right) \partial + 4 \partial^3\rangle (\psi)\nonumber\\
\phantom{\psi_t} {}= \left(3 \widetilde{u} \partial + 3 \partial \widetilde{u} 
+ 4 \partial^3\right) (\psi) = \widetilde{P}_+ (\psi), \label{8.38}
\end{gather}
with
\begin{equation}
\widetilde{P} = 4 \widetilde{L}^{3/2} . \label{8.39}
\end{equation}
Thus, every Lax extension
\begin{equation}
L_t = [P_+, L], \quad \varphi_t = P_+ (\varphi), \quad  P \in Z(L), 
\label{8.40}
\end{equation}
comes from the corresponding modified Lax extension 
\begin{equation}
v_t = \ldots, \quad  \psi_t = \widetilde{P}_+ (\psi) \label{8.41}
\end{equation}
under the extended Miura map 
\begin{equation}
u = -v^2 + \epsilon v,  \quad \varphi = (\epsilon \partial 
 - v) (\psi),  \quad \widetilde{L} = - v^2 - \epsilon v_x.\label{8.42}
\end{equation}
Similar conclusion applies to all scalar Lax equations with the Lax 
operator of order $n \geq 2$:
\begin{equation}
L = \partial^n + \sum^{n-2}_{i=0} u_i \partial^i, \label{8.43}
\end{equation}
with the Miura map being (differential) factorization
\begin{equation}
L = (\partial + v_1) \cdots (\partial + v_n), \quad \sum^n_{j=1} 
v_j =0, \label{8.44}
\end{equation}
and with the B\"ackland transformation being effected 
by a cyclic permutation of the roots $v_j$'s:
\begin{gather}
\widetilde{L} = (\partial + v_2) \cdots (\partial + v_n) (\partial 
+ v_1), \label{8.45}\\
\varphi = (\partial + v_1) (\psi), \label{8.46}\\
\varphi_t = P_+ (\varphi) , \quad \psi_t = \widetilde{\! P}_+ (\psi). \label{8.47}
\end{gather}

\section{KdV and mKdV extensions of order $\pbf{\leq 1}$}

In Section~6, we classified commuting KdV extensions (\ref{6.4}), (\ref{6.5})
\begin{equation}
u_t = \partial \left(3 u^2 + u_{xx}\right), \quad  
\varphi_t = \left(\alpha u_x + \beta u \partial + \gamma \partial^3\right) (\varphi),
 \label{9.1}
\end{equation}
\vspace{-8mm}
\begin{subequations}\label{9.2}
\begin{gather}
u_t = \partial \left(10u^3 + 5u_x^2 + 10 uu_{xx}+u^{(4)}\right), \label{9.2a}\\
\varphi_t = \langle \left(au_{xxx} + buu_x\right) + \left(cu_{xx} + du^2\right) \partial + 
eu_x \partial^2 + fu \partial^3 + g \partial^5 \rangle(\varphi), \label{9.2b}
\end{gather}
\end{subequations}
The assumption there was that $\gamma g \not= 0$.  Let us now consider 
the case
\begin{equation}
\gamma = 0. \label{9.3}
\end{equation}

It's immediate that
\begin{equation}
g = f = e = 0, \label{9.4}
\end{equation}
and then
\begin{subequations}\label{9.5}
\begin{gather}
c = \beta, \quad d = \beta (\beta + 4)/2, \label{9.5a}\\
\beta (\beta - 2) = 0. \label{9.5b}
\end{gather}
\end{subequations}
If
\begin{equation}
\beta = 0 \ \ \Rightarrow \ \  \alpha = 0, \ \ a = b = c = d = 0, \label{9.6}
\end{equation}
and we get the trivial extensions.  So, let
\begin{equation}
\beta = 2 \ \ \Rightarrow \ \ c = 2, \ \ d = 6. \label{9.7}
\end{equation}
Then
\begin{equation}
a = \alpha, \quad b = 6a, \quad \alpha \ \mbox{\rm is arbitrary}, 
\label{9.8}
\end{equation}
and we finally obtain:
\begin{subequations}\label{9.9}
\begin{gather} u_t = \partial \left(3 u^2 + u_{xx}\right),\label{9.9a}\\
\varphi_t = (\alpha u_x + 2u \partial) (\varphi), \label{9.9b}
\end{gather}
\end{subequations}
\vspace{-8mm}
\begin{subequations}\label{9.10}
\begin{gather}
u_t = \partial \left(10u^3 + 5u^2_x + 10uu_{xx} + u^{(4)}\right), \label{9.10a}\\
\varphi_t = \langle \alpha \left(u_{xx} + 3u^2\right)_x + 2\left(u_{xx} + 3u^2\right) \partial 
\rangle (\varphi). \label{9.10b}
\end{gather}
\end{subequations}

We are now going to explain formulae (\ref{9.9}), (\ref{9.10}), and then determine 
similar extensions for the whole KdV hierarchy.

\setcounter{proposition}{10}
\setcounter{equation}{11}

\begin{proposition}\label{proposition:9.11}
 Let $X$ and $Y$ be two commuting 
evolution vector fields, with the motion equations
\begin{subequations}\label{9.12}
\begin{gather}
(X) \quad {\pbf{q}}_t = {\pbf{X}}^\sim, \label{9.12a}\\
(Y) \quad {\pbf{q}}_t = {\pbf{Y}}^\sim. \label{9.12b}
\end{gather}
\end{subequations}
Consider their scalar extensions, of the form
\begin{subequations}\label{9.13}
\begin{gather}
\left(X^{\rm ext}\right) \quad {\pbf{q}}_t = {\pbf{X}}^\sim, \quad  \varphi_t = \left(v + \widehat V\right) 
(\varphi), \label{9.13a}\\
\left(Y^{\rm ext}\right) \quad {\pbf{q}}_t = {\pbf{Y}}^\sim, \quad \varphi_t = \left(w + \widehat W\right) 
(\varphi), \label{9.13b}
\end{gather}
\end{subequations}
where
\begin{equation}
\hat V = \sum_\alpha V^\alpha \frac{\partial}{\partial x^\alpha}, \quad
\widehat W = \sum_\alpha W^\alpha \frac{\partial}{\partial x^\alpha},
\label{9.14}
\end{equation}
are vector fields, and $v$, $w$, $V^\alpha$, $W^\alpha$ are all 
functions of $\{q\}$.

Then the extended systems (\ref{9.13a}) and (\ref{9.13b}) commute iff 
\begin{subequations}\label{9.15}
\begin{gather}
X \left(\widehat W\right) - Y \left(\widehat V\right) = \left[\widehat V, \hat W\right]: 
\label{9.15a}\\
X \left(W^\alpha\right) - Y \left(V^\alpha\right) = \sum_\beta \left(V^\beta \frac{\partial W^\alpha}
{\partial x^\beta} - W^\beta \frac{\partial V^\alpha}{\partial x^\beta}\right), \label{9.15b}
\end{gather}
\end{subequations}
\vspace{-3mm}
\begin{equation} 
X(w) - Y (v) = \widehat V(w) - \widehat W(v). \label{9.16}
\end{equation}
\end{proposition}

\begin{proof} This is just equation (\ref{1.13}):
\begin{equation}
X\left(w+ \widehat W\right) - Y \left(v + \widehat V\right) = \left[v + \widehat V, 
w + \widehat W\right] \label{9.17}
\end{equation}
written out for the summands of order 1 and 0. 
\end{proof}

\setcounter{corollary}{17}
\setcounter{equation}{18}

\begin{corollary}\label{corollary:9.18}
 (i) When $\widehat V$ and $\widehat W$ 
satisfying (\ref{9.15}) are fixed, the solution space $(v, w)$ of equation 
(\ref{9.16}) is a vector space;

(ii) When at least one of $\widehat V$ and $\widehat W$ is not 
divergence-free, the dimension of the vector space in (i) is $\geq 1$.
\end{corollary}

\begin{proof} (i) is obvious; 

(ii) Set first $v = w = 0$, a solution of (\ref{9.16}).  Thus, 
\begin{equation}
{\mathcal A}^X = \widehat V, \quad  {\mathcal A}^Y = \widehat W. \label{9.19}
\end{equation}
Applying the duality involution to expressions (\ref{9.19}), we find
\begin{equation}
\widetilde {{\mathcal A}}^X = - \left({\mathcal A}^X\right)^\dagger = - \left(\sum V^\alpha 
\frac{\partial}{\partial x^\alpha}\right)^\dagger = \sum \frac{\partial}{\partial x^\alpha} V^\alpha 
= \mbox{div}\left(\widehat V\right) + \widehat V. \label{9.20}
\end{equation}
Thus,
\begin{equation}
v = \mbox{div}\, \left(\widehat V\right), \quad w = \mbox{div} \left(\widehat W\right) \label{9.21}
\end{equation}
is a solution of (\ref{9.16}).
\end{proof}

Corollary \ref{corollary:9.18} explains the zero-order terms in equations (\ref{9.9b}) and 
(\ref{9.10b}), $\alpha u_x$ and $\alpha \left(u_{xx} + 3u^2\right)_x$, in terms of 
the 1$^{st}$-degree terms $2u$ and $2\left(u_{xx} + 3u^2\right)$, respectively.  

Let us now interpret the 1$^{st}$-degree terms themselves.  Comparing 
formulae (\ref{9.9}) and (\ref{9.10}), we see that the extended KdV fields are 
of the form:
\begin{equation}
\left(X_n^{\rm ext}\right) \quad u_t = \partial (h_n), \quad \varphi_t = (\alpha h
_{n-1,x} + 2h_{n-1} \partial) (\varphi), \label{9.22}
\end{equation}
where the differential polynomials $h_n$ are those introduced in 
Section~7.  By formulae (\ref{7.9}), (\ref{7.10}), we can rewrite (\ref{9.22}) as
\begin{equation}
\left(\begin{array}{c} u \\ \varphi\end{array}\right)_t  =  2 \left(\begin{array}{cc}
u \partial + \partial u + \partial^3/2 & \displaystyle - \varphi_x + \frac{\alpha}{2} 
\partial \varphi \vspace{2mm}\\
\displaystyle \varphi_x + \frac{\alpha}{2} \varphi \partial & 0 \end{array}\right) 
\left( \begin{array}{c} \delta / \delta u \\ \delta / \delta \varphi\end{array}\right)
(H_{n-1}). \label{9.23}
\end{equation}
Since the Hamiltonians $\{H_n\}$ commute in the 2$^{nd}$ Hamiltonian 
structure of the KdV hierarchy
\begin{equation}
B^{II} = 2 \left(u \partial + \partial u + \partial^3/2\right), \label{9.24}
\end{equation}
the extended KdV flows (\ref{9.22}) commute as well, provided the matrix 
entering (\ref{9.23}) is Hamiltonian.  That it is so indeed, can be seen 
after rewriting one-half of this matrix as
\begin{equation}
{\mathcal B}  = \left(\begin{array}{cc}
u \partial + \partial u & \varphi \partial - \lambda \partial \varphi \vspace{1mm}\\
\partial \varphi - \lambda \varphi \partial & 0 \end{array}\right) + 
\left(\begin{array}{cc}
\partial^{3}/2 & 0 \vspace{1mm}\\
0 & 0 \end{array}\right), \quad \lambda = 1 - \frac{\alpha}{2}. 
\label{9.25}
\end{equation}
The linear part of the matrix ${\mathcal B}$ corresponds to the semidirect sum 
Lie algebra ${\mathcal D}_1 \ltimes V^\lambda$, where ${\mathcal D}_1$ is the 
Lie algebra of vector fields on ${\mathbb{R}}^1$, and $V^\lambda$ is the 
one-dimensional ${\mathcal D}_1$-module of $\lambda$-forms $\{$function $\times \ 
 (dx)^\lambda\}$; the constant part of ${\mathcal B}$ (\ref{9.25}) is a generalized 
2-cocycle on ${\mathcal D}_1$; thus, ${\mathcal B}$ is a Hamiltonian matrix  (see 
[14]).

It remains to show that the thus constructed KdV extensions (\ref{9.22}) 
exhaust completely the centralizer of $X_2^{\rm ext}$ (\ref{9.9}).  So, suppose
\begin{equation}
\left(\bar X_n^{\rm ext}\right) \quad u_t = \partial (h_n), \quad \varphi_t = 
{\mathcal R} (\varphi) \label{9.26}
\end{equation}
commutes with $X_2^{\rm ext}$:
\begin{equation}
X_2 ({\mathcal R}) - X_n (2 u \partial) = [2 u \partial, {\mathcal R}]. \label{9.27}
\end{equation}
This is an inhomogeneous equation on ${\mathcal R}$, with one solution being 
given by formula (\ref{9.22}).  The corresponding homogeneous equation is then 
\begin{equation}
X_2 (\bar {\mathcal R}) = [2u \partial, \bar {\mathcal R}]. \label{9.28}
\end{equation}
If
\begin{equation}
 \bar {\mathcal R} = \sum^N_{s=0} f_s \partial^s, \quad f_s = f_s \{u\}, 
\label{9.29}
\end{equation}
then equation (\ref{9.28}) becomes
\begin{equation}
\sum^N_{s=0} X_2 (f_s) \partial^s = \sum^N_{s=0} 2u f_{s,x} 
\partial^s - 2 \sum^N_{s=0} f_s [\partial^s, u] \partial. \label{9.30}
\end{equation}
Picking out the $\partial^N$-terms, we get for $f_N=f$:
\begin{equation}
X_2 (f) = 2 u \partial (f) - 2 f Nu_x. \label{9.31}
\end{equation}
If $f$ is independent upon $\{u\}$ then $f$ must vanish.  If $f$ 
is not independent upon $\{u\}$, let $\ell$ be the highest 
$x$-derivative of $u$ entering $f$.  Then
\begin{equation}
X_2 (f) = \sum^\ell_{j=0} \frac{\partial f}{\partial u^{(j)}} 
\left(6 uu^{(1)} + u^{(3)}\right)^{(j)} = 2u \sum^\ell_{j=0} \frac{\partial f}{\partial u^{(j)}} 
u^{(j+1)} - 2Nu^{(f)} f. \label{9.32}
\end{equation}
The term
\begin{equation}
\frac{\partial f}{\partial u^{(\ell)}} u^{(\ell +3)} \label{9.33}
\end{equation}
on the LHS of (\ref{9.32}) is not matched by anything else.  Thus, the 
only solution of (\ref{9.31}) is $f=0$, and so $\bar {\mathcal R}=0$.  (The same 
argument works for a linear combination of $\bar X_n^{\rm ext}$'s (\ref{9.26}).)

\setcounter{remark}{33}
\setcounter{equation}{34}

\begin{remark}\label{remark:9.34}
 The Hamiltonian form (\ref{9.23}) provides 
an extension of {\it any} commuting hierarchy with the Hamiltonian 
form (\ref{9.24}), not just the KdV hierarchy.  We have here, thus, an example 
of the alternative ({\it D}) of Section~1.

Let us now examine the case $\gamma = 0$ (\ref{9.3}) from the point of 
view of Miura maps studied in Section~8.  Equation (\ref{8.16}) is now 
automatically satisfied, and we find:
\begin{gather}
\beta =2, \quad \alpha = 2 - \theta; \quad  e = - 2, \quad d = 2 \epsilon, \quad b = 2 
\theta, \quad a = - \epsilon \theta: \label{9.35}\\
u_t = \partial \left(3 u^2 + u_{xx}\right), \quad \varphi_t = \langle(2 - \theta) u_x 
+ 2u \partial \rangle (\varphi), \label{9.36}
\end{gather}
\vspace{-9mm}
\begin{subequations}\label{9.37}
\begin{gather}
v_t = \partial \left(-2v^3 + v_{xx}\right), \label{9.37a}\\
\psi_t = \langle- \theta \left(\epsilon v_x - v^2\right)_x + 2\left(\epsilon v_x - 
v^2\right) \partial \rangle (\psi)  \label{9.37b}\\
\phantom{\psi_t} = \langle- \theta u_x + 2u \partial \rangle (\psi), \label{9.37c}
\end{gather}
\end{subequations}
\vspace{-9mm}
\begin{gather}
u = \epsilon v_x - v^2, \quad  \varphi = (\epsilon \partial + \theta v) (\psi), \label{9.38}\\
\theta \ {\rm is \ arbitrary}. \label{9.39}
\end{gather}
Thus, the reduced KdV extension (\ref{9.36}) comes out of the modified 
KdV extension (\ref{9.37}) under the Miura map (\ref{9.38}).  In the old notation  
(\ref{9.9}), 
\begin{equation}
\alpha = 2 - \theta. \label{9.40}
\end{equation}

We now make a bold leap of faith and declare that the same Miura map 
(\ref{9.38}), when applied to the following mKdV$_n$ extension
\begin{equation}
v_t = \partial (\epsilon \partial + 2 v) (h_{n-1}), \quad \psi_t = (- 
\theta h_{n-1,x} + 2 h_{n-1} \partial) (\psi), \label{9.41}
\end{equation}
produces our KdV$_n$ extension (\ref{9.22}):
\begin{equation}
u_t = \left(\partial^3 + 2 u \partial + 2 \partial u\right) (h_{n-1}), \quad 
\varphi_t = \langle(2 - \theta ) h_{n-1,x} + 2h_{n-1} \partial \rangle (\varphi). 
\label{9.42}
\end{equation}

\end{remark}

\begin{proof} Call $h_{n-1}$ by $h$.  Since the 2$^{nd}$ 
Hamiltonian structure $\partial^3 + 2u \partial + 2 \partial u$ of the 
KdV equations is related to the Hamiltonian structure $-\partial$ of the 
mKdV equations by the Miura map $u = \epsilon v_x - v^2$, the 
motion equations
\begin{equation}
u_t = \left(\partial^3 + 2u \partial + 2 \partial u\right) (h), \quad h = 
\delta H/\delta u, \label{9.43}
\end{equation}
and
\begin{equation}
v_t = - \partial (\delta H/\delta v), \label{9.44}
\end{equation}
are related by the Miura map
\begin{equation}
u = \epsilon v_x - v^2. \label{9.45}
\end{equation}
But
\begin{gather}
\delta H/\delta v = Jac^\dagger (\delta H/\delta u) = (\epsilon 
\partial - 2 u)^\dagger (h) = - (\epsilon \partial + 2 v) (h) 
\label{9.46}\\
{}  \Rightarrow \ v_t = (- \partial) (- \epsilon \partial - 2v) (h) = \partial 
(\epsilon \partial + 2v) (h). \label{9.47}
\end{gather}
This justifies the $v_t$-equation (\ref{9.41}).

Now, 
\begin{subequations}
\begin{gather}
\varphi_t = [(\epsilon \partial + \theta v) (\psi)]_t = \theta 
v_t \psi + (\epsilon \partial + \theta v) (\psi_t) \nonumber\\
\phantom{\varphi_t}= \langle \theta v_t + (\epsilon \partial + \theta v) (- \theta h_x + 
2 h \partial) \rangle (\psi), \label{9.48l}
\end{gather}
while (\ref{9.42}) predicts that
\begin{gather}
\varphi_t = \langle (2 - \theta )h_x + 2 h \partial \rangle  (\varphi) = \langle (2 - 
\theta) h_x + 2 h \partial \rangle (\epsilon \partial + \theta v) (\psi). 
\label{9.48r}
\end{gather}
\end{subequations}
Thus, we have to check that
\begin{equation}
\theta v_t + (\epsilon \partial + \theta v) (- \theta h_x + 2 h 
\partial ) = \langle(2 - \theta) h_x + 2 h \partial \rangle (\epsilon \partial + 
\theta v), \label{9.49}
\end{equation}
and this identity is readily verified.
\end{proof}

Notice that the extended Miura map (\ref{9.38}) is no longer Hamiltonian 
w.r.t. the Hamiltonian structure (\ref{9.23}).  (This can be most easily 
seen by letting $\theta$ vanish.)

\setcounter{remark}{49}
\setcounter{equation}{50}

\begin{remark}\label{remark:9.50}
 Our analysis of KdV extensions, at least 
for the KdV equation itself, is now reasonably complete.  However, 
that analysis was made under the assumptions (a)--(c) (\ref{6.1}).  These 
assumptions may be too restrictive.  Assumption (a), that $\varphi$ 
is scalar, is most evidently so.  The mKdV hierarchy is associated 
with a 2 by 2 eigenvalue problem, so a {\it  pair} of eigenfunctions 
enters there.  It seems that in general, for scalar Lax operators of 
order $n$:
\begin{equation}
L = \partial^n + \sum^{n-2}_{i=0} u_i \partial^i, \label{9.51}
\end{equation}
the independent building blocks of all possible extensions have the 
sizes $1,2, \ldots, n$.  (All intermediate sizes are provided by the 
number of factors of partial factorizations of $L$.)  See more on 
scalar extensions in Section~18.

Assumption (b), that extension operators ${\mathcal A}$ are {\it  polynomial} 
in $\{u\}$ (but not {\it rational}, say), is also suspect.  Recall 
[29] that at the root of the KdV theory lies the 
M\"obius-invariant Ur-KdV equation 
\begin{equation}
\Psi_t = \Psi_{xxx} - \frac{3}{2} \Psi_{xx}^2/\Psi_x, \label{9.52}
\end{equation}
which is connected via the potential map
\begin{equation}
\rho = \Psi_x \label{9.53}
\end{equation}
with the equation
\begin{equation}
\rho_t = \rho_{xxx} - \frac{3}{2} \left(\rho^2_x/ \rho\right)_x, \label{9.54}
\end{equation}
which, in the variable
\begin{equation}
w = \frac{1}{2} \ell n \rho, \label{9.55}
\end{equation}
takes the form
\begin{equation}
w_t = w_{xxx} - 2 w_x^3, \label{9.56}
\end{equation}
which is the potential form
\begin{equation}
v = w_x \label{9.57}
\end{equation}
of the mKdV equation
\begin{equation}
v_t = \left(v_{xx} -2v^3\right)_x, \label{9.58}
\end{equation}
which is mapped via the Miura map
\begin{equation}
u = \epsilon v_x - v^2 \label{9.59}
\end{equation}
into the KdV equation
\begin{equation}
u_t = \left(3 u^2 + u_{xx}\right)_x. \label{9.60}
\end{equation}
The equations (\ref{9.52}) and (\ref{9.54}) are themselves {\it rational}, 
so the polynomial assumption on {\it their} extensions would be 
totally unreasonable; by implication, the polynomial assumption 
(b) for the KdV and mKdV equations may be not as natural as it had 
appeared. 

Assumption (c), that extensions are homogeneous, seems secure at the 
moment.  We shall see later on, in Section~12, that it has to be 
modified to remain true in spirit, if not in form.
\end{remark}

\section{The Burgers hierarchy}

The Burgers equation,
\begin{equation}
u_t + uu_x = \nu u_{xx}, \quad \nu = \mbox{const}, \label{10.1}
\end{equation}
discovered by Bateman [1], can be rescaled into any convenient 
form desirable.  We choose
\begin{equation}
u_t = \partial \left(u^2 + u_x\right). \label{10.2}
\end{equation}
The famous Hopf--Cole transformation makes this equation (\ref{10.2}) out of 
the heat equation 
\begin{gather}
v_t = v_{xx}: \label{10.3}\\
u = v_x/v. \label{10.4}
\end{gather}

We adopt in this Section the following useful language.  
Let ${\mathbf C}\{v\}$ 
be the differential {\it field} generated by an indeterminant 
$v = v^{(0)}$ and its $x$-derivatives $\{v^{(j)}|j \in {\mathbb N}\}$.  This field 
has a differential subfield ${\mathbf C}\{u\} \subset {\mathbf C}\{v\}$, with $u$ (\ref{10.4}) 
being $v^{(1)}/v$.  The evolution derivation
 $X_2:  {\mathbf C}\{v\} \rightarrow {\mathbf C}\{v\}$ (\ref{10.3}), 
\begin{equation}
X_2 (v) = v^{(2)}, \label{10.5}
\end{equation}
restricts onto subfield ${\mathbf C}\{u\}$:
\begin{equation}
X_2 (u) = 2uu^{(1)} + u^{(2)} = \partial \left(u^2 + u^{(1)}\right). 
\label{10.6}
\end{equation}
One, thus, can introduce the whole Burgers hierarchy [3], by 
starting with an infinity of commuting evolution derivations $X_n$ 
of ${\mathbf C}\{v\}$: 
\begin{equation}
X_n (v) = v^{(n)}, \quad n \in {\mathbb N}, \label{10.7}
\end{equation}
and then restricting these derivations onto the subfield 
${\mathbf C}\{u\}$:
\begin{equation}
X_n (u) = \partial (Q_n(u)). \label{10.8}
\end{equation}
Here
\begin{equation}
Q_n (u) = (\partial +u)^{n} (1) \label{10.9}
\end{equation}
are differential polynomials in $u$, satisfying the defining 
relation
\begin{equation}
v^{(n)} = v Q_n \left(v^{(1)}/v\right) = v Q_n (u): \label{10.10}
\end{equation}
If
\begin{equation}
v^{(n)} = v Q_n, \quad Q_n = Q_n (u), \quad  u = v^{(1)}/v, \label{10.11}
\end{equation}
then
\begin{equation}
Q_0 = 1, \quad Q_1 = u, \quad Q_2 = u^2 + u^{(1)}, \quad \ldots \label{10.12}
\end{equation}
and
\begin{gather}
 v^{(n+1)}= \partial (v^{(n)}) = \partial (v Q_n) = v^{(1)} Q_n 
+ v \partial (Q_n) \nonumber\\
\phantom{v^{(n+1)}} {}=  v (u Q_n + \partial (Q_n)) = v Q_{n+1} \label{10.13}\\
 \Rightarrow \ Q_{n+1} = (\partial + u) (Q_n), \label{10.14}
\end{gather}
and formula (\ref{10.9}) follows.

We aim to determine all possible extensions of the Burgers hierarchy 
subject to the restrictions (\ref{6.1}): we are looking at extensions that 
are scalar, polynomial, and homogeneous.

It will later on prove very useful to start with the full $v$-picture 
first.  All extensions of (\ref{10.7}) can of course be immediately written 
down:
\begin{equation}
\left(X_n^{\rm ext}\right) \quad v_t = v^{(n)}, \quad \psi_t = \epsilon_n \psi^{(n)}, \quad 
\epsilon_n = \mbox{const}, \quad n \in {\mathbb N}. \label{10.15}
\end{equation}

Let us estimate what kind of subfields of ${\mathbf C}\{v, \psi\}$ the flows 
(\ref{10.15}) could be restricted onto.  If $\epsilon_n =1$, the flow 
$X_n^{\rm ext}$ can be thought of as a linearization of the flow 
$X_n$ (10.7).  Therefore, the restriction ${\mathbf C} \{v\} \ \supset
{\mathbf C}\{u\}$, 
\begin{equation}
u = \partial (\ell n v) = v^{(1)}/v, \label{10.16}
\end{equation}
can be extended by linearization into
\begin{equation}
\varphi = \partial \frac{D\ell n v}{Dv} (\psi) = \partial (\psi/v). 
\label{10.17}
\end{equation}
To calculate $\varphi_t$, we introduce temporarily a new variable 
$\bar \varphi$ such that
\begin{gather}
\varphi = \partial (\bar \varphi), \label{10.18}\\
\psi = v \bar \varphi. \label{10.19}
\end{gather}
Therefore, 
\begin{gather}
\psi^{(n)} = \sum^n_{k=0} \left(\begin{array}{c} n \\ k\end{array}\right) v^{(n-k)} 
\bar \varphi^{(k)} = v \sum^n_{k=0} \left( \begin{array}{c} n \\ k\end{array}\right) Q_{n-k} 
\bar \varphi^{(k)}  \label{10.20}\\
\Rightarrow\psi^{(n)}/v = Q_n \bar \varphi + \sum^n_{k=1} \left( \begin{array}{c} n \\ k\end{array}
\right) Q_{n-k} \varphi^{(k-1)}. \label{10.21}
\end{gather}
Hence
\begin{gather}
\varphi_t = \partial ((\psi/v)_t) = \partial \left(\frac{\epsilon_n \psi^{(n)}}{v} -
\frac{\psi}{v} \frac{v^{(n)}}{v} \right)\nonumber\\
\phantom{\varphi_t}= \partial \left(\epsilon_n Q_n \bar \varphi + \epsilon_n \sum^n_{k=1} 
\left(\begin{array}{c} n \\ k\end{array}\right) Q_{n-k} \varphi^{(k-1)} - \bar \varphi 
Q_n\right)\label{10.22}\\
 \Rightarrow  \ \varphi_t = \partial \left((\epsilon_n -1) Q_n \bar 
\varphi + \epsilon_n  \sum^n_{k=1} \left( \begin{array}{c} n \\ k\end{array}\right) 
Q_{n-k} \varphi^{(k-1}\right). \label{10.23}
\end{gather}
We see that we must have
\begin{equation}
\epsilon_n =1, \label{10.24}
\end{equation}
and thus, the flows
\begin{equation}
\left(X^{\rm ext}_n\right) \quad  v_t = v^{(n)}, \quad \psi_t = \psi^{(n)}, 
\label{10.25}
\end{equation}
under the restriction
\begin{equation}
u = v^{(1)}/v, \quad  \varphi = \partial (\psi/v), \label{10.26}
\end{equation}
become
\begin{subequations}\label{10.27}
\begin{gather}
\left(X^{\rm ext}_n\right) \quad u_t = \partial (Q_n), \label{10.27a}\\
 \varphi_t = \langle\partial  \sum^n_{k=1}
 \left(\begin{array}{c} n \\ k\end{array}\right) Q_{n-k} \partial
^{k-1}\rangle (\varphi). \label{10.27b}
\end{gather}
\end{subequations}

Since the extension (\ref{10.27}) is in fact linearization, we get:
\begin{subequations}\label{10.28}
\begin{gather}
\frac{DQ_n}{Du} = \sum^n_{k=1} \left(\begin{array}{c} n \\ k\end{array}\right) Q_{n-k} 
\partial^{k-1} \label{10.28a}\\
{}  \Rightarrow 
\frac{D Q_n}{Du} \partial = \sum^n_{k=0} \left(\begin{array}{c} n \\ k\end{array} \right) 
Q_{n-k} \partial^k - Q_n. \label{10.28b}
\end{gather}
\end{subequations}

Analysing the above calculation, one can observe that it will still go 
through if we replace the restriction (\ref{10.17})
\begin{equation}
\varphi = \partial (\psi/v) = \psi^{(1)}/v - \psi v^{(1)}/v^2  \label{10.29}
\end{equation}
by only ``half'' of it.  Set
\begin{equation}
\varphi = \psi^{(1)}/v, \label{10.30}
\end{equation}
so that
\begin{gather}
\psi^{(1)} = v \varphi\nonumber\\
{}  \Rightarrow \
\psi^{(n+1)} = (v \varphi)^{(n)} = \sum^n_{k=0} \left(\begin{array}{c} n \\ k\end{array}
\right) v^{(n-k)} \varphi^{(k)} = v \sum^n_{k=0} \left(\begin{array}{c} n \\ k\end{array}
\right) Q_{n-k} \varphi^{(k)}. \label{10.31}
\end{gather}
Therefore, 
\begin{gather}
\varphi_t = \epsilon_n \psi^{(n+1)}/v - \psi^{(1)} v^{(n)}/v^2 = 
\epsilon_n \sum^n_{k=0} \left(\begin{array}{c} n \\ k\end{array}\right) Q_{n-k} \varphi^{(k)} 
- Q_n \varphi: \nonumber\\
\varphi_t = \langle - Q_n + \epsilon_n \sum^n_{k=0} \left(\begin{array}{c} n \\ k\end{array}
\right) Q_{n-k} \partial^k \rangle (\varphi). \label{10.32}
\end{gather}
This restriction is different from the previous one, (\ref{10.27b}).

If we now go over the derivation of formula (\ref{10.32}), we may notice 
that the restriction $\varphi = \psi^{(1)}/v$ (\ref{10.30}) can be further 
generalized, into
\begin{equation}
\varphi = \psi^{(1)}/v^\rho, \label{10.33}
\end{equation}
where $\rho$ is arbitrary (formal symbol).  We should extend the 
differential fields ${\mathbf C}\{v\}$ and ${\mathbf C}\{v, \psi\}$ by adjoining 
$v^\rho$, but this is a minor matter since
\begin{equation}
\left(v^\rho\right)^{(1)} = \rho v^{\rho} v^{(1)}/v = v^\rho  \rho u. 
\label{10.34}
\end{equation}
Consequently, 
\begin{equation}
\left(v^\rho\right)^{(n)} = v^\rho Q_n (\rho u). \label{10.35}
\end{equation}
Hence, from formula (\ref{10.33}) we obtain:
\begin{gather}
\psi^{(n+1)} = \left(v^\rho \varphi\right)^{(n)} = \sum^n_{k=0} \left(\begin{array}{c} n \\ 
k\end{array}\right) \left(v^\rho\right)^{(n-k)} \varphi^{(k)} = v^\rho \sum^n_{k=0} 
\left(\begin{array}{c} n \\ k\end{array}\right) Q_{n-k} (\rho u) \varphi^{(k)} \label{10.36}\\
{}\Rightarrow \ \varphi_t = \left(\psi^{(1)}/v^\rho\right)_t = \epsilon_n \psi^{(n+1)}/v^\rho 
- \rho v^{- \rho-1} v^{(n)} \psi^{(1)} \nonumber\\
\qquad {} =  \epsilon_n \sum^n_{k=0} \left(\begin{array}{c} n \\ k\end{array}\right) Q_{n-k} 
(\rho u) \varphi^{(k)} - \rho \varphi Q_n (u): \label{10.37}\\
\varphi_t = \langle  - \rho Q_n (u) + \epsilon_n \sum^n_{k=0} 
\left(\begin{array}{c} n \\ k\end{array}\right) Q_{n-k} (\rho u) \partial^k\rangle (\varphi). 
\label{10.38}
\end{gather}

The thus obtained extension (\ref{10.38}) yields, for the 2$^{nd}$ and 
3$^{rd}$ Burgers equations, in the notation
\begin{equation}
\epsilon_2 = \gamma, \quad \epsilon_3 = g, \label{10.39}
\end{equation}
respectively:
\begin{subequations}\label{10.40}
\begin{gather}
u_t = \partial \left(u^2 + u_x\right), \label{10.40a}\\
\varphi_t = \langle\left(\rho (\gamma \rho-1)u^2 + \rho (\gamma -1) u_x\right) 
+ 2 \gamma \rho u \partial + \gamma \partial^2 \rangle (\varphi), \label{10.40b}
\end{gather}
\end{subequations}
\vspace{-8mm}
\begin{subequations}\label{10.41}
\begin{gather}
u_t = \partial \left(u^3 + 3uu_x + u_{xx}\right), \label{10.41a}\\
\varphi_t =\langle \left(\rho \left(g \rho^2 -1\right) u^3 + 3 \rho (g \rho-1) uu_x 
+ \rho (g-1) u_{xx}\right) \nonumber\\
\qquad {}+ 3g \left(\rho^2 u^2 + \rho u_x\right) \partial 
+ 3 g \rho u \partial^2 + g \partial^3 \rangle (\varphi). \label{10.41b}
\end{gather}
\end{subequations}

Are there other nontrivial subfields of ${\mathbf C}\{v, \psi\}$, besides the 
two we have used, (\ref{10.26}) and (\ref{10.33}), that are invariant under extended 
Burgers hierarchy?  Let us see what can be said about the 2$^{nd}$ 
extended Burgers equation
\begin{equation}
v_t = v_{xx}, \quad \psi_t = \gamma \psi_{xx}, \label{10.42}
\end{equation}
restricted by the general form
\begin{equation}
u = v^{(1)}/v, \quad \varphi = (A(v) v_x + B(v) \partial) (\psi), 
\label{10.43}
\end{equation}
onto the system
\begin{equation}
u_t = \partial \left(u^2 + u_x\right), \quad \varphi_t = \langle \left(\alpha u_x + 
\omega u^2\right) + \beta u \partial + \gamma \partial^2\rangle (\varphi). 
\label{10.44}
\end{equation}
Here $A(v)$ and $B(v)$ are unspecified functions of $v$ to be 
determined.  A straightforward calculation shows that there exist 
two types of restrictions:  (\ref{10.33}) for general $\gamma$, and, for 
$\gamma =1$, 
\begin{gather}
\varphi = \left(-v^{- \rho-1} v_x + v^{- \rho} \partial\right) (\psi) 
= v^{-\rho}(\partial - u) (\psi) = v^{1- \rho} \partial (\psi/v), \label{10.45}\\
v_t = v_{xx}, \quad \psi_t = \psi_{xx}, \label{10.46}
\end{gather}
\vspace{-8mm}
\begin{subequations}\label{10.47}
\begin{gather}
u_t = \partial \left(u^2 + u_x\right), \label{10.47a}\\
 \varphi_t = \langle 2 u_x + \rho (\rho - 1) u^2 + 2 \rho u \partial + \partial^2 \rangle (\varphi). 
\label{10.47b}
\end{gather}
\end{subequations}
The linearization map (\ref{10.17}) is the case $\{\rho = 1\}$ of formulae 
(\ref{10.45})--(\ref{10.47}).

To see that the subfield ${\mathbf C}\{u, \varphi\} \subset {\mathbf C}\{v, \psi\}$ 
 (\ref{10.45}) is invariant under {\it arbitrary} flow
\begin{equation} \left(X_n^{\rm ext}\right) \quad v_t = v^{(n)}, \quad  \psi_t = \psi^{(n)}, 
\label{10.48}
\end{equation}
let us temporarily introduce a new variable $\bar \varphi$ such that
\begin{subequations}\label{10.49}
\begin{gather}
\psi = v \bar \varphi  \label{10.49a}\\
\Rightarrow  \ \partial (\bar \varphi) = \partial (\psi/v) = v^{\rho-1} \varphi. \label{10.49b}
\end{gather}
\end{subequations}
Using formula (\ref{10.21}), we obtain 
\begin{subequations}\label{10.50}
\begin{gather}
\varphi_t = \left[v^{1-\rho} \partial (\psi/v)\right]_t = (1-\rho) v^{1-\rho} 
\left(v^{(n)}/v\right) \partial(\psi/v)  + 
 v^{1-\rho} \partial \left(\psi^{(n)}/v - \psi v^{(n)}/v^2\right)\nonumber\\
\qquad {} = (1 - \rho) Q_n \varphi + \label{10.50a}\\
\qquad {}+ v^{1-\rho} \partial \sum^n_{k=1} \left(\begin{array}{c} n \\ k\end{array}\right) Q_{n-k} 
\partial^{k-1} \left(v^{\rho -1} \varphi\right). \label{10.50b}
\end{gather}
\end{subequations}
 
Now,
\begin{gather}
\partial^k v^\sigma = v^\sigma (\partial + \sigma u)^k \label{10.51}\\
\Rightarrow \ \varphi_t = \langle(1 - \rho) Q_n + \sum^n_{k=1} \left(\begin{array}{c} n \\ k\end{array}\right) 
(\partial + (\rho -1) u) Q_{n-k} (\partial + (\rho -1) u)^{k-1} \rangle (\varphi). \label{10.52}
\end{gather}
For $\rho=1$, this formula turns into formula (\ref{10.27b}), while for 
$n=2$, formula (\ref{10.47b}) is recovered.

Our goal now is to determine all commuting extensions of the 2$^{nd}$ 
Burgers equation, (\ref{10.44}), and the 3$^{rd}$ one:
\begin{subequations}\label{10.53}
\begin{gather}
u_t = \partial \left(u^3 + 3 uu_x + u_{xx}\right), \label{10.53a}\\
\varphi_t = \langle\left(au_{xx} + b uu_x + cu^3\right) + \left(du_x + eu^2\right) 
\partial + fu \partial^2 + g \partial^3\rangle (\varphi), \label{10.53b}
\end{gather}
\end{subequations}
where $\alpha$, $\omega$, $\beta$, $\gamma$; $a$, $b$, $c$, $d$, $e$, $f$, $g$ are unknown 
constants.  We shall assume for the time being that $\gamma \not= 0$, 
and defer the nonmaximal case $\gamma=0$ until later.

Writing out in long-hand the equality (\ref{1.13}) for the flows (\ref{10.44}) 
and (\ref{10.53}) and equating the like-terms, we find:
\begin{equation}
\omega = \frac{\beta (\beta -2)}{4 \gamma}, \label{10.54}
\end{equation}
\vspace{-8mm}
\begin{subequations}\label{10.55}
\begin{gather}
f = \frac{3}{2} \frac{g}{\gamma} \beta, \label{10.55f}\\
e = \frac{3}{4} \frac{g}{\gamma^2} \beta^2, \label{10.55e}\\
d = \frac{3}{4} \frac{g}{\gamma} [2 \alpha + \beta (1 + \gamma ^{-1})], \label{10.55d}\\
c = g \left(\frac{\beta}{2 \gamma}\right)^3 - \frac{\beta}{2 \gamma}, \label{10.55c}\\
b = - \frac{3}{2} \frac{\beta}{\gamma} + \frac{3}{4} \frac{g}{\gamma^2}  \alpha (\beta+2) 
+ \frac{3}{2} \frac{g}{\gamma^2} 
\beta \left[-1+ \frac{(\beta + 2)(1 + \gamma^{-1})}{4} \right], \label{10.55b}\\
a = - \frac{\beta}{2 \gamma} + \frac{g}{2 \gamma} (3 \alpha + 
\beta) - \frac{1 - \gamma^{-1}}{2} \frac{3}{4} \frac{g}{\gamma} 
[2 \alpha + \beta (1 + \gamma^{-1})], \label{10.55a}
\end{gather}
\end{subequations}
\begin{equation}
\left(\alpha - \frac{\gamma -1}{2 \gamma} \beta\right) 
\left(g - 1 + \frac{1 - \gamma^2}{\gamma^2} \frac{3}{4} g\right) = 0, \label{10.56}
\end{equation}

\newpage

\vspace*{-8mm}

\begin{subequations}\label{10.57}
\begin{gather}
6g \omega - 3 \gamma b + d \alpha + b + 6a - g \alpha = 0, \label{10.57a}\\
-2b + 6 \omega + d \alpha - 3 \alpha f + 3 \beta a = 0, \label{10.57b}\\
2f \omega - 6 \gamma c - \beta b + 2 d \omega + 4 b - 6 \omega - 
6 \alpha = 0, \label{10.57c}\\
2 f \omega - \beta b + 2e \alpha - 2d \omega + 6c - 6 \omega = 0. 
\label{10.57d}
\end{gather}
\end{subequations}

Equation (\ref{10.56}) says that either
\begin{equation}
\alpha = \frac{\gamma -1}{2 \gamma}\beta \label{10.58}
\end{equation}
or
\begin{equation}
g - 1 + \frac{1 - \gamma^2}{\gamma^2} \frac{3}{4} g = 0
\label{10.59}
\end{equation}
(or both).  Equality (\ref{10.58}) is covered by formula (\ref{10.40b}).

So, we are left to tackle the relation (\ref{10.59}), and of course the 
4 relations (\ref{10.57}); formulae (\ref{10.55}) we treat as defining the 
constants $f$, $e$, $d$, $c$, $b$, $a$ in terms of $\alpha$, $\beta$, $\gamma$, $g$. 

Our strategy in handling these opaque equations is to look at them 
as perturbations around the known solution (\ref{10.40b}).  In other 
words, we set
\begin{subequations}\label{10.60}
\begin{gather}
\rho = \beta/2 \gamma \ \Rightarrow \ \beta = 2 \rho \gamma, \quad  
\omega = \rho (\gamma \rho -1), \label{10.60a}\\
\alpha = \rho (\gamma - 1) + \gamma \Delta, \label{10.60b}
\end{gather}
\end{subequations}
so that $\Delta$ is just $[\alpha - \rho (\gamma -1)]/\gamma$.  The 
values $f$, $e$, $d$, $c$, $b$, $a$ from formula (\ref{10.41b}) we designate as 
``old'':
\begin{subequations}\label{10.61}
\begin{gather}
f^{\rm old} = 3 g \rho, \label{10.61f}\\
e^{\rm old} = 3 g \rho^2, \label{10.61e}\\
d^{\rm old} = 3 g \rho, \label{10.61d}\\
c^{\rm old} = \rho \left(g \rho^2 -1\right), \label{10.61c}\\
b^{\rm old} = 3 \rho (g \rho -1), \label{10.61b}\\
a^{\rm old} = \rho (g -1). \label{10.61a}
\end{gather}\end{subequations}
These old variables satisfy the relations (\ref{10.57}).  The new variables, 
read off formulae (\ref{10.55}) and (\ref{10.60}), are:
\begin{subequations}\label{10.62}
\begin{gather}
a = a^{\rm old} + \frac{3g(1+\gamma)}{4 \gamma} \Delta = \rho (g-1) + 
\frac{3 g(1+\gamma)}{4 \gamma} \Delta, \label{10.62a}\\
b = b^{\rm old} + \frac{3g(\gamma \rho+1)}{2 \gamma} \Delta = 3 \rho 
(g \rho -1) + \frac{3 g(\gamma \rho+1)}{2 \gamma} \Delta, \label{10.62b}\\
c = c^{\rm old} = \rho \left(g \rho^2-1\right), \label{10.62c}\\
d = d^{\rm old} + \frac{3}{2}  g \Delta = 3 g \rho + \frac{3}{2} 
g \Delta, \label{10.62d}\\
e = e^{\rm old} = 3 g \rho^2, \label{10.62e}\\
f = f^{\rm old} = 3 g \rho. \label{10.62f}
\end{gather}
\end{subequations}

Now, equation (\ref{10.57d}) is satisfied identically for any $\Delta$.  
Equation (\ref{10.57c}) yields:
\begin{gather}
0 = (4 - 2 \gamma \rho)  \frac{3g (\gamma \rho +1)}{2 \gamma} 
\Delta + 2 \rho (\gamma \rho-1) \frac{3}{2} g \Delta - 6 \gamma 
\Delta \nonumber\\
\qquad {}= 3 \Delta \left\{\frac{g}{\gamma} \left(2 + \gamma \rho - \gamma^2 \rho^2\right) 
+ g \rho (\rho \gamma-1) - 2 \gamma \right\} = 3 \Delta \left(\frac{2g}{\gamma} - 2 \gamma \right)
\nonumber\\
\qquad {} = 6 \Delta \gamma^{-1} \left(g - \gamma^2\right). \label{10.63}
\end{gather}
Since we are interested in $\Delta \not= 0$, 
\begin{equation}
g = \gamma^2 \label{10.64}
\end{equation}
Substituting this into (\ref{10.59}) we find
\begin{equation}
\gamma^2 = g = 1. \label{10.65}
\end{equation}
With the latter relation granted, formula (\ref{10.57a}) yields:
\begin{gather}
0 = (1 - 3 \gamma) \frac{3}{2} (\rho + \gamma) \Delta + 3 
\rho \Delta \gamma + \frac{3}{2} \Delta \rho (\gamma -1) + 
\frac{3}{2} \Delta \Delta \gamma + 6 \cdot \frac{3}{4} (1 + \gamma) 
\Delta - 9 \Delta \gamma \nonumber\\
\qquad {}= 3 \Delta \left\{\frac{1 - 3 \gamma}{2} \rho + 
\frac{\gamma (1 - 3 \gamma)}{2} + \gamma \rho + \rho \frac{\gamma -1}{2} + 
\frac{\Delta \gamma}{2} + \frac{3}{2} (1+ \gamma) -3 \gamma \right\} 
\nonumber\\
\qquad {}= 3 \Delta \left( - \gamma + \frac{\Delta \gamma}{2} \right) = 
3 \Delta \gamma \left(\frac{\Delta}{2} + 1 \right) \label{10.66}\\
 \Rightarrow \ \Delta = 2. \label{10.67}
\end{gather}
The difference between equations (\ref{10.57a}) and (\ref{10.57b}) is 
linear in $\Delta$, and is identically satisfied for $\gamma^2 
= g = 1$ (\ref{10.65}).

Thus, the complete list of extensions (\ref{10.44}) with nontrivial 
centralizer (\ref{10.53}) is: (\ref{10.40b}); 
\begin{equation}
\gamma = 1, \quad  \alpha = 2, \quad  \beta = 2 \rho, \quad  \omega = \rho 
(\rho -1), \label{10.68}
\end{equation}
and this is the case (\ref{10.47}); and
\begin{equation}
\gamma = -1, \quad  \alpha = 2 \rho -2, \quad \beta = 2 \rho, \quad \omega = 
- \rho (\rho -1), \label{10.69}
\end{equation} 
which is dual to (\ref{10.68}):
\begin{equation}
-\langle 2u_x + \rho (\rho -1) u^2 + 2 \rho u \partial + \partial^2 \rangle^\dagger = 
 (2 \rho - 2) u_x - \rho (\rho -1) u^2 + 2 \rho u \partial -
\partial^2. \label{10.70}
\end{equation}
The extension (\ref{10.69}) is the only one on this list which doesn't 
come from the restriction of an evolution derivation from ${\mathbf C}\{v, \psi\}$ 
onto ${\mathbf C}\{u, \varphi\}$; however, its dual (\ref{10.68}) does.  
This suggests that the family (\ref{10.40b}) is self-dual; indeed, the 
whole family (\ref{10.38}) is:

\setcounter{proposition}{70}
\setcounter{equation}{71}

\begin{proposition}\label{proposition:10.71}
 Denote by ${\mathcal A}_n = {\mathcal A}_n (\rho,   
\epsilon_n)$ the operator entering the RHS of equation (\ref{10.38}): 
\begin{equation}
{\mathcal A}_n (\rho, \epsilon_n) = - \rho Q_n (u) + \epsilon_n \sum^n_{k=0} 
\left(\begin{array}{c} n \\ k\end{array}\right) Q_{n-k} (\rho u) \partial^k. \label{10.72}
\end{equation}
Then
\begin{equation}
-{\mathcal A}_n (\rho, \epsilon_n)^\dagger =  {\mathcal A}_n \left(- \rho, (-1)^{n+1} 
\epsilon_n\right). \label{10.73}
\end{equation}
\end{proposition}

\begin{proof} We need formula ([12, p. 64])
\begin{equation}
\sum^n_{k=0} \left(\begin{array}{c} n \\ k\end{array}\right) Q_{n-k} (\rho u) \partial^k 
= (\partial + \rho u)^n. \label{10.74}
\end{equation}
Then,
\begin{equation}
{\mathcal A}_n (\rho, \epsilon_n) = - \rho Q_n (u) + \epsilon_n (\partial + 
\rho u)^n, \label{10.75}
\end{equation}
and therefore
\begin{equation*}
- {\mathcal A}_n (\rho, \epsilon_n)^\dagger = \rho Q_n (u) - \epsilon_n (- 
\partial + \rho u)^n 
= \rho Q_n (u) + (-1)^{n+1} \epsilon_n (\partial - \rho u)^n.
\tag*{\qed}
  \end{equation*}
  \renewcommand{\qed}{}
\end{proof}

Let us now examine the reduced case $\gamma = 0$: 
\begin{equation}
u_t = \partial \left(u^2 + u_x\right), \quad \varphi_t = \langle \left(\alpha u_x + 
\omega u^2\right) + \beta u \partial \rangle (\varphi). \label{10.77}
\end{equation}
It's immediate that $f = g = 0$ in (\ref{10.53b}), so that
\begin{equation}
\varphi_t = \langle\left(a u_{xx} + b uu_x + cu^3\right) + \left(du_x + eu^2\right) \partial 
\rangle (\varphi). \label{10.78}
\end{equation}
Equating the $\partial^1$-terms in the defining identity (\ref{1.13}), we 
find
\begin{subequations}\label{10.79}
\begin{gather}
e = d = \beta, \label{10.79a}\\
\beta (\beta -1) = 0. \label{10.79b}
\end{gather}
\end{subequations}
We postpone the case $\beta = 0$ 'till later.  So, suppose 
\begin{equation}
\beta = 1 = d = e. \label{10.80}
\end{equation}
By Corollary \ref{corollary:9.18} and formula (\ref{9.21}), we can take
\begin{equation}
\alpha = 0. \label{10.81}
\end{equation}
This will force
\[
a = 0. 
\]
The remaining part of the equation (\ref{1.13}) yields:
\begin{equation}
b \left(u X^\sim\right)_x + 3 cu^2 X^\sim - 2 \omega u Y^\sim =
 u \left(b uu_x + cu^3\right)_x - 
\left(u_x + u^2\right)\cdot 2 \omega uu_x, \label{10.82}
\end{equation}
where
\[
X^\sim = \partial \left(u^2 + u_x\right), \quad Y^\sim = \partial \left(u^3 + 3uu_x 
+ u_{xx}\right).
\]
The terms $u_x u_{xx}$ in (\ref{10.82}) provide $b=0$; then the 
$uu^2_x$-terms force $\omega =0$; and the terms $u^2 u_{xx}$ then 
lead to $c=0$.  Thus, 
\begin{equation}
\alpha = \omega = 0; \quad \beta =1; \quad a = b = c = 0, \quad
d = e = 1 \label{10.83}
\end{equation}
is the essential solution.  It can be generalized for the whole 
Burgers hierarchy:

\setcounter{proposition}{82}
\setcounter{equation}{83}

\begin{proposition}\label{proposition:10.84}
 The extended flows
\begin{equation}
\left(X_n^{ext}\right) \quad u_t = \partial (Q_n), \quad  \varphi_t = (Q_{n-1} 
\partial) (\varphi), \quad n \in {\mathbb N}, \label{10.85}
\end{equation}
all commute.  
\end{proposition}

\begin{proof}
We have to show that
\begin{equation}
X_n (Q_{m-1}) - X_m (Q_{n-1})  = Q_{n-1} Q_{m-1}^{(1)} - 
Q_{m-1} Q_{n-1}^{(1)}. \label{10.86}
\end{equation} 
Now, 
\begin{subequations}\label{10.87}
\begin{gather}
X_n (Q_{m-1}) = \frac{DQ_{m-1}}{Du} \left(X_n (u) \right) = 
\frac{DQ_{m-1}}{Du} \partial (Q_n) \nonumber\\
\qquad {} \overset{{\rm[by \ (\ref{10.28b}), \ (\ref{10.74})]}}{=} \ 
 \left((\partial +u)^{m-1} - Q_{m-1}\right) (Q_n), \label{10.87l}
\end{gather}
while
\begin{equation}
Q_{n-1} Q_{m-1}^{(1)} = Q_{n-1} ((\partial + u) -u) (Q_{m-1} ) = 
Q_{n-1} (Q_m - u Q_{m-1}). \label{10.87r}
\end{equation}
\end{subequations}
Hence, the equality (\ref{10.86}) becomes
\begin{equation}
(\partial + u)^{m-1} (Q_n) - (\partial +u)^{n-1} (Q_m) = 0 
\label{10.88}
\end{equation}
which is obviously true since
\begin{equation*}
(\partial + u)^{m-1} (Q_n) = (\partial + u)^{m-1} (\partial+u)^n 
(1) = (\partial + u)^{n+m-1} (1).
\tag*{\qed}
  \end{equation*}
  \renewcommand{\qed}{}
\end{proof}

Thus, we have found the general 1$^{st}$-order extensions
\begin{equation}
u_t = \partial (Q_n), \quad \varphi_t = \langle \alpha Q_{n-1,x} 
+ Q_{n-1} \partial \rangle  (\varphi). \label{10.89}
\end{equation}
 
The remaining case of (\ref{10.79b}), $\beta = 0$, leads after a 
quick calculation with $\alpha = 0$ to the trivial extension
\begin{equation}
\omega = \beta = 0; \quad a = b = c = d = e = 0. \label{10.90}
\end{equation}
It appears that there exist no nontrivial extensions of the 
Burgers hierarchy given by extension operators of order zero.  
Nevertheless, we have:

\setcounter{proposition}{89}
\setcounter{equation}{90}

\begin{proposition}\label{proposition:10.91}
 The extended flows 
\begin{equation}
u_t = \partial (Q_n), \quad \varphi_t = \gamma Q_n \varphi, \quad
n \in {\mathbb N}, \quad \gamma = \mbox{\rm const}, \label{10.92}
\end{equation}
all commute.
\end{proposition}

\begin{proof}
We have to show that
\begin{equation}
X_n (Q_m) - X_m (Q_n) = 0. \label{10.93}
\end{equation}
But, by (\ref{10.87l}), 
\begin{equation}
X_n (Q_m) = \left(\partial + u)^m - Q_m\right) (Q_n) = Q_{n+m} - Q_n Q_m, 
\label{10.94}
\end{equation}
and this is symmetric in $n$, $m$.
\end{proof}

\setcounter{remark}{93}
\setcounter{equation}{94}

\begin{remark}\label{remark:10.95}
 The system (\ref{10.92}) is a restriction 
\begin{equation}
u = v^{(1)}/v, \quad \varphi = v^\gamma \psi, \label{10.96}
\end{equation}
of a larger system
\begin{equation}
v_t = v^{(n)}, \quad \psi_t = 0 \label{10.97}
\end{equation}
from ${\mathbf C}\{v, \psi\}$.  The dual to (\ref{10.92}) system has the same 
form (\ref{10.92}), with $\gamma^{\rm new} = - \gamma^{\rm old}$.  Whether the 
system (\ref{10.85}) could be extended from ${\mathbf C}\{u, \varphi\}$ into 
${\mathbf C}\{v, \psi\}$ or not, I was not able to determine.
\end{remark}

\section{The equation $\pbf{u_t + uu_x =0}$ and its hierarchy}

In the quasiclassical limit, the Burgers hierarchy 
\begin{equation}
u_t = \partial (Q_n), \quad  n \in {\mathbb N}, \label{11.1}
\end{equation}
becomes
\begin{equation}
(X_n) \quad u_t  = \partial \left(u^n\right), \quad n \in {\mathbb N}. \label{11.2}
\end{equation}
These are quasilinear equations.  Let's determine all their 
quasilinear extensions, of the form
\begin{equation}
\left(X^{\rm ext}\right) \quad u_t = \partial \left(u^n\right), \quad \varphi_t = 
\langle \alpha_n \left(u^{n-1}\right)_x + \beta_n u^{n-1} \partial \rangle (\varphi). 
\label{11.3}
\end{equation}
The linearization of (\ref{11.2}) yields
\begin{equation}
\varphi_t = \partial \left(n u^{n-1} \varphi\right) = \langle n \left(u^{n-1}\right)_x + 
nu^{n-1} \partial \rangle (\varphi), \label{11.4}
\end{equation}
so that
\begin{equation}
\alpha_n = n, \quad \beta_n = n, \label{11.5}
\end{equation}
while its dual provides
\begin{equation}
\varphi_t = \left(n u^{n-1} \partial\right) (\varphi), \label{11.6}
\end{equation}
so that
\begin{equation}
\alpha_n = 0, \quad \beta_n = n. \label{11.7}
\end{equation}

The $\partial^1$-terms of the equation (\ref{1.13}), 
\begin{equation}
X_n ({\mathcal A}_m) - X_m ({\mathcal A}_n) = [{\mathcal A}_n, {\mathcal A}_m], \label{11.8}
\end{equation}
yield:
\begin{subequations}\label{11.9}
\begin{gather}
\beta_m (m-1) u^{m-2} nu^{n-1} u_x - \beta_n (n-1) u^{n-2} 
mu^{m-1} u_x = \beta_n \beta_m u^{n+m-3} u_x (m-n) \nonumber\\
\Rightarrow \ \beta_m (m-1) n - \beta_n (n-1) m = (m-n) \beta_n \beta_m. \label{11.9a}
\end{gather}
Regrouping, we find
\begin{equation}
\beta_m (m-1) (n-\beta_n) = \beta_n (n-1) (m-\beta_m). 
\label{11.9b}
\end{equation}
\end{subequations}
The degenerate case
\begin{equation}
\beta_n = 0, \quad \forall\; n, \label{11.10}
\end{equation}
we shall examine later on.  The case
\begin{equation}
\beta_n = n, \quad \forall \; n, \label{11.11}
\end{equation}
corresponds to the linerization (\ref{11.5}) and its dual (\ref{11.7}).  So, 
suppose
\begin{equation}
\beta_n \not= 0, \; n. \label{11.12}
\end{equation}
Rewriting the equation (\ref{11.9}) as
\begin{equation}
\frac{\beta_m (m-1)}{m - \beta_m} = \frac{\beta_n (n-1)}{n - \beta_n}, \label{11.13}
\end{equation}
we see that
\begin{gather}
\beta_m (m-1) = \theta (m-\beta_m), \quad \theta = \mbox{const}\nonumber\\
 \Rightarrow \ \beta_m = \frac{\theta m}{\theta + m-1} = 
\frac{m}{1 + \omega (m-1)}, \quad \omega = \mbox{const}. \label{11.14}
\end{gather}
When $\omega = \infty$, we recover $\beta_m =0$ (\ref{11.10}), and when 
$\omega = 0$, we obtain $\beta_m = m$ (\ref{11.11}).

Let us now tackle the $\partial^0$-terms:
\begin{gather}
\alpha_m \left[X_n \left(u^{m-1}\right)\right]_x - \alpha_n \left[X_m \left(u^{n-1}\right)\right]_x 
\nonumber\\
\qquad {}= \beta_n \alpha_m u^{n-1} \left(u^{m-1}\right)_{xx} 
- \beta_m \alpha_n u^{m-1} \left(u^{n-1}\right)_{xx}. \label{11.15}
\end{gather}
I claim that the $\alpha_n$'s are just the div-terms $\{\mbox{const}\;
\beta_n\}$.  Indeed, fix $n > 1$.  By Corollary~\ref{corollary:9.18} and formula 
(\ref{9.21}), we can make $\alpha_n$ vanish for this particular $n$.  
Equation (\ref{11.15}) then becomes
\begin{equation}
\alpha_m \left[X_n \left(u^{m-1}\right)\right]_x = \alpha_m \beta_n u^{n-1} \left(u^{m-1}\right)
_{xx}. \label{11.16}
\end{equation}
Since $\beta_n \not= 0$, $\alpha_m$ must vanish because the RHS of 
(\ref{11.16}) is not a total derivative while the LHS is:
\begin{equation}
u^{n-1} \left(u^{m-1}\right)_{xx} \sim - \left(u^{n-1}\right)_x \left(u^{m-1}\right)_x = - 
(n-1) (m-1) u^{n+m-4} u^2_x \not\sim 0. \label{11.17}
\end{equation}
Thus, all extensions of (\ref{11.2}) with $\beta_n \not= 0$ are of the 
form 
\begin{equation}
\left(X_n^{\rm ext}\right) \quad u_t = \partial \left(u^n\right), \quad \varphi_t = 
\frac{n}{1+\omega (n-1)} \partial \left(u^{n-1} \varphi\right) \label{11.18}
\end{equation}
and their dual
\begin{equation}
\left(X_n^{\rm ext}\right) \quad u_t = \partial (u_n), \quad \varphi_t = 
\frac{n}{1 + \omega (n-1)} u^{n-1} \partial (\varphi). \label{11.19}
\end{equation}
For general $\omega$, these extensions are rational in $n$ and 
certainly do not come under quasiclassical limit from any Burgers 
extensions of Section~10.  

It remains to tackle the case $\beta_n \equiv 0$ (\ref{11.10}).  The 
equation (\ref{11.8}), in the form (\ref{11.15}), becomes, for $n, m>1$: 
\begin{equation}
\alpha_m X_n \left(u^{m-1}\right) = \alpha_n X_m \left(u^{n-1}\right), \label{11.20}
\end{equation}
 which is 
\begin{equation}
\alpha_m (m-1) u^{m-2} nu^{n-1} u_x = \alpha_n (n-1) u^{n-2} 
mu^{m-1} u_x, \label{11.21}
\end{equation}
so that
\begin{equation}
\alpha_m (m-1) n = \alpha_n (n-1) m. \label{11.22}
\end{equation}
Since $n, m >1$, we find
\begin{equation}
\alpha_m (m-1) = \mu m, \quad \mu = \mbox{const}, \label{11.23}
\end{equation}
and we obtain the extensions
\begin{equation}
u_t = \partial \left(u^n\right), \quad  \varphi_t = \mu n u^{n-2} u_x \varphi. 
\label{11.24}
\end{equation}
In {\it this} form, not the form (\ref{11.3}), it is applicable also to 
the case $n=1$:
\begin{equation}
u_t = u_x, \quad \varphi_t = \mu u^{-1} u_x \varphi, \label{11.25}
\end{equation}
which is a {\it rational} extension, not a {\it polynomial}
one.  As such, it is not a quasiclassical limit of any of the Burgers 
extensions of Section~10.

\section{Long wave equations}

The $1 + 1 - d$ system
\begin{equation}
u_t = uu_x + h_x, \quad h_t = (uh)_x \label{12.1}
\end{equation}
for a pair of functions $u(x,t)$ and $h(x,t)$ is the oldest one 
considered in this paper.  Riemann analyzed it in 1860~[25].  
More recently it was realized that this system [17, 18, 12] (and its 
various generalizations) is a part of an infinite hierarchy of 
commuting  Hamiltonian systems of the form
\begin{equation}
\left(\begin{array}{c} u \\ h\end{array}\right)_t = \left(\begin{array}{cc} 0 &  \partial \\ 
\partial &  0\end{array}\right) \left(\begin{array}{c} \delta/\delta u \\ \delta/\delta h\end{array}
\right) (H), \label{12.2}
\end{equation}
with 
\begin{equation}
H_2 = \frac{h u^2 + h^2}{2} \label{12.3}
\end{equation}
being the Hamiltonian of the system (\ref{12.1}).  The next flow, $\#3$, 
\begin{equation}
u_t = u^2 u_x + (2uh)_x, \quad  h_t = \left(hu^2 + h^2\right)_x, \label{12.4}
\end{equation}
has the Hamiltonian 
\begin{equation}
H_3 = \frac{hu^3}{3} + h^2 u. \label{12.5}
\end{equation}

Let us determine which extensions, if any, of the flows (\ref{12.1}) and 
(\ref{12.4}), still commute; we assume that the conditions (\ref{6.1}) are in 
force.  So, let the extensions of (\ref{12.1}) and (\ref{12.4}) be, respectively:
\begin{subequations}\label{12.6}
\begin{gather}
\varphi_t = (\alpha u_x + \beta u \partial) (\varphi), \label{12.6a}\\
\varphi_t = \langle (\gamma uu_x + \delta h_x) + \left(\mu u^2 + \nu h\right) \partial 
\rangle (\varphi), \label{12.6b}
\end{gather}
\end{subequations}
where we have utilized the natural grading
\begin{equation}
rk(h) = 2rk(u). \label{12.7}
\end{equation}
First, 
\begin{equation}
\beta = 0 \ \ \Rightarrow \ \ \mu = \nu = 0 \ \ \Rightarrow \ \ \alpha = \gamma = 
\delta = 0. \label{12.8}
\end{equation}
So, assume
\begin{equation}
\beta \not= 0. \label{12.9}
\end{equation}
By Corollary \ref{corollary:9.18} and formula (\ref{9.21}), we can then make $\alpha$ 
vanish, and this forces $\gamma =  \delta = 0$ as well.  Thus, our 
extensions (\ref{12.6}) become
\begin{equation}
\varphi_t = \beta u \varphi_x, \quad \varphi_t = \left(\mu u^2 + \nu h\right) 
\varphi_x . \label{12.10}
\end{equation}
Equation (\ref{1.13}) then yields:
\begin{gather}
2 \mu u(uu_x + h_x) + \nu (uh)_x - \beta \left[u^2 u_x + 2 (uh)_x\right]\nonumber\\
\qquad {}= \beta u \left(\mu u^2 + \nu h\right)_x - \left(\mu u^2 + \nu h\right) \beta u_x. 
\label{12.11}
\end{gather}
Equating the like-terms, we obtain:
\begin{subequations}\label{12.12}
\begin{gather}
2 \mu - \beta = \mu \beta, \label{12.12a}\\
\nu - 2 \beta = - \nu \beta, \label{12.12b}\\
2 \mu + \nu - 2 \beta = \beta \nu, \label{12.12c}
\end{gather}
\end{subequations}
or
\begin{subequations}\label{12.13}
\begin{gather}
\mu (2 - \beta) = \beta, \label{12.13a}\\
\nu (1+\beta) = 2 \beta, \label{12.13b}\\
\mu - \nu \beta = 0. \label{12.13c}
\end{gather}
\end{subequations}
Eliminating $\mu$, $\nu$, we get:
\begin{gather}
0 = \frac{\beta}{2 - \beta} - 2 \beta \frac{\beta}{1+\beta} 
= \frac{\beta}{(2-\beta)(1+\beta)} \{1+\beta -2 \beta (2 - \beta) \}\nonumber\\
\Rightarrow \ 0 = (2 \beta^2 - 3 \beta +1) = 2(\beta -1) (\beta - 1/2). \label{12.14}
\end{gather}
Thus, we got {\it two different} extensions:
\begin{subequations}\label{12.15} 
\begin{gather}
\beta = 1, \quad \mu = \nu = 1: \nonumber\\
\varphi_t = u \varphi_x,\label{12.15a}\\
\varphi_t = (u^2 + h) \varphi_x; \label{12.15b}
\end{gather}
\end{subequations}
\vspace{-8mm}
\begin{gather}
\beta = 1/2, \quad \mu = 1/3, \quad \nu = 2/3: \nonumber\\
\varphi_t = \frac{u}{2} \varphi_x, \quad \varphi_t = \frac{1}{3} 
\left(u^2 + 2h\right) \varphi_x. \label{12.16}
\end{gather}

The question now is if these two extensions apply to the whole 
hierarchy of commuting flows $-$ or not. 

We start with the $\beta=1$-extension (\ref{12.15}) first.  Noticing that
\begin{equation}
u = \frac{1}{h} \frac{\partial H_2}{\partial u}, \quad u^2 + h 
= \frac{1}{h} \frac{\partial H_3}{\partial u}, \label{12.17}
\end{equation}
We arrive at the following explanation of formulae (\ref{12.15}).

\setcounter{proposition}{17}
\setcounter{equation}{18}

\begin{proposition}\label{proposition:12.18}
 The following matrix is Hamiltonian 
for arbitrary constant $\alpha$:
\begin{equation}
B = \bordermatrix{&u&h&\varphi \cr
u&0&\partial&\displaystyle -\frac{1}{h} \varphi_x + \alpha 
\frac{1}{h} \partial 
\varphi \cr
h&\partial&0&0 \cr
\varphi&\displaystyle \frac{1}{h} \varphi_x + \alpha \varphi 
\partial \frac{1}{h} & 
0 & 0 \cr} \label{12.19}
\end{equation}
\end{proposition}

\begin{proof}
 Proof is a tedious but straightforward verification of the 
Hamiltonian criterion [14, p. 47]
\begin{equation}
B \delta \left[{\pbf{Y}}^t B ({\pbf{X}})\right] = D[B({\pbf{Y}})] B({\pbf{X}}) - D [B ({\pbf{X}})] 
B ({\pbf{Y}}), \qquad \forall \; {\pbf{X}}, {\pbf{Y}}. \label{12.20}
  \end{equation}
\end{proof}

The matrix $B$ (\ref{12.19}) provides a Hamiltonian solution to our 
extension problem.  As a bonus, this same matrix automatically 
supplies an extension of the {\it Dispersive Water Waves}
(DWW) hierarchy [12]
\begin{subequations}\label{12.21}
\begin{gather}
u_t = uu_x + h_x + \theta u_{xx}, \label{12.21a}\\
h_t = (uh)_x - \theta h_{xx}, \quad  \theta = \mbox{const}, \label{12.21b}
\end{gather}
\end{subequations}
because the system (\ref{12.21}) is also of the Hamiltonian type (\ref{12.2}), 
with the Hamiltonian
\begin{equation}
\widetilde{H}_2 = \frac{hu^2 + h^2}{2} + \theta u_x h. 
\label{12.22}
\end{equation}

Let us now analyse the case $\beta = 1/2$ (\ref{12.16}).  Recall that 
each of the systems we are looking at, the hierarchy of commuting 
systems including (\ref{12.1}), is not just a Hamiltonian system but a 
{\it tri}-Hamiltonian system [12].  In particular, the 2$^{nd}$ 
Hamiltonian structure is given by the Hamiltonian matrix
\begin{equation}
B^{II} = \begin{pmatrix}
2 \partial  &  \partial u \\
  u \partial  & h \partial + \partial  h \end{pmatrix}.
\label{12.23}
\end{equation}
The corresponding Hamiltonians of the flows (\ref{12.1}) and (\ref{12.4}) are, 
respectively
\begin{subequations}\label{12.24}
\begin{gather}
{\mathcal H}_2 = \frac{uh}{2} = \frac{1}{2} H_1, \label{12.24a}\\
{\mathcal H}_3 = \frac{hu^2 + h^2}{3} = \frac{2}{3} H_2. \label{12.24b}
\end{gather}
\end{subequations}
In particular, 
\begin{equation}
\frac{\delta {\mathcal H}_2}{\delta h} = \frac{u}{2}, \quad 
\frac{\delta {\mathcal H}_3}{\delta h} = \frac{u^2 + 2h}{3}. \label{12.25}
\end{equation}
Comparing the latter expressions with formulae (\ref{12.16}), we arrive 
at the following explanations of the latter:

\setcounter{proposition}{25}
\setcounter{equation}{26}

\begin{proposition}\label{proposition:12.26}
 The following matrix is Hamiltonian 
for arbitrary constant $\alpha$:
\begin{equation}
{\mathcal B}^{II} = \bordermatrix{&u&h&\varphi \cr
u&2 \partial & \partial u& 0 \cr
h & u \partial& h \partial + \partial h & - \varphi_x + 
\alpha \partial \varphi \cr
\varphi& 0 & \varphi_x + \alpha \varphi \partial & 0 \cr} .
\label{12.27}
\end{equation}
\end{proposition}

\begin{proof} Apart from the noninterfering $u$-part of the 
matrix (\ref{12.27}), this matrix is identical to the Hamiltonian matrix 
(\ref{9.23}).
\end{proof}

\setcounter{remark}{27}
\setcounter{equation}{28}

\begin{remark}\label{remark:12.28} The matrix (\ref{12.27}) provides the $\beta = 
\frac{1}{2}$-extension for the whole hierarchy of dispersiveless 
long wave systems.  What about the full {\it{dispersive}} version, 
the DWW hierarchy, of which (\ref{12.21}) is just one flow?  Nothing to it. 
The matrix $B^{II}$ (\ref{12.23}) is the quasiclassical limit of the full 
2$^{nd}$ DWW Hamiltonian matrix [12]
\begin{equation}
 \widetilde{\! B}^{II} = \begin{pmatrix}
2 \partial & \partial u + 2\theta \partial^2 \\
u \partial - 2 \theta \partial^2 & h \partial + \partial h 
\end{pmatrix}, 
\label{12.29}
\end{equation}
which differs from the Hamiltonian matrix $B^{II}$ (\ref{12.23}) just by 
an extra 2-cocycle $2 \theta \partial^2$ at the $u$-$h$ place.  Therefore, 
the $\varphi$-extension of $\widetilde{\! B}^{II}$, as that of 
${\mathcal B}^{II}$, goes through in exactly the same way:
\begin{equation}
\widetilde{\!{\mathcal B}}^{II} = \begin{pmatrix}
2 \partial & \partial u+ 2 \theta \partial ^2 & 0 \\
u \partial - 2 \theta \partial ^2 & h \partial + \partial h & 
- \varphi_x + \alpha \partial \varphi \\
0 & \varphi_x + \alpha \varphi \partial & 0 \end{pmatrix}. 
\label{12.30}
\end{equation}
\end{remark}

We now return to our original system (\ref{12.1}).  Because it's quasilinear, 
the weight $r k (\partial)$ is undetermined.  This forces us to the 
extension form (\ref{12.6}).  However, the full dispersive system (\ref{12.21}), 
of which (\ref{12.1}) is the quasiclassical limit, fixes the weight of 
$\partial$ as
\begin{equation}
r k (\partial ) = r k (u). \label{12.31}
\end{equation}
This allows for a far greater apriori form of extensions than (\ref{12.6}).  
Let us examine one such extension, of zeroth order.

We assume it be of the form
\begin{equation}
\varphi_t = \bar H_n (u,h) \varphi, \label{12.32}
\end{equation}
where $\bar H_n(u,h)$ is a polynomical in $u$ and $h$, homogeneous 
of weight
\begin{equation}
rk (\bar H_n) = rk (H_n) - 2 rk (u) = n r k (u). \label{12.33}
\end{equation}
Thus, the flows (\ref{12.1}) and (\ref{12.4}) are extended as
\begin{equation}
\varphi_t = \left(\alpha_1 h + \alpha_2 u^2 \right) \varphi, \quad \varphi_t = 
\left(\alpha_3 hu + \alpha_4 u^3\right) \varphi, \label{12.34}
\end{equation}
respectively.  The commutativity of the extended flows leads to 
\begin{subequations}\label{12.35}
\begin{gather}
\bar H_2 = \alpha \left(h + \frac{u^2}{2}\right), \label{12.35a}\\
\bar H_3 = \alpha \left(2h u+\frac{u^3}{3}\right), \quad \alpha = \mbox{const}. 
\label{12.35b}
\end{gather}
\end{subequations}
If we notice that formulae (\ref{12.35}) can be rewritten as
\begin{equation}
\bar H_2 = \frac{\delta H_2}{\delta h}, \quad \bar H_3 = 
\frac{\delta H_3}{\delta h}, \label{12.36}
\end{equation}
we can extend formulae (\ref{12.35}) to the whole long wave hierarchy:

\setcounter{proposition}{36}
\setcounter{equation}{37}
\begin{proposition}\label{proposition:12.37}
 The following matrix is Hamiltonian 
for arbitrary constant $\alpha$:
\begin{equation}
\bar B = \bordermatrix{& u & h & \varphi \cr
u & 0 & \partial & 0 \cr
h & \partial & 0 & - \alpha \varphi \cr
\varphi & 0 & \alpha \varphi & 0 \cr} . \label{12.38}
\end{equation}
\end{proposition}

\begin{proof} The $h$-$\varphi$ part of $\bar B$ is a skewsymmetric 
non-differential 2 by 2 matrix, and so is Hamiltonian (and Lie-algebraic). 
The remaining $u$-$h$ part represents a 2-cocycle on a trivial Lie 
algebra.
\end{proof}

Again, the Hamiltonian matrix $\bar B$ (\ref{12.38}) extends the full 
dispersive DWW hierarchy, not just its quasiclassical limit.

\setcounter{remark}{38}
\setcounter{equation}{39}

\begin{remark}\label{remark:12.39}
 The Hamiltonian extensions (\ref{12.19}) and (\ref{12.38}) 
of the Hamiltonian matrix (\ref{12.2}), are far from being the only ones 
possible.  For example, if we for a moment forget about the 
homogeneuity requirement and notice that $u$ and $h$ enter into the 
Hamiltonian matrix (\ref{12.2}) on equal footing, we can interchange 
$u$ and $h$ in each of the Hamiltonian matrices (\ref{12.19}) and (\ref{12.38}).  
In the case of the latter, we get the Hamiltonian matrix
\begin{equation}
B^{{\mathcal N}} = \bordermatrix{&u & h & \varphi \cr
u& 0 & \partial & -\alpha \varphi \cr
h & \partial & 0 & 0 \cr
\varphi& \alpha \varphi & 0 & 0 \cr}. \label{12.40}
\end{equation}
However, this Hamiltonian matrix provides an extension that is no 
longer homogeneous for the long wave system (\ref{12.1}) and every other 
system in the hierarchy, because the $\varphi_t$-equation generated 
by the Hamiltonian matrix $B^{{\mathcal N}}$ (\ref{12.40}) yields
\begin{subequations}\label{12.41}
\begin{equation}
rk (\partial_t) = rk (H) - rk (u), \label{12.41a}
\end{equation}
while the $h_t$-equation provides
\begin{equation}
rk (\partial_t) = rk (H) - rk (u) + rk (\partial) - rk (h), 
\label{12.41b}
\end{equation}
\end{subequations}
and these weights {\it are not equal}.  However, the remedy 
is clear:  $\varphi$ is not a good variable, but
\begin{equation}
\bar \varphi = \log (\varphi) \label{12.42}
\end{equation}
is.  In the variables $\{u, h, \bar \varphi\}$, the Hamiltonian matrix 
$B^{{\mathcal N}}$ (\ref{12.40}) becomes 
\begin{equation}
B^{{\mathcal H}} = \bordermatrix{& u & h & \bar \varphi \cr
u & 0 & \partial & - \alpha \cr
h & \partial & 0 & 0 \cr
\bar \varphi & \alpha & 0 & 0 \cr}. \label{12.43}
\end{equation}
{\it Now} the $\bar \varphi_t$-equation yields
\begin{equation}
rk (\partial_t) = rk (H) - rk (u) - rk (\bar \varphi), \label{12.44}
\end{equation}
so that setting
\begin{equation}
rk (\bar \varphi) = rk (h) - rk (\partial), \label{12.45}
\end{equation}
the homogenuity is restored $-$ in the $\bar \varphi$-language.  
Obviously, formula (\ref{12.45}) can not be transferred back to the original 
$\varphi$-language, because
\begin{equation}
\varphi = \exp (\bar \varphi). \label{12.46}
\end{equation}
\end{remark}

We shall increasingly deal with similar quasi-homogeneous extensions 
in what follows.

\setcounter{remark}{46}
\setcounter{equation}{47}

\begin{remark}\label{remark:12.47}
 The hierarchies of long wave equations, 
both dispersive and dispersiveless, allow the reduction
\begin{equation}
\{h=0\}. \label{12.48}
\end{equation}
Let's see what happens with the dark extensions under this reduction 
in the zero-dispersion case.  Here
\begin{gather}
H_n = hu^n + O \left(h^2\right), \quad n \in {\mathbb Z}_+, \label{12.49}\\
\frac{\partial H_n}{\partial u} = hnu^{n-1} + O \left(h^2\right), \quad 
\frac{\partial H_n}{\partial h} = u^n + O (h), \label{12.50}
\end{gather}
and therefore the $n^{th}$ flow is
\begin{subequations}\label{12.51}
\begin{gather}
u_t = \partial \left(\frac{\partial H_n}{\partial h}\right) = 
\partial \left(u^n\right) + O (h), \label{12.51a}\\
h_t = \partial \left(\frac{\partial H_n}{\partial u} \right) = 
\partial \left(hnu^{n-1}\right) + O \left(h^2\right). \label{12.51b}
\end{gather}
\end{subequations}
On the constraint $\{h=0\}$ (\ref{12.48}) thus becomes 
\begin{equation}
u_t = \partial \left(u^n\right), \label{12.52}
\end{equation}
which is the flow (\ref{11.2}).  The $\varphi$-equation from the Hamiltonian 
matrix (\ref{12.19}) (discarding $\alpha$ as inessential) is
\begin{equation}
\varphi_t = \varphi_x h^{-1} \frac{\delta H_n}{\delta u} = 
\varphi_x nu^{n-1} + O (h), \label{12.53}
\end{equation}
so that on the constraint $\{h =0\}$ this becomes
\begin{equation}
\varphi_t = nu^{n-1} \varphi_x. \label{12.54}
\end{equation}
This is the case $\beta_n=n$ (\ref{11.6}), (\ref{11.7}), (\ref{11.11}).  
The 2$^{nd}$ dark extension, 
provided by the Hamiltonian matrix (\ref{12.27}), yields:
\begin{subequations}\label{12.55}
\begin{gather}
u_t = 2 \partial \left(\frac{\partial H_{n-1}}{\partial u}
\right) + \partial u \left(\frac{\partial H_{n-1}}{\partial h}\right) =
\partial \left(uu^{n-1}\right) + O (h), \label{12.55a}\\
h_t = u \partial \left(\frac{\partial H_{n-1}}{\partial u} \right) +
(h \partial + \partial h) \left(\frac{\partial H_{n-1}}{\partial h}
\right) = O (h), \label{12.55b}\\
\varphi_t = \varphi_x \frac{\partial H_{n-1}}{\partial h} = 
\varphi_x u^{n-1} + O (h). \label{12.55c}
\end{gather}
\end{subequations}
On the constraint $\{h=0\}$ this becomes
\begin{equation}
u_t = \partial \left(u^n\right), \quad \varphi_t = u^{n-1} \varphi_x. \label{12.56}
\end{equation}
Here $\beta_n=1$, which corresponds to the case $\omega =1$ of 
formula (\ref{11.14}).
\end{remark}

\setcounter{remark}{56}
\setcounter{equation}{57}

\begin{remark}\label{remark:12.57}
 The nonlinear Hamiltonian matrix $B$ (\ref{12.19}) 
looks rather unpretensious but is in fact very remarkable and stands 
at the intersection of many interesting fluid dynamical and Hamiltonian 
theories.  As a simple illustration, pick an arbitrary function $E(u)$, 
and exchange the variable $\varphi$ into the variable
\begin{equation}
\psi = \varphi + E(u). \label{12.58}
\end{equation}
For $\alpha = 0$, the Hamiltonian matrix $B$ (\ref{12.19}) becomes, in the 
new variables $(u, h, \psi)$:
\begin{equation}
B = \bordermatrix{& u & h & \psi \cr
u & 0 & \partial & \displaystyle - \frac{\lambda}{h} (\psi - E)_x \cr
h& \partial & 0 & \partial E_u \cr
\psi& \displaystyle\frac{\lambda}{h} (\psi - E)_x & E_u \partial & 0 \cr}, 
\label{12.59}
\end{equation}
where 
\begin{equation}
\lambda = 1. \label{12.60}
\end{equation}
If we now set (without any visible justification)
\begin{equation}
\lambda = 0, \label{12.61}
\end{equation}
the matrix (\ref{12.59}) becomes
\begin{equation}
B = \begin{matrix} u \\ h \\ \psi \end{matrix}
\begin{pmatrix}  0 & \partial & 0 \\
\partial & 0 & \partial E_u \\
 0 & E_u \partial & 0 \end{pmatrix}. \label{12.62}
\end{equation}
This matrix is still Hamiltonian!  It can be interpreted as follows.  
Start with the $(u, h)$-system (\ref{12.2}): 
\begin{equation}
u_t = \partial \left(\frac{\delta H}{\delta h} \right), \quad h_t 
= \partial \left(\frac{\delta H}{\delta u}\right). \label{12.63}
\end{equation}
Now introduce the new variable
\begin{equation}
\psi = E(u). \label{12.64}
\end{equation}
Then
\begin{equation}
\psi_t = E_u u_t=E_u \partial \left(\frac{\delta H}{\delta u}
\right). \label{12.65}
\end{equation}
The third row of the Hamiltonian matrix $B$ (\ref{12.62}) produces this 
exact motion equation (\ref{12.65}).  However, the third row of the 
Hamiltonian matrix (\ref{12.59}) yields the same result (\ref{12.65}), since 
$\psi$ is in fact $E(u)$.  Thus, Hamiltonian formalism doesn't discriminate 
between the matrices (\ref{12.59}) and (\ref{12.62}).  (However, the fluid-dynamical 
considerations show that the more complex Hamiltonian matrix (\ref{12.59}) 
is the correct one.)
\end{remark}

\section{The Benney hierarchy}

In 1973 Benney [2] derived the following remarkable $2+1-d$ 
generalization of the classical $1+1-d$ long wave system (\ref{12.1}):
\begin{equation}
u_t = uu_x + h_x - u_y \int^y_0 u_x dy, \quad h_t = \left(\int^h_0 u dy\right)
_x. \label{13.1}
\end{equation}
Here $h = h (x, t)$ denotes the height of the fluid free surface, 
counted from the bottom $\{y = 0\}$, and $u = u (x, y, t)$ is the 
horizontal component of the fluid velocity; as always, the sign of time $t$ is 
changed for aesthetic reasons.  When $u$ is independent of $y$,
\begin{equation}
u_y = 0, \label{13.2}
\end{equation}
the Benney system (\ref{13.1}) turns into the classical system (\ref{12.1}).  
What makes the Benney system so remarkable are the following 
amazing properties of it that Benney had found:

1)~Introduce the moments of $u$:
\begin{equation}
A_i = A_i (x, t) = \int^h_0 u^i dy, \quad  i \in {\mathbb Z}_+. \label{13.3}
\end{equation}
Then the system (\ref{13.1}) implies:
\begin{equation}
A_{i,t} = A_{i +1, x} + i A_{i-1} A_{0,x}, \ \ \ i \in {\mathbb Z}_+; 
\label{13.4}
\end{equation}

2)~For each $n \in {\mathbb Z}_+$, the moments system (\ref{13.4}) has an 
infinite number of conserved densities
\begin{equation}
H_0 = A_0, \quad  H_1 = A_1, \quad  H_2 = A_2 + A_0^2, \  \ldots , \ H_n \in A_n +
{\mathbf{Q}}[A_0, \ldots, A_{n-2}]. \label{13.5}
\end{equation}

Later on [17, 18], Manin and myself found that:

3)~(\ref{13.4}) is a Hamiltonian system, in the Hamiltonian structure 
\begin{equation}
B_{nm} = n A_{n+m-1} \partial + \partial m A_{n + m-1}, \quad n, m \in {\mathbb Z}_+, 
\label{13.6}
\end{equation}
and with the Hamiltonian
\begin{equation}
H_2/2 = \left(A_2 + A_0^2\right)/2; \label{13.7}
\end{equation}

4)~All the Hamiltonians $H_n$ (\ref{13.5}) commute with respect to 
the Hamiltonian structure (\ref{13.6}); 

5)~Under the reduction the map (\ref{13.2}):
\begin{equation}
A_i = hu^i, \quad i \in {\mathbb Z}_+, \label{13.8}
\end{equation}
the Hamiltonian structure (\ref{13.6})  collapses into the Hamiltonian 
structure (\ref{12.2}); 

6)~For any Hamiltonian $H = H\{A\}$, the Hamiltonian system 
in the moments space produced by the Hamiltonian structure (\ref{13.6}):
\begin{equation}
A_{n,t} = \sum_{m \geq 0} (n A_{n+m-1} \partial + \partial_m 
A_{n+m-1}) (H_{|m}), \quad H_{|m} = \frac{\delta H}{\delta A_m}, \label{13.9}
\end{equation}
is lifted up by the moments map (\ref{13.3}) from the following system 
in the physical $\{u; h\}$-space:
\begin{subequations}\label{13.10}
\begin{gather}
u_t = \sum_m \left\{\left(u^m H_{|m}\right)_x - u_y \int^y_0 dy \left(u^{m-1} m H_{|m}\right) 
_x \right\}, \label{13.10a}\\
h_t = \sum_m \left(A_{m-1} m H_{|m}\right)_x. \label{13.10b}
\end{gather}
\end{subequations}

We aim in this Section to generalize into the moments language those 
results of the preceding Section which deal with the dispersiveless 
long wave equations; the full dispersive case will be discussed in 
Section~14. 

We start with the $\{\beta =1\}$-case, covered by the Hamiltonian 
matrix $B$ (\ref{12.19}).  We'd like to extend the Kupershmidt--Manin 
Hamiltonian matrix $B$ (\ref{13.6}) into a matrix ${\mathcal B}$ in such a way 
that the reduction $\{A_i = hu^i\}$ of ${\mathcal B}$ reproduces the matrix 
$B$ (\ref{12.19}).  How to approach this problem?

The extension motion equation for $\varphi$, generated by the matrix 
$B$ (\ref{12.19}), is
\begin{equation}
\varphi_t = \left(\frac{1}{h} \varphi_x + \alpha \varphi \partial 
\frac{1}{h}\right) \left(\frac{\delta H}{\delta u}\right). \label{13.11}
\end{equation}
Now, under the reduction $\{A_i = hu^i\}$ (\ref{13.8}), 
\begin{subequations}\label{13.12}
\begin{gather}
\frac{\delta H}{\delta u} = \sum_m \frac{\delta H}{\delta A_m} 
\frac{\partial A_m}{\partial u} = \sum_m H_{|m} hu^{m-1} m = 
\sum_m m A_{m-1} H_{|m}, \label{13.12a}\\
\frac{\delta H}{\delta h} = \sum_m \frac{\delta H}{\delta A_m} 
\frac{\partial A_m}{\partial h} 
= \sum_m H_{|m} u^m = \frac{1}{A_0} \sum_m A_m H_{|m}, \label{13.12b}
\end{gather}
\end{subequations}
and it is {\it plausible} that the $\varphi_t$-equation (\ref{13.11}) comes
out of reduction of the equation
\begin{equation}
\varphi_t = \left(\frac{1}{A_0} \varphi_x + \alpha \varphi 
\partial \frac{1}{A_0} \right) \sum_m m A_{m-1} H_{|m}. 
\label{13.13}
\end{equation}
I say ``plausible'' because the RHS's of formulae (\ref{13.12}) are hugely 
nonunique, so the equation (\ref{13.13}) should be considered at the moment 
as a guess in need of verification.

\setcounter{proposition}{13}
\setcounter{equation}{14}

\begin{proposition}\label{13.14}
The following matrix ${\mathcal B}$ is Hamiltonian 
for arbitrary constant $\alpha$:
\begin{subequations}\label{13.15}
\begin{gather}
{\mathcal B}_{nm} = n A_{n+m-1} \partial + \partial m A_{n+m-1}, \quad  n, 
m \in {\mathbb Z}_+, \label{13.15a}\\
{\mathcal B}_{\varphi m} = (\varphi_x + \alpha \varphi \partial ) 
\frac{m A_{m-1}}{A_0}, \quad {\mathcal B}_{n \varphi} = \frac{n A_{n-1}}{A_0} 
(\alpha \partial \varphi - \varphi_x), \label{13.15b}\\
{\mathcal B}_{\varphi \varphi}=0. \label{13.15c}
\end{gather}
\end{subequations}
\end{proposition}

\begin{proof} We need to verify the identity (\ref{12.20}).  A part of 
this verification, corresponding to the submatrix $\{B_{nm}\}$ (\ref{13.6}) 
in ${\mathcal B}$ (\ref{13.15}), can be omitted, since $\{B_{nm}\}$ is known to be
Hamiltonian.  The remaining verification is long, mind-numbing, not 
illuminating, and so is better left out.  I had done it, and the 
identity (\ref{12.20}) is true.
\end{proof}

\setcounter{example}{15}
\setcounter{equation}{16}

\begin{example}\label{example:13.16}
For the Benney system (\ref{13.7}), 
\begin{equation}
H_{|m} = \frac{1}{2} \delta^2_m + A_0 \delta^0_m, \label{13.17}
\end{equation}
so we get
\begin{subequations}\label{13.18}
\begin{gather}
\varphi_t = (\varphi_x + \alpha \varphi \partial) \left(\frac{A_1}{A_0}\right) \label{13.18a}\\
\phantom{\varphi_t }{}= (\varphi_x + \alpha \varphi \partial) \left(\frac{1}{h} \int^h_0
udy \right). \label{13.18b}
\end{gather}
\end{subequations}
This equation becomes $\varphi_t = u \varphi_x$ (\ref{12.15a}) when 
$0 = u_y = \alpha$. 
\end{example}

Next, let's take up the Hamiltonian matrix $\bar B$ (\ref{12.38}), which, 
in the variable
\begin{equation}
\bar \varphi = \log (\varphi), \label{13.19}
\end{equation}
takes the form
\begin{equation}
\bar B = \bordermatrix{& u & h & \bar \varphi \cr
u & 0 & \partial & 0 \cr
h & \partial & 0 & - \alpha \cr
\bar \varphi & 0 & \alpha & 0 \cr}. \label{13.20}
\end{equation}
Since the $\bar \varphi_t$-equation
\begin{equation}
\bar \varphi_t = \alpha \frac{\delta H}{\delta h} \label{13.21}
\end{equation}
can be rewritten, by formula (\ref{13.12b}), as 
\begin{equation}
\bar \varphi_t = \frac{\alpha}{A_0} \sum_m A_m H_{|m}, \label{13.22}
\end{equation}
we arrive at the following matrix $\bar {\mathcal B}$:
\begin{subequations}\label{13.23}
\begin{gather}
\bar {\mathcal B}_{nm} = n A_{n+m+1} \partial + \partial m A_{n+m-1}, 
\quad n, m, \in {\mathbb Z}_+, \label{13.23a}\\
\bar {\mathcal B}_{\bar \varphi m} = \alpha \frac{A_m}{A_0}, \quad  
\bar {\mathcal B}_{n \bar \varphi} = - \alpha \frac{A_n}{A_0}, \label{13.23b}\\
{\mathcal B}_{\bar \varphi \bar \varphi} = 0. \label{13.23c}
\end{gather}
\end{subequations}

\setcounter{proposition}{23}
\setcounter{equation}{24}

\begin{proposition}\label{proposition:13.24} The matrix $\bar {\mathcal B}$ (\ref{13.23}) 
is {\it not} 
Hamiltonian for $\alpha \not= 0$.
\end{proposition}

\begin{proof} The $\bar \varphi$-entry of the criterion (\ref{12.20}), 
vanishes on the LHS and doesn't on the RHS.
\end{proof}

Thus, the extension (\ref{12.35a})
\begin{equation}
u_t = uu_x + h_x, \quad h_t = (u h)_x, \quad \bar \varphi_t = 
\alpha \left(h + u^2/2\right), \label{13.25}
\end{equation}
and similar ones for the higher flows, remain unlifted into the 
moments space.  The source of the difficulty is easy to pinpoint:  
it is the RHS of the formula (\ref{13.12b}), where $u^m$ was represented 
as $A_m/A_0$ (when $u_y=0$), but could have been equally well 
represented as $A_{m+1}/A_1$ or just about any other rational 
function of $\{A\}$.

We would encounter exactly the same problem with the Hamiltonian 
matrix ${\mathcal B}^{II}$ (\ref{12.27}), since there 
\begin{equation}
\varphi_t = (\varphi_x + \alpha \varphi \partial ) \left({\delta H}
{\delta h}\right). \label{13.26}
\end{equation}
The $\varphi$-independent part of the matrix ${\mathcal B}^{II}$ (\ref{12.27}), 
the Hamiltonian matrix $B^{II}$  (\ref{12.23}), comes out of the second 
Hamiltonian structure of the Benny hierarchy [10, p.~373].  
It is equation (\ref{13.26}), like the equation (\ref{13.21}) before it, that 
resists momentous understanding.  
The same problem appears in the fully 
dispersive case, exemplified by the Hamiltonian matrix 
$\widetilde{\! {\mathcal B}}^{II}$ (\ref{12.30}).

The last remaining Hamiltonian matrix of Section~12, $B^{\mathcal H}$ (\ref{12.43}), 
has no non-zero $\delta H/\delta h$-entries in the 
$\bar \varphi$-row; there
\begin{equation}
\bar \varphi_t = \alpha \frac{\delta H}{\delta u}, \label{13.27}
\end{equation}
and by analogy with formula (\ref{13.13}), we can utilize formula (\ref{13.12a})
 to guess the following generalization of the matrix $B^{\mathcal H}$ 
(\ref{12.43}):
\begin{subequations}\label{13.28}
\begin{gather}
{\mathcal B}^{{\mathcal H}}_{nm} = n A_{n + m +} \partial + \partial m A_{n+m-1}, 
\quad  n, m \in {\mathbb Z}_+, \label{13.28a}\\
{\mathcal B}^{{\mathcal H}}_{\bar \varphi m} = \alpha m A_{m-1}, \quad
 {\mathcal B}^{{\mathcal H}}_{n \bar \varphi} = - \alpha n A_{n-1}, \label{13.28b}\\
{\mathcal B}^{{\mathcal H}}_{\bar \varphi \bar \varphi} = 0. \label{13.28c}
\end{gather}
\end{subequations}

\setcounter{proposition}{28}
\setcounter{equation}{29}

\begin{proposition}\label{proposition:13.29}
 The matrix ${\mathcal B}^{{\mathcal H}}$ (\ref{13.28}) is 
Hamiltonian for any constant $\alpha$.
\end{proposition}

We defer a Proof until Section~15, as the matrix ${\mathcal B}^{{\mathcal H}}$ (\ref{13.28}) 
hides some interesting mathematics behind it.  At the moment let's 
look at how this matrix could have been derived (and in fact was) 
by considerations entirely within the moments space.

The Kupershmidt--Manin Hamiltonian matrix $B$
 (\ref{13.6}) has two distinguished Hamiltonians: 
${\mathcal P} = A_1$, a momentum:
\begin{equation}
X_{{\mathcal P}} = \frac{d}{dx}, \label{13.30}
\end{equation}
and $H = \alpha A_0$, a Casimir:
\begin{equation}
X_{\alpha A_{0}} = 0. \label{13.31}
\end{equation}
If we extend the momentum flow $X_{{\mathcal P}}$ by the Casimir $\alpha A_0$: 
\begin{subequations}\label{13.32}
\begin{gather}
A_{i,t} = A_{i,x}, \quad i \in {\mathbb Z}_+, \label{13.32a}\\
\bar \varphi_t = \alpha A_0 \quad (\Leftrightarrow \varphi_t = 
\alpha A_0 \varphi), \label{13.32b}
\end{gather}
\end{subequations}
and ask for such an extension of the arbitrary flow (\ref{13.9}) with an 
$x$-independent Hamiltonian $H$, of the form
\begin{equation}
\bar \varphi_t = ? (H), \label{13.33}
\end{equation}
that it commutes with the flow (\ref{13.32}), we shall get
\begin{equation}
X_{{\mathcal P}} (? (H)) = [?(H)]_x = X_H (\alpha A_0) = \left(\sum_m \alpha m 
A_{m-1} H_{|m}\right)_x, \label{13.34}
\end{equation} 
so that
\begin{equation}
? (H) = \sum_m \alpha m A_{m-1} H_{|m}, \label{13.35}
\end{equation}
and we thus recover the ${\mathcal B}^{{\mathcal H}}_{\bar \varphi m}$-entry (\ref{13.28b}) 
of the matrix ${\mathcal B}^{{\mathcal H}}$ (\ref{13.28}).

\section{The KP hierarchy}

The Benney hierarchy of Section~13 is the quasiclassical limit of the 
KP hierarchy [27, 4]
\begin{gather}
L_t = [P_+, L], \label{14.1}\\
L = \partial + \sum^\infty_{i=0} \partial^{-i-1} A_i, \label{14.2}\\
P=L^N, \quad N \in {\mathbb N}. \label{14.3}
\end{gather}
The (1$^{st}$) Hamiltonian structure $B_{nm}$ (\ref{13.6}) of the Benney
hierarchy, 
\begin{equation}
B_{nm} = n A_{n+m-1} \partial + \partial m A_{n+m-1}, \quad 
n, m \in {\mathbb Z}_+, \label{14.4}
\end{equation}
is the quasiclassical limit of the (1$^{st}$) Hamiltonian structure 
of the KP hierarchy
\begin{equation}
\bar B_{nm} = \sum_\mu \left[\left(\begin{array}{c} n \\ \mu\end{array}\right) A_{n+m- \mu}
 \partial^\mu - (- \partial)^\mu \left(\begin{array}{c} m \\ \mu\end{array}\right) 
A_{n + m - \mu} \right], \quad n, m \in {\mathbb Z}_+. \label{14.5}
\end{equation}

We aim in this Section to find dispersive analogs of the two 
zero-disperson results of the preceding Section, the Hamiltonian 
extensions ${\mathcal B}$ (\ref{13.15}) and ${\mathcal B}^{{\mathcal H}}$ (\ref{13.28}).

We begin with the Hamiltonian matrix ${\mathcal B}$ (\ref{13.15}), an 1-dimensional 
extension of the Hamiltonian  matrix $B$ (\ref{14.4}).  Recall that the 
latter Hamiltonian matrix is connected via the reduction map
\begin{equation}
A_i = hu^i, \quad i \in {\mathbb Z}_+, \label{14.6}
\end{equation}
with the Hamiltonian matrix $b$ (\ref{12.2}) 
\begin{equation}
b = \bordermatrix{& u & h \cr
u & 0 & \partial \cr
h & \partial & 0 \cr} . \label{14.7}
\end{equation}
Similarly [12, p.~66], the full KP Hamiltonian matrix $\bar B$ 
(\ref{14.5}) is connected to this same Hamiltonian matrix $b$ (\ref{14.7}) via 
the differential reduction map
\begin{equation}
A_i = h Q_i (u), \quad i \in {\mathbb Z}_+, \quad Q_i (u) = (\partial + 
u)^i (1), \label{14.8}
\end{equation}
a dispersive analog of the zero-dispersion reduction map (\ref{14.6}).  
Now, the motion equation for $\varphi$ generated by the Hamiltonian 
matrix $\{b$ (\ref{14.7}) extended into the Hamiltonian matrix (\ref{12.19})$\}$, 
is:
\begin{equation}
\varphi_t = (\varphi_x + \alpha \varphi \partial) \frac{1}{h} 
\frac{\delta H}{\delta u}, \label{14.9}
\end{equation}
and under the reduction map (\ref{14.8}), 
\begin{gather}
\frac{\delta H}{\delta u} = \sum_m \left(\frac{DA_m}{Du} \right)
^\dagger \left(\frac{\delta H}{\delta A_m}\right) \nonumber\\
\phantom{\frac{\delta H}{\delta u}} = \sum \left(h \frac{DQ_m}{Du}\right)^\dagger \left(H_{|m}\right) = 
\sum \left(\frac{DQ_m}{Du} \right)^\dagger h(H_{|m}). \label{14.10}
\end{gather}
By formula (\ref{10.28a}), 
\begin{subequations}\label{14.11}
\begin{gather}
\left(\frac{D Q_m}{Du}\right)^\dagger = \left(\sum_{s \geq 1} \left(\begin{array}{c} m \\ s\end{array}\right)
 Q_{m-s} \partial^{s-1}\right)^\dagger = \sum_{s \geq 1} 
(- \partial)^{s-1} \left( \begin{array}{c} m \\ s\end{array}\right) Q_{m-s}\label{14.11a}\\
 \Rightarrow \ \left(\frac{D Q_m}{Du}\right)^\dagger h = \sum_{s \geq 0} 
(- \partial)^s \left(\begin{array}{c} m \\ s+1\end{array}\right) Q_{m-1-s} h \nonumber\\
\qquad {} = \sum_{s \geq 0} (- \partial)^s \left(\begin{array}{c} m \\ s+1\end{array}\right) A_{m-1-s}\label{14.11b}
\end{gather}
\end{subequations}
\begin{equation}
\Rightarrow \ \frac{\delta H}{\delta u} = \sum_{s \geq 0} (-\partial)^s 
\left(\begin{array}{c} m \\ s+1\end{array}\right) A_{m-1-s} (H_{|m}). \label{14.12}
\end{equation}
We thus arrive at the equation
\begin{equation}
\varphi_t = (\varphi_x + \alpha \varphi \partial) \frac{1}{A_0} 
\sum_{s,m} (- \partial)^s \left(\begin{array}{c} m \\ s+1\end{array}\right) (A_{m-1-s} 
H_{|m}). \label{14.13}
\end{equation}
Hence, a plausible extension of the KP matrix $\bar B$ (\ref{14.5}) is:
\begin{subequations}\label{14.14}
\begin{gather}
\bar {\mathcal B}_{nm} = \bar B_{nm} = \sum_\mu \left[\!\left(\!\begin{array}{c} n \\ \mu\end{array}\!\right) 
\! A_{n+m-\mu} \partial^u - (- \partial)^\mu \left(\!\begin{array}{c} m \\ \mu\end{array}\!\right) \!
A_{n+m-\mu}\!\right]\!, \ n, m \in {\mathbb Z}_+,\! \label{14.14a}\\
\bar {\mathcal B}_{\varphi m} = (\varphi_x + \alpha \varphi \partial) 
\frac{1}{A_0} \sum_s (- \partial)^s \left(\begin{array}{c} m \\ s+1\end{array}\right) 
A_{m-1-s}, \label{14.14b}\\
\bar {\mathcal B}_{n \varphi} = \sum_s \left(\begin{array}{c} n \\ s+1\end{array}\right) A_{n-1-s} 
\partial^s \frac{1}{A_0} (\alpha \partial \varphi - \varphi_x), 
\quad \bar {\mathcal B}_{\varphi\varphi}=0. \label{14.14c}
\end{gather}
\end{subequations}
This matrix $\bar {\mathcal B}$ has the following 2 properties:

1) Its quasiclassical limit is the Hamiltonian matrix 
${\mathcal B}$ (\ref{13.15}); 

2) Under the reduction $A_i = h Q_i (u)$ (\ref{14.8}), the matrix 
$\bar {\mathcal B}$ (\ref{14.14}) reduces to the Hamiltonian matrix (\ref{12.19}).

\setcounter{proposition}{14}
\setcounter{equation}{15}

\begin{proposition}\label{proposition 14.15} The matrix $\bar {\mathcal B}$ (\ref{14.14}) is 
Hamiltonian for any constant $\alpha$.
\end{proposition}

\begin{proof}
We have to verify the identity (\ref{12.20}) for the 
matrix $\bar {\mathcal B}$:
\begin{equation}
\bar {\mathcal B} \delta [Y^t \bar {\mathcal B} (X)] = D[\bar {\mathcal B} (Y)] \bar {\mathcal B} 
(X) - D [\bar {\mathcal B} (X)] \bar {\mathcal B} (Y), \label{14.16}
\end{equation}
with $X = ({\pbf{X}}, u)$ and $Y = ({\pbf{Y}}, v)$. 

To say that this verification is long, mind-numbing, and not 
illuminating, would be to indulge in coy understatements.  It is 
all that, but in the end one arrives at some non-evident 
identities to be checked, so I provide below a guide to the 
verification sequence.

1) Introduce the variable $\bar \varphi = \log (\varphi)$ 
instead of $\varphi$.  The operators $(\varphi_x + \alpha \varphi 
\partial)$ in $\bar{\mathcal B}_{\varphi m}$ and $(\alpha \partial \varphi - 
\varphi_x)$ in $\bar {\mathcal B}_{n \varphi}$ become  
$(\alpha \partial + \bar \varphi_x)$ and $(\alpha \partial - 
\bar \varphi_x)$, respectively; 

2) The variables $u$ and $v$, after grouping and cancellations, 
should enter only in combinations
\begin{equation}
{\mathcal U} = \frac{\alpha u_x - \bar \varphi_x u}{A_0}, \quad
 {\mathcal V} = \frac{\alpha v_x - \bar \varphi_x v}{A_0}; \label{14.17}
\end{equation}

3) In checking the $\bar \varphi$-entry of the vectors 
in (\ref{14.16}), the ${\mathcal U}$- and ${\mathcal V}$-terms cancel out, and the remaining 
expressions can be divided out from the left by the operator 
$(\alpha \partial + \varphi_x) \frac{1}{A_0}$.  What remains 
is a trilinear differential identity in $X = X_m$, $Y = Y_n$, $A = A_L$,
 for all fixed ($m, n, L)$.  To prove it, pass to the 
symbols by substituting 
\begin{equation}
\partial \approx \partial/\partial z, \quad  X \approx e^{x z}, \quad
Y \approx e^{yz}, \quad A \approx e^{az}, \label{14.18}
\end{equation}
and then use Newton's binomial summation repeatedly; 

4) In the $A_N$-entry of the vectors in (\ref{14.16}), the terms 
bilinear in $(u,v)$ check out; the terms bilinear in $(X,Y)$ are 
accounted for by the known Hamiltonian character of the submatrix 
$\bar B$ (\ref{14.5}) of $\bar {\mathcal B}$ (\ref{14.14}); the remaining terms are 
bilinear in $(X, {\mathcal V})$ and $(Y, {\mathcal U})$, and due to the skewsymmetry of 
${\mathcal B}$, only the $(X, {\mathcal V})$-entries need to be checked out.  After 
repeated cancellations and simplifications, one arrives at the 
identity
\begin{gather}
- \sum_{\nu + k = L} \left[\left(\begin{array}{c} N \\ \nu\end{array}\right) A \partial^\nu
- \left(\begin{array}{c} m - k - 1 \\ \nu\end{array}\right) (- \partial)^\nu A\right]
\left[X \left(\begin{array}{c} m \\ k+1\end{array} \right) {\mathcal V}^{(k)} \right] \nonumber\\
\qquad {}= - \sum_{\nu+k=L} \left[\left(\begin{array}{c} N \\ \nu\end{array}\right) X^{(\nu)} - 
\left(\begin{array}{c} m \\ \nu\end{array}\right) (- \partial)
^\nu X \right] \left[A \left(\begin{array}{c} N+m - \nu \\ k+1\end{array}\right) 
{\mathcal V}^{(k)}\right] \nonumber\\
\qquad {}+ \sum_{k+\nu=L} \left(\!\begin{array}{c} N \\ k+1\!\end{array}\right) {\mathcal V}^{(k)} \left[\left(
\!\begin{array}{c} N-k-1 \\ \nu\end{array}\!\right) AX^{(\nu)} - \left(\!\begin{array}{c} m \\ \nu
\end{array}\!\right) 
(-\partial)^\nu (AX)\right],\label{14.19}
\end{gather}
for all fixed $N$, $m$, $L$.  Passing to symbols, multiplying by 
$v\tau^{L+1}$, and summing on $L$, we arrive at the identity
\begin{gather}
- [1 + \tau (x+v)]^N \left[(1 + \tau v)^m-1\right] + [1 - \tau (a + x + v)]^m 
\nonumber\\
\quad {}\times
\left\{\left[1+\frac{\tau v}{1-\tau (a+x+v)}\right]^m -1\right\}=  (1 + \tau x)^N - [1 - \tau (a + x + v)]^m
 \nonumber\\
\quad {}- (1 + \tau v)^{N+m}  \left[1 + \frac{\tau x}{1 + \tau v} \right]^N 
 + (1 + \tau v)^{N+m} \left[1 - \frac{\tau (a + x + v)}{1 + \tau v} 
\right]^m \nonumber\\
\quad {}-[1 - \tau (a+x)]^m \left[(1 + \tau v)^N -1\right] + (1+\tau x)^N \left[\left(
1 + \frac{\tau v}{1 + \tau x}\right)^N-1 \right], \label{14.20}
\end{gather}
which is obviously true.
\end{proof}

Our next extended Hamiltonian matrix is ${\mathcal B}^{{\mathcal H}}$ (\ref{13.28}).  To 
find its dispersive analog we use the motion equation (\ref{13.27})
\begin{subequations}\label{14.21}
\begin{equation}
\bar \varphi_t = \alpha \frac{\delta H}{\delta u} \label{14.21a}
\end{equation}
and formula (\ref{14.21}).  We find
\begin{equation}
\bar \varphi_t = \alpha \sum_{s, m} (- \partial)^s \left(\!\begin{array}{c} m \\ 
s+1\end{array}\!\right) A_{m-1-s} (H_{|m}). \label{14.21b}
\end{equation}
\end{subequations}
The resulting matrix $\bar {\mathcal B}^{{\mathcal H}}$ is 
\begin{subequations}\label{14.22}
\begin{gather}
\bar {\mathcal B}^{{\mathcal H}}_{nm} =  
\bar B_{nm} = \sum_\mu \left[\!\left(\!\begin{array}{c} n \\ \mu\end{array}\!\right) \!
A_{n+m- \mu}
 \partial^\mu - (- \partial)^\mu \left(\!\!\begin{array}{c} m \\ \mu\end{array}\!\!\right) \!
A_{n + m - \mu}\! \right]\!, \ n, m \in {\mathbb Z}_+,  \label{14.22a}\\
\bar {\mathcal B}^{{\mathcal H}}_{\bar \varphi m} = \alpha \sum_s (- \partial)^s 
\left(\begin{array}{c} m \\ s+1\end{array}\right) A_{m-1-s}, \label{14.22b}\\
\bar {\mathcal B}^{{\mathcal H}}_{n \bar \varphi} = - \alpha \left(\begin{array}{c} n \\ s+1\end{array}
\right) A_{n-1-s} \partial^s, \quad \bar {\mathcal B}^{{\mathcal H}}_{\bar \varphi \bar
\varphi} = 0. \label{14.22c}
\end{gather}
\end{subequations}
This matrix $\bar {\mathcal B}^{{\mathcal H}}$ (\ref{14.22}), like the matrix $\bar {\mathcal B}$ 
(\ref{14.14}) before it, has the following 2 properties:

1) Its quasiclassical limit is the matrix ${\mathcal B}^{{\mathcal H}}$ (\ref{13.28}); 

2) Under the reduction $A_i = h Q_i (u)$ (\ref{14.8}), the 
matrix $\bar {\mathcal B}^{{\mathcal H}}$ reduces to the Hamiltonian matrix 
${\mathcal B}^{{\mathcal H}}$ (\ref{12.43}).

\setcounter{proposition}{22}
\setcounter{equation}{23}

\begin{proposition}\label{proposition:14.23}
 The matrix $\bar {\mathcal B}^{{\mathcal H}}$ (\ref{14.22}) is 
Hamiltonian for any constant $\alpha$.
\end{proposition}

Since this matrix is {\it linear} in the variables $\{A\}$, there is 
a Lie algebra lurking behind it.  We examine the mathematics 
responsible for this matrix to be Hamiltonian in the next Section.  

At the moment, let us derive the matrix $\bar {\mathcal B}^{{\mathcal H}}$ (\ref{14.22}) 
in purely $\{A\}$-language, like we did at the end of Section~13 with 
the quasiclassical limit ${\mathcal B}^{{\mathcal H}}$ (\ref{13.28}) of this matrix.

First, $H = A_0$ is a Casimir of the Hamiltonian matrix $\bar B$ 
(\ref{14.5}), because
\begin{gather}
\bar B_{N0} = \sum_{\mu >0} \left(\begin{array}{c} N \\ \mu\end{array}\right) A_{N-\mu} 
\partial^\mu = \sum_\mu \left(\begin{array}{c} N \\ \mu + 1\end{array} \right) A_{N-1-\mu} 
\partial^{\mu+1} \label{14.24}\\
\Rightarrow  \ 
X_{A_{0}} (A_N) = 0, \quad \forall \; N \in {\mathbb Z}_+. \label{14.25}
\end{gather}
Next, ${\mathcal P} = A_1$ is a momentum:
\begin{gather}
\bar B_{N1} = \sum_\mu \left(\begin{array}{c} N \\ \mu\end{array}\right) A_{N+1-\mu} 
\partial^\mu - A_{N+1} + \partial A_N \nonumber\\
\phantom{\bar B_{N1}} = 
\sum_\mu \left(\begin{array}{c} N \\ \mu+1\end{array}\right) A_{N-\mu} \partial^{\mu+1} 
+ \partial A_{N} \label{14.26}
\end{gather}
\vspace{-3mm}
\begin{subequations}\label{14.27}
\begin{equation}
\Rightarrow \ X_{A_{1}} (A_N) = \partial (A_N), \quad  \forall \; N \in {\mathbb Z}_+. \label{14.27a}
\end{equation}
Therefore, if we extend the flow $X_{\mathcal P}$ by the equation 
\begin{equation}
\bar \varphi_t = \alpha A_0, \label{14.27b}
\end{equation}
\end{subequations}
and every other flow $X_H$ with $x$-independent Hamiltonian 
$H$, by the equation
\begin{equation}
\bar \varphi_t = ? (H), \label{14.28}
\end{equation}
and require that the thus extended flows still commute with the 
flow (\ref{14.27}), we find:
\begin{gather}
X_{A_{1}} [? (H)] = [?(H)]_x = X_H (\alpha A_0) = \alpha \sum_m 
\bar B_{0m} (H_{|m}) \nonumber\\
\qquad {} \overset{\rm [by \ adjoint \ of \ (\ref{14.24})]}{=} \ 
 - \alpha \sum_{m, \mu} (- \partial)^{\mu +1} A_{m-1-\mu} 
\left(\begin{array}{c} m \\ \mu + 1\end{array}\right) \left(H_{|m}\right)\label{14.29}\\
\Rightarrow \ ? (H) = \alpha \sum_{\mu, m} (- \partial)^\mu \left(\begin{array}{c} m \\ 
\mu +1\end{array} \right) A_{m-1-\mu} \left(H_{|m}\right). \label{14.30}
\end{gather}
Substituting this in (\ref{14.28}), we recover precisely the equation 
(\ref{14.21b}) and hence the matrix $\bar {\mathcal B}^{{\mathcal H}}$ (\ref{14.22}).

\setcounter{remark}{30}
\setcounter{equation}{31}

\begin{remark}\label{remark:14.31}
 The KP Hamiltonian matrix $\bar B$ (\ref{14.5}), 
under the differential reduction $A_i = h Q_i (u)$ (\ref{14.8}), 
reduces to the $1+1-d$ Hamiltonian matrix $b$ (\ref{14.7}); the KP hierarchy 
is thereby reduced to the dispersive $DWW$ hierarchy of Section~12.  
Similar conclusion applies to the quasiclassical limit of this picture, 
with the Benney hierarchy being reduced to the zero-dispersion long 
wave hierarchy of Section~12.  {\it However}, the Benney hierarchy also 
comes out of a genuine $2+1-d$ hierarchy under the moments map 
$A_i = \int^h_0 u^i dy$ (\ref{13.3}).  There must exist a common picture 
uniting both these properties, dispersion and $2+1-d$, something like 
the map 
\begin{equation}
A_i = \int^h_0 Q_i (u) dy, \label{14.32}
\end{equation}
reproducing the KP hierarchy out of some $2+1-d$ dispersive 
hierarchy.  None has been found so far, and this remains one of the 
most important unsolved problems in the theory of integrable 
systems, the true Holy Grail.
\end{remark}

\setcounter{remark}{32}
\setcounter{equation}{33}

\begin{remark}\label{remark 14.33}
 In this Section we have constructed two 
1-component Hamiltonian extensions of the KP hierarchy, those 
given by the Hamiltonian matrices $\bar {\mathcal B}$ (\ref{14.14}) and $\bar {\mathcal B}
^{{\mathcal H}}$ (\ref{14.22}).  These extensions are {\it{different}} from still 
another 1-component Hamiltonian extension of the KP hierarchy, the 
one provided by the mKP hierarchy ([15] and [16, \S~6.2]).  Formula 
(6.2.3) in [16] shows that that third extension is in fact of the 
extended Lax type of Section~5; the unexpected fact proven in [16] is 
that that Lax-type extension is Hamiltonian --- we have found nothing 
similar in this paper, even though the Hamiltonian property of 
extended Lax systems could not be ruled out with certainty.
\end{remark}

\section{One-dimensional linear extensions\\ of linear 
Hamiltonian matrices}

In Sections~13 and 14 we derived two conjecturally Hamiltonian 
matrices, ${\mathcal B}^{{\mathcal H}}$ (\ref{13.28})
 and $\bar {\mathcal B}^{{\mathcal H}}$ (\ref{14.22}):
\begin{subequations}\label{15.1}
\begin{gather}
{\mathcal B}^{{\mathcal H}}_{nm} = n A_{n+m-1} \partial + \partial m A_{n+m-1}, 
\quad n, m \in {\mathbb Z}_+, \label{15.1a}\\
{\mathcal B}^{{\mathcal H}}_{\bar \varphi m} = \alpha m A_{m-1}, \quad
 {\mathcal B}^{{\mathcal H}}_{n \bar \varphi} = - \alpha n A_{n-1}, \label{15.1b}\\
{\mathcal B}^{{\mathcal H}}_{\bar \varphi \bar \varphi} = 0, \label{15.1c}
\end{gather}
\end{subequations}
\vspace{-5mm}
\begin{subequations}\label{15.2}
\begin{gather}
\bar {\mathcal B}^{{\mathcal H}}_{nm} = \sum_\mu \left[\left(\begin{array}{c} n \\ \mu\end{array}\right) 
A_{n+m- \mu} \partial^\mu - (- \partial)^\mu \left(\begin{array}{c} m \\ \mu\end{array}\right) 
A_{n+m-\mu}\right], \quad n, m \in {\mathbb Z}_+, \label{15.2a}\\
\bar {\mathcal B}^{{\mathcal H}}_{\bar \varphi m} = \alpha \sum_s (-\partial)^s 
\left(\begin{array}{c} m \\ s+1\end{array} \right) A_{m-1-s}, \label{15.2b}\\
\bar {\mathcal B}^{{\mathcal H}}_{n \bar \varphi} = - \alpha \sum_s \left(\begin{array}{c} n \\ 
s+1\end{array}\right) A_{n-1-s} \partial^s, \quad \bar {\mathcal B}^{{\mathcal H}}_{\bar \varphi \bar 
\varphi} = 0. \label{15.2c}
\end{gather}
\end{subequations}
The 1$^{st}$ of these matrices, ${\mathcal B}^{{\mathcal H}}$ (\ref{15.1}), is supposed to be 
the quasiclassical limit of the 2$^{nd}$, $\bar {\mathcal B}^{{\mathcal H}}$ (\ref{15.2}), 
although some work is required to make this supposition precise.

Each of these two matrices has the same general structure:  a 
linear Hamiltonian matrix; extended linearly by one extra component; 
with extended entries involving only old variables; and with no 
self-interaction  (i.e., $B_{\bar \varphi \bar \varphi}=0.)$ 

Since a linear Hamiltonian matrix $B = B\{q\}$ corresponds uniquely 
to (the dual space of) a Lie algebra, say ${\mathcal G}$, via the relation 
(see [14]) 
\begin{equation}
X^t B(Y) \sim q^t [X, Y], \quad \forall \; X, Y, \label{15.3}
\end{equation}
where $\sim$ stands for equality modulo trivial elements (``divergencies''), 
let us see exactly what kind of Lie algebras we get with these 
types of one-dimensional extensions.

So, suppose we start with a Lie algebra ${\mathcal G}$, which we assume for the 
time being to be finite-dimensional-like:  this means that no 
{\it functional} operations $-$ like derivations and shifts $-$ are 
involved in the commutators $-$ both in ${\mathcal G}$ and in the 1-dimensional 
extension $\widehat{{\mathcal G}}$ of ${\mathcal G}$.  We choose a basis $(e_i, e)$ in 
$\widehat{{\mathcal G}}$, with $(e_i)$ being a basis in ${\mathcal G}$.  Denote by $q$ 
the coordinates on ${\mathcal G}^*$, so that the Hamiltonian matrix $b=b({\mathcal G})$ 
attached to ${\mathcal G}^*$ has the form
\begin{equation}
b_{ij} = \sum_k c_{ij}^k q_k \label{15.4}
\end{equation}
where $\left\{c_{ij}^k\right\}$ are the structure constants of ${\mathcal G}$ in the basis 
$(e_i)$:
\begin{equation}
X^t b (Y) = \sum_{ij} X^i b_{ij} Y^j = \sum q_k c_{ij}^k X^i Y^i = 
\sum_k q_k \left[\sum_i X^i e_i, \sum_j X^j e_j\right]^k. \label{15.5}
\end{equation}
Thus, our extended matrix $\hat b = \hat b (\widehat{{\mathcal G}})$ 
looks like 
\begin{equation}
\hat b = \bordermatrix{& q_j & \bar \varphi \cr
q_i & \sum\limits_k c_{ij}^k q_k & - \sum\limits_k \theta_i^k q_k \cr
\bar \varphi & \sum\limits_k \theta_j^k q_k & 0 \cr} , \label{15.6}
\end{equation}
where $\theta_j^k$ are some constants.  Now, by formula (\ref{15.3}), the 
commutator in $\widehat{{\mathcal G}}$ associated with the matrix $\hat b$ 
(\ref{15.6}) can be extracted from the following calculation
\begin{gather}
({\pbf{X}}, u)^t \hat b ({\pbf{Y}}, v) = \sum X^i Y^j c_{ij}^k q_k + 
\sum q_k \sum_j \theta^k_j \left(u Y^j - v X^j\right) \nonumber\\
\qquad {}= \sum_k q_k \left\{ \sum_{ij} c_{ij}^k X^i Y^i + \sum_{j} \theta_j^k 
\left(uY^j - v X^j\right)\right\} . \label{15.7}
\end{gather}
Therefore, if we denote by $\widehat{\theta}: {\mathcal G} \rightarrow {\mathcal G}$ the 
operator acting on ${\mathcal G}$ by the rule
\begin{equation}
\hat \theta (Y)^k = \sum_j \theta_j^k Y^j, \label{15.8}
\end{equation}
the commutator in $\widehat{{\mathcal G}}$  has the form
\begin{equation}
\left[\left(\begin{array}{c} X \\ u\end{array}\right), \left(\begin{array}{c} Y \\ v\end{array}\right)\right] = 
\left(\begin{array}{c} [X,Y] + u \widehat{\theta} (Y) - v \widehat{\theta} 
 (X) \\ 0\end{array}\right). \label{15.9}
\end{equation}
The Jacobi identity for the commutator (\ref{15.9}) is equivalent to the 
equality
\begin{equation}
[ \widehat{\theta}(X), Y] - [\widehat{\theta}(Y),X]
 = \widehat{\theta}([X, Y]), \quad \forall \; X, Y \in  {\mathcal G}. \label{15.10}
\end{equation}
Thus, the matrix $\hat b$ (\ref{15.6}) is Hamiltonian iff the operator 
$\widehat{\theta}: {\mathcal G} \rightarrow  {\mathcal G}$ is a 
derivation of ${\mathcal G}$.  In other words, there is a one-to-one correspondence 
between derivations of a Lie algebra ${\mathcal G}$ and linear one-dimensional 
extensions of the Hamiltonian matrix $b({\mathcal G})$ attached to ${\mathcal G}$.

Although it may be not immediately obvious, this natural one-to-one 
correspondence between derivations and one-dimensional extensions is a 
purely finite-dimensional-like phenomenon which doesn't carry over to 
functional Lie algebras.  We shall see about this breakdown 
presently.  Notice that both Lie algebras ${\mathcal G}$ entering formulae (\ref{15.1a}) 
and (\ref{15.2a}) are {\it differential}, not finite-dimensional-like, 
so our analysis above doesn't apply to the two matrices we are 
interested in, ${\mathcal B}^{{\mathcal H}}$ 
(\ref{15.1}) and $\bar {\mathcal B}^{{\mathcal H}}$ (\ref{15.2}).  
Thus, we need to analyze each one of these matrices directly.

For the matrix ${\mathcal B}^{{\mathcal H}}$, (\ref{15.1}) formula (\ref{15.3}) yields:
\begin{subequations}\label{15.11}
\begin{gather}
[(X, u), (Y, v)]^k = \sum_{n+m=k+1} (n X^n Y^m_x - m Y^m X^n_x)\label{15.11a}\\
\qquad {}+ \alpha (k+1) \left(u Y^{k+1} - v X^{k+1}\right). \label{15.11b}
\end{gather}
\end{subequations}
To get a better handle on this commutator, let's pass on to the 
generating functions, considering instead of the vector 
$X = (X^n (x))$ 
the function $\widetilde{X} = \widetilde{X} (x, p) = 
\sum\limits^\infty_{n=0} X^n p^n$.  Formula (\ref{15.11}) then becomes
\begin{gather}
\left[\left(\widetilde{X}, u\right), \left(\widetilde{Y}, v\right)\right] = \sum^\infty_{k=0} 
p^k \left[(X, u), (Y, v)\right]^k \nonumber\\
\qquad {} = \sum_{nm} \left[n X^n p^{n-1} \left(Y^m p^m\right)_x - m Y^m p^m 
\left(X^n p^n\right)_x\right]\nonumber\\
\qquad {} + \alpha \sum_k \left[u(k+1) Y^{k+1} p^k - v (k+1) X^{k+1} p^k\right]
\nonumber\\
\qquad{}= \left(\widetilde{X}_p \widetilde{Y}_x - \widetilde{X}_x \widetilde{Y}_p\right) 
+ \alpha \left(u \widetilde{Y}_p -  v \widetilde{ X}_p\right)  \label{15.12}\\
\qquad {}= \left\{ \widetilde{X}, \widetilde{Y}\right\} + \alpha \left(\left\{ X, - \int
v\right\} + \left\{ - \int u, Y \right\}\right) \nonumber\\
\qquad {}=  \left\{ \widetilde{X} - \alpha \int u, \widetilde{Y} - 
\alpha \int v\right\}, \label{15.13}
\end{gather}
where
\begin{equation}
\{a, b\} = a_p b_x - a_x b_p \label{15.14}
\end{equation}
is the ususal Poisson bracket on the $(x, p)$-plane.  Thus, finally, 
our commutator is
\begin{equation}
\left[\left(\begin{array}{c} \widetilde{X} \\ u\end{array}\right), 
\left(\begin{array}{c} \widetilde{Y} \\ v\end{array}\right)\right] = \left(\begin{array}{c}\left\{\widetilde{X} 
- \alpha \int u, \widetilde{Y} - \alpha \int v\right\} \\ 0\end{array} \right). 
\label{15.15}
\end{equation}
The Jacobi identity for this commutator follows at once from that 
for the Poisson bracket (\ref{15.14}) on the $(x, p)$-plane.

Formula (\ref{15.15}) suggests the following general 

\setcounter{proposition}{15}
\setcounter{equation}{16}

\begin{proposition}\label{proposition:15.16}
Let ${\mathcal G}$ be a Lie algebra over  
 (functional) ring ${\mathcal R}$ and let ${\mathcal H} \subset {\mathcal G}$ be an {\it  abelian}  
subalgebra of ${\mathcal G}$.  Let ${\mathcal O}: {\mathcal R} \rightarrow {\mathcal H}$
 be a linear 
operator.  Then the following commutator defines a Lie algebra 
structure on ${\mathcal G} + {\mathcal R}$:
\begin{equation}
\left[\left(\begin{array}{c} X \\ u\end{array}\right), \left(\begin{array}{c} Y \\ v\end{array}\right)
\right] = \left(\begin{array}{c} [X + {\mathcal O} (u), Y + {\mathcal O} (v)] \\ 0\end{array}
\right). \label{15.17}
\end{equation}
\end{proposition}

{\bf Proof} is obvious.

Formula (\ref{15.15}) is the particular case
\begin{equation}
[{\mathcal O}(u)]^k = - \alpha \delta^k_0 \int u \label{15.18}
\end{equation}
of the Proposition \ref{proposition:15.16}, for formula (\ref{15.11a}) shows that 
${\mathcal H} = {\mathcal R} \left\{ \delta^k_0\right\}$ is an abelian subalgebra of ${\mathcal G}$ 
(\ref{15.11a}).  (More generally, if $M$ is a manifold and ${\mathcal G} =
 C^{\infty}(T^* M)$, the Lie algebra of smooth functions with respect to 
the standard Poisson bracket on $T^*M$, we can take ${\mathcal H} = 
C^\infty (M)$.)

 The 2$^{nd}$ case, $\bar {{\mathcal B}}^{{\mathcal H}}$ (\ref{15.2}), is rather similar.  
First, for the Lie algebra ${\mathcal G}$ itself, formula (\ref{15.2a}) yields:
\begin{equation}
[X, Y]^k = \sum_{n+m=k+\mu} \left[\left(\begin{array}{c} n \\ \mu\end{array}\right) 
X^n Y^{m(\mu)} - \left(\begin{array}{c} m \\ \mu\end{array}\right) X^{n (\mu)} Y^m\right]. 
\label{15.19}
\end{equation}
Passing to the generating functions $\widetilde{X} = \sum\limits_k 
X^k p^k,$ etc., we can rewrite formula (\ref{15.19}) as
\begin{subequations}\label{15.20}
\begin{gather}
\left[\widetilde{X}, \widetilde{Y}\right] = \sum \left\{X^n p^{n-\mu} 
\left(\begin{array}{c} n \\ \mu\end{array}\right) \left(Y^m p^m\right)^{(\mu)} - Y^m p^{m-\mu} 
\left(\begin{array}{c} m \\ \mu\end{array}\right) \left(X^n p^n\right)^{(\mu)}\right\} \label{15.20a}\\
\phantom{\left[\widetilde{X}, \widetilde{Y}\right]}= \widetilde{X} \circ \widetilde{Y} - \widetilde{Y}\circ
\widetilde{X}, \label{15.20b}
\end{gather}
\end{subequations}
where $\circ$ is the associative multiplication of (symbols of) 
differential operators:
\begin{equation}
\widetilde{X} \circ \ \widetilde{Y} = \sum_\mu \frac{1}{\mu} 
\left[\left(\frac{\partial}{\partial p}\right)^\mu \left(\widetilde{X}\right)
\right] \left[\left(\frac{\partial}{\partial x} \right)^\mu 
\left(\widetilde{Y}\right)\right]. \label{15.21}
\end{equation}
Second, for the extended algebra $\widehat{{\mathcal G}}$ we get from 
formulae (\ref{15.2}) and (\ref{15.3}):
\begin{subequations}\label{15.22}
\begin{gather}
[(X, u), (Y, v)]^k = \left\{{\rm old}\; [X, Y]^k\right\} \label{15.22a}\\
\qquad {}- \alpha \sum_s \left(\begin{array}{c} k+s+1 \\ k\end{array}\right) \left(X^{k+s+1} v^{(s)} - 
Y^{k+s+1} u^{(s)}\right). \label{15.22b}
\end{gather}
\end{subequations}
Passing to the generating functions and noticing that 
\begin{equation}
\sum_{k|s} \left(\begin{array}{c} k+s+1 \\ s+1\end{array}\right) X^{k+s+1} p^k v^{(s)} 
= \left[\widetilde{X}, \int v \right] = \widetilde{X} \circ \left(\int v\right) - 
\left(\int v\right) \circ \widetilde{X}, \label{15.23}
\end{equation}
we can rewrite formulae (\ref{15.22}) as
\begin{gather}
\left[\left(\begin{array}{c} \widetilde{X} \\ u\end{array}\right), \left(\begin{array}{c} \widetilde{Y}
\\ v\end{array}\right) \right] = \left(\begin{array}{c} \left[\widetilde{X}, \widetilde{Y}\right] - 
\alpha \left[\widetilde{X}, \int v\right] + \alpha \left[\widetilde{Y}, \int u\right]
\\ 0\end{array} \right)\nonumber\\
\qquad {}= \left(\begin{array}{c} \left[\widetilde{X} - \alpha \int u, \widetilde{Y} - 
\alpha \int v\right] \\ 0\end{array} \right). \label{15.24}
\end{gather}
This formula is identical in form to formula (\ref{15.15}), and therefore 
we get a Lie algebra for the same reason as before, with the map 
${\mathcal O}: {\mathcal R} \rightarrow {\mathcal H}$ of Proposition 
\ref{proposition:15.16} being exactly the 
same (\ref{15.18}).

Thus, both matrices ${\mathcal B}^{{\mathcal H}}$ (\ref{15.1}) 
and ${\mathcal B}^{{\mathcal H}}$ (\ref{15.2}) 
are Hamiltonian.

Let us notice that in a finite-dimensional-like framework, formula 
(\ref{15.17}) reduces to the following particular case of formula (\ref{15.9}):
\begin{subequations}\label{15.25}
\begin{gather}
[{\mathcal O} (u), Y] = [u {\mathcal O} (1), Y] = u [{\mathcal O} (1), Y] \label{15.25a}\\
 \Rightarrow \ \widehat{\theta} (Y) = [ {\mathcal O} (1), Y] = {\mbox{ad}}_{{\mathcal O} (1)} (Y). 
\label{15.25b}
\end{gather}
\end{subequations}

We now address the problem of linear one-dimensional extensions in 
general.  Let ${\mathcal G}$ be a Lie algebra with a fixed basis, and let $B = 
B ({\mathcal G})$ be the associated Hamiltonian matrix (\ref{15.3}), so that 
\begin{equation}
\sum_k q_k [X, Y]^k \sim X^t B(Y) = \sum_{ij} X^i B_{ij} 
\left(Y^i\right). \label{15.26}
\end{equation}
Let the extended matrix $\hat B$ be of the general form 
\begin{equation}
\widehat{B} = \bordermatrix{& q_j & \bar \varphi \cr
q_i & B_{ij} & - \sqcap_i^\dagger \cr
\bar \varphi & \sqcap_j & 0 \cr}, \label{15.27}
\end{equation}
where $\sqcap_j$'s are some $\varphi$-independent 
operators {\it linearly-dependent} 
upon the $q$'s.  Then
\begin{gather}
(X, u)^t \widetilde{B} (Y, v) = X^t B (Y) + u \sum_j \sqcap_j 
\left(Y^j\right) - v \sum_i \sqcap_i (X^i) \label{15.28}\\
\qquad {}\sim \sum_k q_k [X, Y]^k + \sum_k q_k 
\left({\mathcal M}^k (u, Y) - {\mathcal M}^k (v, X)\right), 
\label{15.29}
\end{gather}
where ${\mathcal M}^k (u, Y)$ are some bilinear operators satisfying the 
relations
\begin{equation}
\sum_k q_k {\mathcal M}^k (u, Y) \sim u \sum_j \sqcap_j \left(Y^j\right) \label{15.30}
\end{equation}
Thus, the multiplication in an extended Lie algebra 
$\widehat{{\mathcal G}} \supset {\mathcal G}$ corresponding to the matrix 
$\widehat{B}$ (\ref{15.27}) is
\begin{equation}
\left[\left(\begin{array}{c} X \\ u\end{array}\right), \left(\begin{array}{c} Y \\ v\end{array}\right)\right] = 
\left(\begin{array}{c} [X, Y]+ {\mathcal M} (u, Y) - {\mathcal M} (v, X) \\ 0\end{array} \right). 
\label{15.31}
\end{equation}
Set
\begin{equation}
{\mathcal M}_u (Y) = {\mathcal M} (u, Y). \label{15.32}
\end{equation}
Then, for each $u$, ${\mathcal M}_u$ is a linear map of ${\mathcal G}$ into ${\mathcal G}$, and 
the finite-dimensional-like formula (\ref{15.9}) is the particular case 
\begin{equation}
{\mathcal M}_u (Y) = u \widehat{\theta} (Y). \label{15.33}
\end{equation}
The Jabobi identity for the commutator (\ref{15.31}) amounts to:
\begin{subequations}\label{15.34}
\begin{gather}
\!0 = \{[[X, Y] + {\mathcal M} (u, Y) \!- \!{\mathcal M} (v, X), Z] \!- \!
{\mathcal M} (w, [X, Y] + {\mathcal M} (u, Y) \!- \!{\mathcal M} (u, X))\} 
+ \mbox{c.p.}\nonumber\\
\qquad {}= \{[{\mathcal M}_w (X), Y] - [{\mathcal M}_w (Y), X] - {\mathcal M}_w ([X, Y]\} +
\mbox{c.p.} \label{15.34a}\\
\qquad {}+ \{ - {\mathcal M}_u {\mathcal M}_v (Z) + {\mathcal M}_v {\mathcal M}_u (Z)\} 
+ \mbox{c.p.}, \label{15.34b}
\end{gather}
\end{subequations}
so that the matrix $\widehat{B}$ (\ref{15.27}) is Hamiltonian iff 
the commutator (\ref{15.31}) defines a Lie algebra iff
\begin{subequations}\label{15.35}
\begin{gather}
{\mathcal M}_u ([X, Y]) = [{\mathcal M}_u (X), Y] + [X, {\mathcal M}_u (Y)], \quad \forall \;
X, Y, u, \label{15.35a}\\
{\mathcal M}_u {\mathcal M}_v = {\mathcal M}_v {\mathcal M}_u, \quad \forall \; u, v. \label{15.35b}
\end{gather}
\end{subequations}
Thus, for each {\it fixed} $u$, ${\mathcal M}_u: {\mathcal G} \rightarrow {\mathcal G}$ is a 
{\it derivation} of ${\mathcal G}$, but as $u$ varies, the derivations 
$\{{\mathcal M}_u\}$ sweep a commutative family.  (Formula (\ref{15.17}) is the 
case ${\mathcal M}_u = {\mbox{ad}}_{{\mathcal O}(u)}$, 
${\mathcal O} (u) \subset {\mathcal H}$, 
${\mathcal H}$ being abelian.)  
We see that, in general, linear 1-dimensional extensions of linear 
Hamiltonian matrices correspond to much much more than  just derivations 
of Lie algebras; instead, one must have a commuting one-parameter 
family of such derivations, the parameter ranging over the ring(s) 
over which our Lie algebras are free modules.

Notice that if the family ${\mathcal M}_u$ satisfies the defining relations 
(\ref{15.35}) and $\alpha$ is a constant then the family 
\begin{equation}
\widetilde{{\mathcal M}}_u = \alpha {\mathcal M}_u \label{15.36}
\end{equation}
also satisfies the relations (\ref{15.35}).  This accounts for the appearance 
of the free parameter $\alpha$ in formulae (\ref{15.1}) and (\ref{15.2}).

\section{Nonlinear derivations of Lie algebras}

In working out various {\it scalar} extensions
\begin{equation}
u_t = X^{\sim}, \quad \varphi_t = {\mathcal A}^X (\varphi), \label{16.1}
\end{equation}
we have met in this paper a few such where the extension operators
${\mathcal A}^X$ are of {\it order zero}, i.e., {\it functions}:
\begin{equation}
u_t = X^{\sim}, \quad \varphi_t = \varphi {\mathcal A} (X). \label{16.2}
\end{equation}
In this case, and in this case only, one can pass to the variable
\begin{equation}
\bar{\varphi} = \log (\varphi), \label{16.3}
\end{equation}
so that the extensions (\ref{16.2}) become
\begin{equation}
u_t = X^{\sim}, \quad \bar{\varphi}_t = {\mathcal A} (X). \label{16.4}
\end{equation}
The examples of this sort include the system (\ref{10.92}):
\begin{equation}
u_t = \partial (Q_n), \quad \varphi_t = \alpha Q_n \varphi, \quad 
n \in {\mathbb N}, \quad \alpha = \mbox{const}, \label{16.5}
\end{equation}
the system (\ref{11.24}):
\begin{equation}
u_t = \partial \left(u^n\right), \quad \varphi_t = \alpha n u^{n-2} u_x 
\varphi, \quad n \in {\mathbb N},  \quad \alpha = \mbox{const}, \label{16.6}
\end{equation}
the system (\ref{12.38}):
\begin{subequations}\label{16.7}
\begin{gather}
u_t = \partial \left(\frac{\delta H}{\delta h} \right), \quad h_t =
\partial \left(\frac{\delta H}{\delta u} \right) - \alpha \varphi
\frac{\delta H}{\delta \varphi}, \label{16.7a}\\
\varphi_t = \alpha \varphi \frac{\delta H}{\delta h}, 
\quad \alpha = \mbox{const}, \label{16.7b}
\end{gather}
\end{subequations}
the system (\ref{12.40}):
\begin{subequations}\label{16.8}
\begin{gather}
u_t = \partial \left(\frac{\delta H}{\delta h} \right) - \alpha
\varphi \frac{\delta H}{\delta \varphi}, \quad  h_t = 
\partial \left(\frac{\delta H}{\delta u} \right), \label{16.8a}\\
u_t = \alpha \varphi \frac{\delta H}{\delta u}, \quad
\alpha  = \mbox{const}, \label{16.8b}
\end{gather}
\end{subequations}
and the Hamiltonian matrices ${\mathcal B}^{{\mathcal H}}$ (\ref{13.28}) 
and $\bar{\mathcal B}^{{\mathcal H}}$  (\ref{14.22})
studied in the preceding Section.

Looking the above examples over, we may notice that none of them is of a
universal character ({\it C}) of Section~1:  the first two, (\ref{16.5}) and (\ref{16.6}), are
of the type ({\it B}), and the last four are of the Hamiltonian type ({\it D}).  As the
analysis in the preceding Section suggests, this absence of 
({\it C})-type examples is surely a consequence of infinite-dimemsionality~--- 
in other words,
of working with PDE's rather that ODE's, or with Field Theory rather than
Mechanics. So let us consider the finite-dimensional-ODE-Mechanics case now.

We start with a few examples, of the following sort: given a 
finite-dimensional
Lie algebra ${\mathcal G}$ whose elements are represented as vector 
fields on a finite-dimensional space $V$, find a linear function
 ${\mathcal A}: {\mathcal G} \rightarrow C^{\infty} (V)$ 
satisfying the equation (\ref{1.11}):
\begin{equation}
{\mathcal A}  ([X, Y]) = \widehat{X} ({\mathcal A} (Y)) - \widehat{Y} ({\mathcal A} (X)), \quad 
\forall \; X, Y \in {\mathcal G}, \label{16.9}
\end{equation}
where $\widehat{X}$ is the vector field on $C^\infty (V)$ corresponding to the element
$X \in {\mathcal G}$.
 
Our first example is ${\mathcal G} = s 0 (3)$, with the basis $(e_i)$, 
$i \in {\mathbb Z}_3$, satisfying
\begin{equation}
[e_i, e_{i+1}] = e_{i+2}, \quad i \in {\mathbb Z}_3. \label{16.10}
\end{equation}
Set
\begin{equation}
\widehat{e}_1 = z \partial_y - y \partial_z, \quad \widehat{e}_2 = 
x \partial_z - z \partial_x, \quad  
\widehat{e}_3 =  y \partial_x - x \partial_y, \quad  \partial_x = 
\frac{\partial}{\partial x}, \ \ldots
\label{16.11}
\end{equation}
Let ${\mathcal O} : so (3) \rightarrow so (3)$ be the following automorphism:
\begin{equation}
{\mathcal O} (e_i) = e_{i+1}, \quad i \in {\mathbb Z}_3.  \label{16.12}
\end{equation}
Extend ${\mathcal O}$ to act on ${\mathbf C} [x, y, z]$ in such way that
\begin{gather}
{\mathcal O} (\widehat{X} (f)) = {\mathcal O} (\widehat{X}) ({\mathcal O} (f)), 
\quad \forall \; X \in {\mathcal G}, \quad  \forall \; f \in {\mathbf C} [x,  y, z]: \label{16.13}\\
{\mathcal O} ( \{x; y; z \}) = \{y; z; x \}. \label{16.14}
\end{gather}
If we demand that solutions of the equation (\ref{16.9}) be 
${\mathcal O}$-invariant --- and if we don't,
we get too many of them --- we arrive at the single equation
\begin{equation}
\widehat{e}_1 ({\mathcal O} ({\mathcal A}) ) - \widehat{e}_2 ({\mathcal A}) = 
{\mathcal O}^{-1} ({\mathcal A}), \quad {\mathcal A}: = {\mathcal A} (e_1), \label{16.15}
\end{equation}
which is
\begin{equation}
\left(z \frac{\partial}{\partial y} - y \frac{\partial}{\partial z} \right)
({\mathcal A} (y,  z,  x)) - \left(x
\frac{\partial}{\partial z} - z \frac{\partial}{\partial x}\right) 
({\mathcal A} (x,  y,  z)) = {\mathcal A} (z, x, y). \label{16.16}
\end{equation}
A quick inspection shows that
\begin{subequations}\label{16.17}
\begin{gather}
\mbox{deg}\, ({\mathcal A}) = 1 \ \Rightarrow \ {\mathcal A} = \mbox{const}\, (y - z), \label{16.17a}\\
\mbox{deg}\, ({\mathcal A}) = 2 \ \Rightarrow \ {\mathcal A} = \mbox{const}\, (y - z) 
(x + y + z), \label{16.17b}
\end{gather}
\end{subequations}
and that
\begin{equation}
{\mathcal A} (x, y, z) = (y - z) f (x + y + z) \label{16.18}
\end{equation}
is a solution of the equation (\ref{16.16}) for any function $f$.  It seems likely that all
{\it polynomial} solutions of the equation (\ref{16.16}) are of the form (\ref{16.18}).

Our next example is the 3-dimensional Lie algebra $s \ell_2$ 
acting on $C^\infty \left({\mathbb R}^1\right)$ :
\begin{subequations}\label{16.19}
\begin{gather}
\widehat{X} = \partial_t, \quad  \widehat{Y} = t \partial_t, \quad 
\widehat{Z} = t^2 \partial_t,\label{16.19a}\\
{}[X, Y] = X, \quad  [Z, Y] = - Z, \quad [X, Z] = 2 Y. \label{16.19b}
\end{gather}
\end{subequations}
The solution space of the equation (\ref{16.9}) is infinite-dimensional, with a basis
\begin{equation}
{\mathcal A} (X) = t^{\alpha}, \quad {\mathcal A} (Y) = t^{\alpha +1}, 
 \quad  {\mathcal A} (Z) = t^{\alpha+2}, \quad  \forall \; \alpha. \label{16.20}
\end{equation}
More generally, consider the centerless Virasoro algebra
\begin{equation}
[e_n, e_m] = (n - m) e_{n + m}, \quad   n, m \in {\mathbb Z}, \label{16.21}
\end{equation}
of which the Lie algebra (\ref{16.19b}) above is the subalgebra 
$X = e_1$, $Y = e_0$, $Z =e_{-1}$.  The Lie algebra (\ref{16.21}) acts 
on ${\mathbf C} [t,  t^{-1}]$ as
\begin{equation}
\widehat{e}_n = t^{1-n} \partial_t, \label{16.22}
\end{equation}
and a basis of the solution space of the equation (\ref{16.9}) is
\begin{equation}
{\mathcal A} (e_n) = t^{\alpha-n}, \quad n \in {\mathbb Z}, \quad  
\forall \; \alpha.  \label{16.23}
\end{equation}

Having the multitude of examples at our disposal, we can now interprete them.
 
If ${\mathcal G}$ is a (finite-dimensional) Lie algebra with a basis $(e_i)$,
\begin{equation}
[e_i, e_j] = \sum_k c_{ij}^k e_k, \label{16.24}
\end{equation}
then there exists a natural representation of ${\mathcal G}$
 by vector fields on $C^\infty ({\mathcal G}^*)$:
\begin{equation}
\widehat{e}_i = \sum_{ks} c^k_{is} q_k \frac{\partial}{\partial
q_s}, \label{16.25}
\end{equation}
based on the identification of any element 
$X \in {\mathcal G}$ with a linear function $\langle~ ,X \rangle$
on ${\mathcal G}^*$ whose Hamiltonian vector field is (\ref{16.25}).  
This is the origin of formula
(\ref{16.11}) for ${\mathcal G} = so (3)$.  
A {\it linear function} on ${\mathcal G}^*$ is the same as an
element of ${\mathcal G}$; therefore, a linear map 
${\mathcal A}: {\mathcal G} \rightarrow  \{{\rm linear \ functions \ on} \
{\mathcal G}^*\}$ is just a linear operator ${\mathcal A}: 
{\mathcal G}  \rightarrow {\mathcal G};$ the equation (\ref{16.9})
for such ${\mathcal A}$ simply says that ${\mathcal A}$ is a 
{\it derivation of} ${\mathcal G}$.  (The solution
(\ref{16.17a}) corresponds to ${\mathcal A} (X) = \mbox{const}\,
 [e_i + e_2 + e_3, X]$.)  We can
interprete the equation (\ref{16.9}) for the case 
${\mathcal A}: {\mathcal G} \rightarrow C^{\infty}
({\mathcal G}^*)$ as defining a 
{\it nonlinear derivation} of ${\mathcal G}$.

Moreover, the Hamiltonian structure on $C^\infty 
({\mathcal G}^*)$ is not too important.  Let
$\rho : {\mathcal G} \rightarrow \mbox{End}\, ({\mathcal W})$ be a representation 
of ${\mathcal G}$ on a vector space
${\mathcal W}$. Now, any linear operator $\sigma : {\mathcal U}  
\rightarrow {\mathcal U}$ on a vector
space ${\mathcal U}$ defines a linear vector field 
$\widehat{\sigma}$ on $C^\infty ({\mathcal U}^*)$,
by the rule 
\begin{equation}
\widehat{\sigma} ( \langle u, ~\rangle) = \langle \sigma (u), ~\rangle, \quad  \forall \;
u \in {\mathcal U}, \label{16.26}
\end{equation}
where $\langle u, ~\rangle$ is a linear function on ${\mathcal U}^*$.  Thus, 
formula (\ref{16.26}) defines the action of
the vector field $\widehat{\sigma}$ on linear functions, 
and the Leibniz rule then
extends the action of $\widehat{\sigma}$ onto the 
whole $C^\infty ({\mathcal U}^*)$.  Formula (\ref{16.25})
is a particular case $\{{\mathcal U} = {\mathcal G}, 
\;\sigma = [e_i, ~] \}$ of formula (\ref{16.26}).
Therefore, in general, if we set
\begin{equation}
\widehat{X} = \widehat{\rho (X)}, \quad  
\forall \; X \in {\mathcal G}, \label{16.27}
\end{equation}
the equation (\ref{16.6}) is an equation for an unknown linear map
${\mathcal A}: {\mathcal G} \rightarrow C^\infty ({\mathcal W}^*)$.  
If $\mbox{Im}\, ({\mathcal A})\subset {\mathcal W} \subset C^\infty
({\mathcal W}^*)$, the equation (\ref{16.6}) becomes
\begin{equation}
{\mathcal A} ([X, Y]) = \rho (X) ({\mathcal A} (Y)) - \rho (Y) ({\mathcal A} (X)), \quad 
 \forall \; X, Y \in {\mathcal G}. \label{16.28}
\end{equation}
Such a linear map ${\mathcal A}: {\mathcal G} \rightarrow {\mathcal W}$ can 
be thought of as a relative
derivation of ${\mathcal G}$ with the values
 in a ${\mathcal G}$-module.  We get the usual
derivation when the representation
 $\rho : {\mathcal G} \rightarrow \mbox{End}\, ({\mathcal W})$ is just the adjoint
representation.  (An analog of an interior 
derivation of ${\mathcal G}$, ${\mathcal A} (X) = [X, a]$
for a fixed $a \in {\mathcal G}$, is the following 
relative interior derivation:
\begin{equation}
{\mathcal A} (X) = \rho (X). v, \quad {\rm fixed}  \ v \in {\mathcal W}. \label{16.29}
\end{equation}
Equality (\ref{16.9}) then follows from the representation 
property of $\rho$.)
The general equation (\ref{16.9}) for the linear map
 ${\mathcal A} : {\mathcal G} \rightarrow C^\infty ({\mathcal W}^*)$ describes
then {\it nonlinear} relative derivations.  
Formula (\ref{16.9}), rewritten as
\begin{equation}
\widehat{X} ({\mathcal A} (Y)) -
 \widehat{Y} ({\mathcal A} (X) ) - {\mathcal A} ( [ X, Y]) = 0
\label{16.30}
\end{equation}
suggests that one can get this equation as a 2-cocycle 
condition in a suitably constructed
complex.  This is left as an Exercise to the reader.

\section{Linear extensions as Symbols of nonlinear ones}

In this Section we return to the unfinished business of the mysterious 
and the strange extensions of Section~6.  We prove that such extensions 
exist for the whole KdV and mKdV hierarchies.

We start with the mysterious extension (\ref{6.33}):
\begin{equation}
u_t = \partial \left(3 u^2 + u_{xx}\right), \quad \varphi_t = \langle - \partial \left(6 u 
+ 2 \partial^2\right)\rangle (\varphi). \label{17.1}
\end{equation}
This can be recognized as $\varphi$-linearized part of the Hirota--Satsuma 
equation [5].  The following explanation is inspired by the Wilson 
succinct analysis [28].

Let ${\mathcal L}$ be an even-order scalar Lax operator:
\begin{equation}
{\mathcal L} = \partial^{2N} + \sum^{2N-2}_{0} u_i \partial^i. \label{17.2}
\end{equation}
The commuting hierarchy of Lax equations
\begin{equation}
{\mathcal L}_t = [{\mathcal P}_+, {\mathcal L}], \quad \mbox{ord} ({\mathcal P}) = 2n-1, \quad
n \in {\mathbb N}, 
\label{17.3}
\end{equation}
is compatible with the self-adjointness constraint
\[
{\mathcal L}^\dagger = {\mathcal L}. 
\]
So let ${\mathcal L}$ be self-adjoint:
\begin{equation}
{\mathcal L} = \partial^{2N} + \sum^{N-1}_{i=0} \left(U_i \partial^{2i} + 
\partial^{2i} U_i\right). \label{17.4}
\end{equation}
Set
\begin{equation}
L = \partial^2 + u, \quad u = \frac{2}{N} U_{N-1}. \label{17.5}
\end{equation}
Then $L$ is self-adjoint, and so are $L^N$ and 
\begin{equation}
{\mathcal L} - L^N = \sum^{N-2}_{i=0} \left(f_i \partial^{2i} + \partial^{2i} 
f_i\right), \quad {\rm some} \ f_i\mbox{\rm {}'s}. \label{17.6}
\end{equation}
Since
\begin{equation}
L_t = [P_+, L] \ \Rightarrow \ \left(L^N\right)_t = \left[P_+, L^N\right], \label{17.7}
\end{equation}
the motion equations (\ref{17.3}):
\begin{equation}
{\mathcal L}_t = [{\mathcal P}_+, {\mathcal L}], \quad
 {\mathcal L}^\dagger = {\mathcal L}, \quad {\mathcal P}^\dagger = - 
{\mathcal P}, \quad \mbox{ord} ({\mathcal P}) = 2n+1, \label{17.8}
\end{equation}
have the form
\begin{subequations}\label{17.9}
\begin{gather}
u_t = X^\sim_{n} \{u\} + O (f), \label{17.9a}\\
f_{i,t} = {\mathcal O}_i (f) + O \left(f^2\right), \quad  i = 0, \ldots, N-2, \label{17.9b}
\end{gather}
\end{subequations}
where $u_t = X^\sim_n \{u\}$ is the $n^{th}$ KdV equation, 
and ${\mathcal O}_i$'s 
are some linear differential operators whose coefficients depend upon~$u$.  

We now take the $f$-Symbol of the system (\ref{17.9}) by setting 
\begin{equation}
f_i = \epsilon \varphi_i, \quad i = 0, \ldots, N -2, \label{17.10}
\end{equation}
and then letting $\epsilon \rightarrow 0$.  We obtain:
\begin{subequations}\label{17.11}
\begin{gather}
u_t = X^\sim_n \{u\}, \label{17.11a}\\
\varphi_{i,t} = {\mathcal O}_i (\varphi), \quad i = 0, \ldots, N -2. \label{17.11b}
\end{gather}
\end{subequations}
This is an $(N-1)$-component extension of the KdV hierarchy, which 
should be called the dual-to-mysterious $(N-1)$-extension.  This name 
is justified by the following observation:

Take $N=2$, so that $N-1=1$, $\mbox{ord} ({\mathcal L})=4$:
\begin{equation}
{\mathcal L} = \partial^4 + U_0 \partial^2 + \partial^2 U_0 + V_1 = 
\left(\partial^2 + u\right)^2 + f. \label{17.12}
\end{equation}
With $\mbox{ord}({\mathcal P}) = 3$, ${\mathcal P} = 4 {\mathcal L}^{3/4}$:
\begin{equation}
{\mathcal P}_+ = 4 \partial^3 + 3 u \partial + 3 \partial u, \label{17.13}
\end{equation}
we find:
\begin{subequations}\label{17.14}
\begin{gather}
{\mathcal L}_t = LL_t + L_t L + f_t = \left(\partial^2 + u\right) u_t + u_t \left(\partial^2 
+ u\right) + f_t = u_t \partial^2 + \partial^2 u_t + 2 uu_t + f_t \nonumber\\
= [{\mathcal P}_+ , L^2 + f] = [{\mathcal P}_+, L] L + L [{\mathcal P}_+, L] + [{\mathcal P}_+, f] \ 
\{{\rm with} \ \bar u_t = [{\mathcal P}_+, L] = 6uu_x + u_{xxx}\}\nonumber\\
= \bar u_t L + L \bar u_t + 4 (f_{xxx} + 3 f_{xx} \partial + 
3f_x \partial^2) + 6u f_x \nonumber\\
= \bar u_t \partial^2 + \partial^2 \bar u_t + 2 u \bar u_t + 
6 f_x \partial^2 + \partial^2 6f_x - 6 (f_{xxx} + 2f_{xx} \partial) 
+ 4 f_{xxx} + 12 f_{xx} \partial + 6u f_x \nonumber\\
= (\bar u_t + 6f_x) \partial^2 + \partial^2 (\bar u_t + 6 f_x) 
+ 2 u \bar u_t + 6u f_x - 2f_{xxx} \nonumber\\
\Rightarrow \ u_t = \bar u_t + 6 f_x = 6 uu_x + u_{xxx} + 6f_x, \label{17.14a}\\
f_t = - 2 uu_t + 2 u \bar u_t + 6 u f_x - 2f_{xxx} = - 6u f_x - 2 f_{xxx}. \label{17.14b}
\end{gather}
\end{subequations}
Taking the $f$-Symbol of the above, by setting
\[
f = \epsilon \varphi
\]
and letting $\epsilon$ go, we find
\begin{equation}
u_t = \partial\left(3 u^2 + u_{xx}\right), \quad \varphi_t = - \left(6u \partial + 2 
\partial^3\right) (\varphi), \label{17.15}
\end{equation}
which is the dual to the mysterious system (\ref{17.1}).  As Wilson shows 
[28], the system (\ref{17.14}) and the corresponding hierarchy are bi-Hamiltonian 
and they possess a Miura map which is canonical.  Taking the Symbol 
of the corresponding modified systems, we get the hierarchy of the 
dual-to-mysterious mKdV extensions.  For $N \geq 2$, the corresponding 
Miura map comes out of the factorization.
\begin{equation}
{\mathcal L} = (\partial - v_1) \cdots (\partial - v_N) (\partial + v_N)\cdots 
(\partial + v_1). \label{17.16}
\end{equation}

Having thus realized the mysterious extension as a Symbol of a 
coupled 2-component KdV system, one may hope to find the remaining 
unresolved extension, the strange one, among known 2-component 
integrable systems extending the KdV.  This was suggested to me 
by G.~Wilson, and his advice was right on the mark.

There are two known integrable super-extensions of the KdV hierarchy 
[22]: the super-KdV system [9]
\begin{subequations}\label{17.17}
\begin{gather}
u_t = \partial \left(3 u^2 + u_{xx} + 3 \xi \xi_x\right), \label{17.17a}\\
\xi_t = 3 u_x \xi + 6u \xi_x + 4 \xi_{xxx}, \label{17.17b}
\end{gather}
\end{subequations}
and the supersymmetric KdV equation of Manin and Radul [21]:
\begin{subequations}\label{17.18}
\begin{gather}
u_t = \partial \left(3 u^2 + u_{xx} + 3 \xi \xi_x\right), \label{17.18a}\\
\xi_t = 3 u_x \xi + 3u \xi_x + \xi_{xxx}, \label{17.18b}
\end{gather}
\end{subequations}
where $\xi$ is now an {\it{odd}} variable (``fermion'').

Taking the Symbol of system (\ref{17.18}) we arrive at the strange system 
(\ref{6.30}):
\begin{equation}
u_t = \partial \left(3u^2 + u_{xx}\right), \quad \varphi_t = \langle\partial \left(3u + 
\partial^2 \right)\rangle (\varphi). \label{17.19}
\end{equation}
The fact that $\varphi$ is a descendant of an odd variable is no 
longer material.

The system (\ref{17.18}) is a Lax system, with the Lax operator [21]
\begin{subequations}\label{17.20}
\begin{gather}
L = {\mathcal D}^4 + \Gamma {\mathcal D}, \label{17.20a}\\
\Gamma = \xi - \theta u, \quad  {\mathcal D} = \frac{\partial}{\partial \theta} 
+ \theta \partial, \quad  \theta^2 = 0. \label{17.20b}
\end{gather}
\end{subequations}
This provides us, upon extracting the Symbol, with the whole 
strangely-extended KdV hierarchy.  Also, the system (\ref{17.18}) 
is Hamiltonian, as 
is the corresponding hierarchy, and they possess a Miura map that 
is canonical [23]; the strangely-extended mKdV hierarchy results 
thereby. (The strange system (\ref{17.19}) for {\it bosonic} $\varphi$ 
appears as an universal fluid-dynamical system, and also as a continuous 
limit of certain discrete mechanical spring models [24].  In the paper 
[24], Miller and Clarke consider the system
\begin{equation}
u_t = \partial \left(3u^2 + u_{xx}\right), \quad \varphi_t = \langle\partial \left(6ku + 
\partial^2\right) \rangle(\varphi), \label{17.21}
\end{equation}
and conclude after a detailed analytical study that only for $k=1$ 
or $k=1/2$ does this system exhibit behavior that can be reasonably 
described as regular.)
 
If we now extract the Symbol from the super-KdV system (\ref{17.17}), we 
arrive at the Lax-type extension (\ref{6.22}):
\begin{equation}
u_t = \partial \left(3 u^2 + u_{xx}\right), \quad \varphi_t = \left(3u \partial + 
3 \partial u + 4 \partial^3\right) (\varphi). \label{17.22}
\end{equation}
Since the super-KdV system (\ref{17.17}) and the associated hierarchy 
are bi-Hamiltonian, with a canonical Miura map [9], we obtain 
the corresponding extension of the mKdV hierarchy; this also 
explains the unexpected at the time formula (\ref{8.28}).

The three examples above show that with the single exception 
of the linearized KdV extension, every other nontrivial 
KdV extension of Section~6 is either a Symbol of a nonlinear coupled 
integrable KdV system~--- or else its dual is.  This suggests that 
our subject~--- linear extensions~--- in some large parts should be 
considered as a first approximation to the much more complex, difficult, 
and amorphous problem of nonlinear extensions.

To test this point of view, let us examine the additional KdV 
extension (\ref{9.9}):
\begin{equation}
u_t = \partial \left(3u^2 + u_{xx}\right), \quad \varphi_t = (\alpha u_x + 
2u \partial) (\varphi). \label{17.23}
\end{equation}
For $\alpha = 2$, we get
\begin{equation}
u_t = \partial \left(3 u^2 + u_{xx}\right), \quad \varphi_t = 2 (u \varphi)_x, 
\label{17.24}
\end{equation}
and this is the Symbol of the Ito system [6]
\begin{equation}
u_t = \partial \left(3u^2 + u_{xx} + \varphi^2\right), \quad \varphi_t = 2 
(u \varphi)_x. \label{17.25}
\end{equation}
This system is bi-Hamiltonian [6]; the 2$^{nd}$ Hamiltonian structure 
is given by the Hamiltonian matrix (\ref{9.23}) for $\alpha = 2$, and 
the $1^{st}$ Hamiltonian structure is
\[
\begin{pmatrix} \partial & 0 \\ 0 & \partial \end{pmatrix}. 
\]
Since the Miura map (\ref{9.38}) is no longer Hamiltonian, the existence 
of {\it nonlinear} mKdV extensions of which (\ref{9.41}) is the Symbol  
(and which map into the Ito hierarchy), remains an open problem.

It would be interesting to find nonlinear analogs of other linear 
extensions constructed in this paper, in particular for the Burgers, 
long waves, Benney, and KP hierarchies.  It wouldn't be surprising if 
different nonlinear extensions have the same Symbol, as integrable 
systems are often realized as ``invariant submanifolds'' of many 
different larger ones.  For example, the Burgers hierarchy results 
upon setting $\{h=0\}$ in the Dispersive Water Waves (DWW) hierarchy, 
mentioned in Section~12, while that same DWW hierarchy results in the 
KdV hierarchy upon setting $\{u=0\}$ for every second flow [12].  

\setcounter{remark}{25}
\setcounter{equation}{26}

\begin{remark}\label{remark:17.26}
 The search for nonlinear extensions can be 
made systematic.  For the Hamiltonian case, most likely one has to 
take the Hamiltonian structure of linear extension and deform the Hamiltonian 
function itself.  For the general case, one has to use the ansaltz
\begin{subequations}\label{17.27}
\begin{gather}
u_t = \sum^\ell_{s=0} {\mathcal O}_s (\varphi) + O \left(\varphi^{\ell +1}\right), 
\label{17.27a}\\
\varphi_t = \sum^{\ell +1}_{s=1} A_s (\varphi) + O \left(\varphi^{\ell+2}\right), 
\quad \ell \in {\mathbb Z}_+ \label{17.27b}
\end{gather}
\end{subequations}
where ${\mathcal O}_s (\varphi)$ and ${\mathcal A}_s (\varphi)$ are $s$-linear in 
$\varphi$; for each such fixed $\ell$, one can define the 
commutativty of the flows modulo appropriate $O$-terms.  The 
homogeneuity 
requirement then guarantees that the sums on $s$ in (\ref{17.27}) are 
finite, so the whole problem can be automated.  (For systems of the 
type considered in Section~9, with the lowest order extensions of the form
\begin{subequations}\label{17.28}
\begin{gather}
u_t = X^\sim (u), \label{17.28a}\\
\bar \varphi_t = {\mathcal A} (u), \label{17.28b}
\end{gather}
\end{subequations}
the analog of an $\ell^{th}$ order extension (\ref{17.27}) is 
\begin{subequations}\label{17.29}
\begin{gather}
u_t = \sum^\ell_{s=0} {\mathcal O}_s (\varphi) + O \left(\bar \varphi^{\ell +1}\right), 
\label{17.29a}\\
\bar \varphi_t = \sum^\ell_{s=0} {\mathcal A}_s (\bar \varphi) + O 
\left(\bar \varphi ^{\ell+1}\right), \quad \ell \in {\mathbb Z}_+. \label{17.29b}
\end{gather}
\end{subequations}
As a somewhat funny example, consider the dual to the linearized 
KdV extension
\begin{subequations}\label{17.30}
\begin{equation}
u_t = \partial \left(3u^2 + u_{xx}\right), \quad \varphi_t = \left(6 u \partial 
+ \partial^3\right)(\varphi). \label{17.30a}
\end{equation}
It can be completed to the fully nonlinear system
\begin{equation}
u_t = \partial \left(3u^2 + u_{xx}\right), \quad \varphi_t = \left(6 u \partial + 
\partial^3\right) (\varphi) + 3 \varphi_x \varphi_{xx} + \varphi_x^3, \label{17.30b}
\end{equation}
which appears as the addition to the KdV equation of another copy of 
it, but in the variable
\begin{equation}
\varphi = \log (u). \label{17.30c}
\end{equation}
\end{subequations}
It's very likely that this particular doubling device (which can be 
made unique by the requirement that $\varphi_t$ is a differential 
polynomial in $\varphi$ and a (nondifferential) rational function 
 in $u$ 
which vanishes together with $\varphi$) applies to the whole 
KdV hierarchy.
\end{remark}

\setcounter{remark}{30}
\setcounter{equation}{31}

\begin{remark}\label{remark:17.31}  The Ito example (\ref{17.24}) suggests that 
``nonlinearization'' may be
possible, if at all, only for some {\it special values} 
of the parameters of a linear
extension.  Formulae (\ref{17.28}) and (\ref{17.30}) offer an opportunity 
to construct and evaluate
some sort of obstructions to nonlinearization; we won't pursue 
this route here.  Instead,
here is another instance of such quantization.  Consider the 
extended Hamiltonian matrix
(\ref{12.29}) for $\alpha = 0$:
\begin{equation}
B = \bordermatrix{& u & h & \varphi \cr
u & 0 & \partial & - \varphi_x/h \cr
h & \partial & 0 & 0 \cr
\varphi & \varphi_x/h & 0 & 0 \cr} \label{17.32}
\end{equation}
The Hamiltonian $h$ is still a Casimir of the extended matrix, but the Hamiltonian
$u$ is no longer so; instead, there is a new Casimir
\begin{equation}
H = h \varphi^2. \label{17.33}
\end{equation}
In addition, the matrix $B$ (\ref{17.32}) admits the following Hamiltonian automorphism:
\begin{equation}
U = u + \lambda \varphi, \quad h = h, \quad \varphi = \varphi, 
\quad \lambda = \mbox{const}, \label{17.34}
\end{equation}
a $\varphi$-extension of the identity automorphism of the unextended Hamiltonian matrix (\ref{12.2}).
This implies that the old {\it linear} extended hierachy in $U$, $h$, $\varphi$ variables
becomes new fully {\it nonlinear} extended hierarchy in $u$, $h$, 
$\varphi$ variables.  For example,
the original 1$^{st}$ flow (\ref{12.1}), (\ref{12.15}) now takes the form
\begin{subequations}\label{17.35}
\begin{gather}
u_t = uu_x + h_x + \lambda u_x \varphi, \quad  
h_t = [h (u + \lambda \varphi)]_x, \label{17.35a}\\
\varphi_t = ( u + \lambda \varphi) \varphi_x. \label{17.35b}
\end{gather}
\end{subequations}
This is so because the old Hamiltonian $\left(h^2 u + h^2\right) /2$
 turns into
\begin{equation}
H_2 = \frac{h U^2 + h^2}{2} = \frac{h u^2 + h^2}{2} + \lambda 
h u \varphi + \left(\frac{\lambda^2}{2} h
\varphi^2 \ {\rm is \ a \ Casimir}\right). \label{17.36}
\end{equation}
All this is only for the value $\alpha = 0$. Whether other values 
of $\alpha$ allow
nonlinearization,  I~don't know.  Moreover, the above picture 
can be generalized into the
moments space:
\end{remark}

\setcounter{proposition}{36}
\setcounter{equation}{37}

\begin{proposition}\label{proposition:17.37}  Let ${\mathcal B}$ be the Hamiltonian 
matrix (\ref{13.15}) for $\alpha = 0$:
\begin{subequations}\label{17.38}
\begin{gather}
{\mathcal B}_{nm} = n A_{n + m-1} \partial + \partial m A_{n + m-1}, \quad 
 n,  m \in {\mathbb Z}_+,\label{17.38a}\\
{\mathcal B}_{\varphi m} = \varphi_x m A_{m-1}/A_0, \quad {\mathcal B}_{n \varphi} 
= - \varphi_x n A_{n-1}/A_0,\label{17.38b}\\
{\mathcal B}_{\varphi \varphi} = 0. \label{17.38c}
\end{gather}
\end{subequations}

(i) The map
\begin{equation}
{\mathcal A}_n = \sum^n_{k=0} \left(\begin{array}{c} n \\ k\end{array}\right) A_k 
(\lambda \varphi)^{n-k}, \quad  n \in {\mathbb Z}_+; \quad \varphi = \varphi, \label{17.39}
\end{equation}
is a Hamiltonian automorphism of the Hamiltonian matrix 
${\mathcal B}$ (\ref{17.38});

(ii) In addition to the old Casimir $H = A_0$, 
the extended Hamiltonian matrix
${\mathcal B}$ (\ref{17.38}) has also a new one;
\begin{subequations}\label{17.40}
\begin{equation}
H = A_0 \varphi^2; \label{17.40a}
\end{equation}

(iii) The automorphism (\ref{17.39}) covers the 
automorphism (\ref{17.34}) with respect to
the Hamiltonian embeddings ${\mathcal A}_n = h 
{\mathcal U}^n, A_n = h u^n$:
\begin{equation}
{\mathcal A}_n = h {\mathcal U}^n, \quad  A_n = h u^n \ \Rightarrow \ {\mathcal U} = u +
\lambda \varphi. \label{17.40b}
\end{equation}
\end{subequations}
\end{proposition}

\begin{proof}
(i) A lengthy but straightforward verification,

(ii) From formula (\ref{17.39}) for $n = 1,$ we have 
${\mathcal U} = {\mathcal A}_1/h = (A_1 +
\lambda \varphi A_0)/A_0 = u + \lambda \varphi, $
and then
\begin{equation*}
\sum^n_{k=0} \left(\begin{array}{c} n \\ k\end{array}\right) A_k (\lambda \varphi)^{n-k} =
\sum \left(\begin{array}{c} n \\ k\end{array}\right)
h u^k (\lambda \varphi)^{n-k} = h (u + \lambda \varphi)^n = h 
{\mathcal U}^n = {\mathcal A}_n.
\tag*{\qed}
\end{equation*}
\renewcommand{\qed}{}
\end{proof}

As a Corollary, the old {\it linearly} extended Benney 
system (\ref{13.14}), (\ref{13.18}):
\begin{subequations}\label{17.41}
\begin{gather}
A_{n,t} = A_{n + 1, x} + n A_{n-1}, A_{0,x}, \quad n \in {\mathbb Z}_+, \label{17.41a}\\
\varphi_t = \varphi_x A_1/A_0, \label{17.41b}
\end{gather}
\end{subequations}
generates a fully {\it nonlinear} extension
\begin{subequations}\label{17.42}
\begin{gather}
A_{n,t} = A_{n + 1,x} + n A_{n - 1} A_{0, x} + \lambda \varphi 
A_{n,x} \nonumber\\
\phantom{A_{n,t} =} + \lambda \varphi_x [(n + 1) A_n - n A_{n - 1}
 A_1/A_0], \quad n \in {\mathbb Z}_+, \label{17.42a}\\
\varphi_t = \varphi_x (A_1 + \lambda \varphi A_0)/A_0; \label{17.42b}
\end{gather}
\end{subequations}
this is so because the old Hamiltonian $H_2 = \left(A_2 + A^2_0\right)/2$ 
turns into
\begin{equation}
\left({\mathcal A}_2 + {\mathcal A}_0^2\right)/2 = \left(A_2 + A^2_0\right)/2 + \lambda A_1 \varphi + 
\left(\lambda^2 A_0 \varphi^2/2 \ {\rm is \ a \ Casimir}\right).  \label{17.43}
\end{equation}
Naturally, when $u_y = 0$, so that $A_n = \int^h_0 u^n d y$ 
becomes
$A_n = h u^n$ and ${\mathcal A}_n = \int^h_0 (u + \lambda
\varphi)^n d y$
becomes ${\mathcal A}_n = h (u + \lambda \varphi)^n$, equations 
(\ref{17.42}) reduce to the system (\ref{17.35}).

Moreover still, the preceding picture can be generalized for 
the fully dispersive case:

\setcounter{proposition}{43}
\setcounter{equation}{44}

\begin{proposition}\label{proposition:17.44}
Let $\bar {\mathcal B}$ be the 
extended KP Hamiltonian matrix
(\ref{14.14}) for the $\alpha = 0$:
\begin{subequations}\label{17.45}
\begin{gather}
\bar{{\mathcal B}}_{nm} = \sum_\mu \left[\left(\begin{array}{c} n \\ \mu\end{array}\right) A_{n + m - \mu}
 \partial^\mu - (- \partial)^\mu \left(\begin{array}{c} m \\ \mu\end{array}\right) 
 A_{n + m - \mu}\right], \quad n, m \in {\mathbb Z}_+,\label{17.45a}\\
\bar{{\mathcal B}}_{\varphi m} = \frac{1}{A_0} \varphi_x \sum_s 
(- \partial)^s \left(\begin{array}{c} m \\ s+1\end{array}\right) A_{m-1-s}, \label{17.45b}\\
{\bar{{\mathcal B}}}_{n \varphi} = - \sum_s \left(\begin{array}{c} n \\ s + 1\end{array}\right) 
A_{n-1-s} \partial^s
\frac{1}{A_0} \varphi_x, \quad {\bar{{\mathcal B}}}_{\varphi \varphi} = 0. 
\label{17.45c}
\end{gather}
\end{subequations}

(i) The map
\begin{equation}
{\mathcal A}_n = \sum^n_{k=0} \left(\begin{array}{c} n \\ k\end{array}\right) 
A_k Q_{n-k} (\lambda \varphi), \quad n \in {\mathbb Z}_+; 
\quad \varphi = \varphi \label{17.46}
\end{equation}
is a Hamiltonian automorphism of the Hamiltonian matrix 
$\bar{{\mathcal B}}$ (\ref{17.45});

(ii) In addition to the old Casimir $H = A_0$, the extended 
Hamiltonian matrix
$\bar{{\mathcal B}}$ (\ref{17.45}) has also a new one,
\begin{equation}
H = A_0 \varphi^2; \label{17.47}
\end{equation}

(iii) The automorphism (\ref{17.46}) covers the automorphism 
(\ref{17.34}) with respect to the
Hamiltonian embeddings
\begin{gather}
{\mathcal A}_n = h Q_n ({\mathcal U}), \quad A_n = h Q_n (u):\nonumber\\
{\mathcal A}_n = h Q_n ({\mathcal U}), \quad A_n = h Q_n (u) \  \Rightarrow \  
{\mathcal U} = u + \lambda \varphi.
\label{17.48}
\end{gather}
\end{proposition}

\begin{proof} (i) A lengthy but straightforward 
verification, requiring at the last
step the identity ([12, p.~66])
\begin{equation}
\sum_\alpha \left(\begin{array}{c} m \\ \alpha\end{array}\right) 
Q_{m-\alpha} \partial^\alpha Q_r = \sum_\alpha
\left(\begin{array}{c} m \\ \alpha\end{array}\right) 
Q_{m+r-\alpha} \partial^\alpha; \label{17.49}
\end{equation}

(ii) Is obvious;

(iii) Is based on the addition formula (a differential analog of the Newton
binomial)
\begin{equation}
Q_n (u + v) = \sum^n_{k=0} \left(\begin{array}{c} n \\ k\end{array}\right) 
Q_k (u) Q_{n-k} (v), \quad n \in {\mathbb Z}_+.
\label{17.50}
\end{equation}
which is easily proved by induction on $n$.
\end{proof}

As a Corollary, the old linearly extended KP system (for 
the flow $L_t = [P_+, L],$
$ P = L^2/2$, $L = \partial + \sum\limits^\infty_{i=0} \partial^{-i-1} A_i$):
\begin{subequations}\label{17.51}
\begin{gather}
A_{n,t} = A^{(1)}_{n+1} - \frac{1}{2} A^{(2)}_{n} +
\sum_{\mu \ge 0} \left(\begin{array}{c} n \\ \mu + 1\end{array}\right) 
A_{n-1-\mu} A^{(\mu + 1)}_0, \quad n \in {\mathbb Z}_+, \label{17.51a}\\
\varphi_t = \varphi_x \left(A_1 - A_0^{(1)}/2\right)/A_0, \label{17.51b}
\end{gather}
\end{subequations}
becomes the fully nonlinear system
\begin{subequations}\label{17.52}
\begin{gather}
A_n, t = A^{(1)}_{n+1} - \frac{1}{2} A^{(2)}_n + 
\sum_{\mu=0} \left(\begin{array}{c} n \\ \mu + 1\end{array}\right)
A_{n-1-\mu} \left(A_0 + \lambda \varphi_x/2\right)^{(\mu+1)} + \lambda (A_n
 \varphi)^{(1)} \nonumber\\
\qquad {}+  \lambda \sum_{\mu \ge 0} \left(\!\begin{array}{c} n \\ \mu+1\end{array}\!\right)
 \left\{A_{n-\mu}\varphi^{(\mu +1)} - A_{n-1-\mu} \left[\varphi_x A_1 - A^{(2)}_0 (2)/A_0\right]^{(\mu)}
 \right\}, \label{17.52a}\\
\varphi_t = \varphi_x \left(A_1 - A_0^{(1)}/2 + \lambda 
\varphi A_0\right)/A_0; \label{17.52b}
\end{gather}
\end{subequations}
this is so because the old Hamiltonian $\left(A_2 + A^2_0\right)/2$ turns into
\begin{gather}
\left({\mathcal A}_2 + {\mathcal A}^2_0\right)/2 = \left[A_2 + 2 A_1 \lambda \varphi + A_0
\left(\lambda^2 \varphi^2 + \lambda \varphi_x\right) + A^2_0\right]/2 \nonumber\\
\qquad {}=  \left(A_2 + A^2_0\right)/2 + \lambda
(A_1 \varphi + A_0 \varphi_x/2) +
\left(\lambda^2 A_0 \varphi^2/2 \ {\rm is \ a \ Casimir}\right). \label{17.53}
\end{gather}

\section{Scalar extensions associated with scalar Lax 
operators of order $\pbf{\geq 4}$}

Let $L$ be a scalar Lax operator of order $N \geq 4$:
\begin{equation}
L = {\partial^N} + \sum^{N-2}_{i=0} u_i \partial^{N-2-i}. 
\label{18.1}
\end{equation}
Consider the 2$^{nd}$ and the $3^{rd}$ flow of the hierarchy 
\begin{gather}
(L_t = ) \quad X_n (L) = [P_+, L], \quad P = \left(L^{1/N}\right)^n: \label{18.2}\\
\left(L^{2/N}\right)_+ = \partial^2 + \frac{2}{N} u_0, \label{18.3}\\
\left(L^{3/N}\right)_+ = \partial^3 +\frac{3}{N} u_0 \partial + \frac{3}{N} 
\left(\frac{3-N}{2} u_{0,x} + u_1\right). \label{18.4}
\end{gather}
(The last expression results from the condition
\begin{equation}
\mbox{ord}([P_+, L]) \leq N-2. \label{18.5}
\end{equation}
The motion equations (\ref{18.2}), at least those we shall need below, 
are:
\begin{subequations}\label{18.6}
\begin{gather}
X_2 (u_0) = u_{0,t} = (2-N) u_0^{(2)} + 2 u_1^{(1)}, \label{18.6a}\\
X_2 (u_1) = u_{1,t} = -\frac{(N-1) (N-2)}{3} u_0^{(3)} + 
u_1^{(2)} + 2u_2^{(1)} - \frac{2(N-2)}{N} u_0 u_0^{(1)}, \label{18.6b}
\end{gather}
\end{subequations}
\vspace{-5mm}
\begin{equation}
X_3 (u_0) = u_{0,t} = \left(\frac{N-3}{2}\right)^2 u_0^{(3)} 
- \frac{3(N-3)}{2} u_1^{(2)} + 3u_2^{(1)} - \frac{3(N-3)}{N} 
u_0 u_0^{(1)}. \label{18.7}
\end{equation}

Let's look for {\it scalar} extensions of the flows $X_2$ and $X_3$: 
\begin{subequations}\label{18.8}
\begin{gather}
\left(X_2^{\rm ext}\right) \quad \varphi_t = {\mathcal A}_2 (\varphi) = \left(\alpha u_0 + 
\gamma \partial^2\right) (\varphi), \label{18.8a}\\
\left(X_3^{\rm ext}\right) \quad \varphi_t = {\mathcal A}_3 (\varphi) = \langle
\left(a u_0^{(1)} + bu_1\right) + f u_0 \partial + g \partial^3 \rangle (\varphi), \label{18.8b}
\end{gather}
\end{subequations}
where $\alpha$, $\gamma$; $a$, $b$, $f$, $g$ are unknown constants.  The 
commutativity condition (\ref{1.13})
\begin{equation}
X_2 \left(a u_0^{(1)} + bu_1 + f u_0 \partial\right) - X_3 (\alpha u_0) = 
\ [{\mathcal A}_2, {\mathcal A}_3], \label{18.9}
\end{equation}
yields the relations:
\begin{subequations}\label{18.10}
\begin{gather}
2 \gamma f = 3 \alpha g, \label{18.10a}\\
\alpha \left[\frac{3(N-3)}{N} + f\right] = b\frac{2(N-2)}{N}, 
\label{18.10b}\\
\gamma f - 3 \alpha g + 2 \gamma a = f (2-N), \label{18.10c}\\
\gamma b = f, \label{18.10d}\\
2b = 3 \alpha, \label{18.10e}\\
- \alpha g + \gamma a = a (2 - N) - b \frac{(N-1) (N-2)}{3} - 
\alpha \left(\frac{N-3}{2}\right)^2, \label{18.10f}\\
\gamma b = 2a + b + 3 \alpha \frac{N-3}{2}. \label{18.10g}
\end{gather}
\end{subequations}
If
\begin{equation}
\alpha = 0 \ \Rightarrow \ a = b = f = 0 \label{18.11}
\end{equation}
is the decomposed case.  So let us consider
\begin{equation}
\alpha \not= 0. \label{18.12}
\end{equation}

If $\gamma = 0$ then (\ref{18.10a}) yields $g=0$, (\ref{18.10d}) yields 
$f=0$, and $\{$(\ref{18.10b})  \& (\ref{18.10e})$\}$ force a contradiction.

So, let
\begin{equation}
\alpha \not= 0,  \quad \gamma \not= 0. \label{18.13}
\end{equation}
Then (\ref{18.10e}) yields
\begin{equation}
b = \frac{3 \alpha}{2}, \label{18.14}
\end{equation}
and from (\ref{18.10d}) we get
\begin{equation}
f = \frac{3 \alpha \gamma}{2}. \label{18.15}
\end{equation}
Substituting this into (\ref{18.10a}), we find
\begin{equation}
g = \gamma^2. \label{18.16}
\end{equation}
Then (\ref{18.10c}) yields
\begin{equation}
a = \frac{3 \alpha}{4} (2 + \gamma - N), \label{18.17}
\end{equation}
while (\ref{18.10b}) provides
\begin{equation}
\alpha = \frac{2}{\gamma N}. \label{18.18}
\end{equation}
Combining (\ref{18.18}) with the preceding formulae (\ref{18.14})--(\ref{18.17}), we 
find
\begin{equation}
\alpha = \frac{2 \gamma}{N}; \quad  g = \gamma^2, \quad  f = \frac{3}{N}, \quad  
b = \frac{3 \gamma}{N}, \quad a = \frac{3}{2 \gamma N} (2 + \gamma - N). 
\label{18.19}
\end{equation}
Substituting this into (\ref{18.10g}), we get
\[
(\gamma - 1) \frac{3 \gamma}{N} = \frac{3}{\gamma N} (2 + \gamma -N) 
+ \frac{3 \cdot 2}{\gamma N} \frac{N - 3}{2},
\]
or 
\begin{gather}
(\gamma -1) \gamma^2 = 2 + \gamma - N + N - 3 = \gamma -1 \nonumber\\
\Leftrightarrow \ 
(\gamma - 1) \left(\gamma^2 -1\right) = 0. \label{18.20}
\end{gather}
Thus, 
\begin{equation}
\gamma = \pm 1. \label{18.21}
\end{equation}
For $\gamma = 1$, formulae (\ref{18.19}) yield
\begin{equation}
\alpha = \frac{2}{N}; \quad g = 1, \quad f = \frac{3}{N}, \quad b = \frac{3}{N}, 
\quad a = \frac{3}{N} \cdot \frac{3 - N}{2}, \label{18.22}
\end{equation}
and formulae (\ref{18.3}), (\ref{18.4}) show that this is precisely the Lax-type 
extension.  The case $\gamma = -1$ is the dual extension.

Thus, scalar Lax operators of order $\geq 4$ have just two 1-component 
extensions:  the Lax-type one and its dual.  Extensions with 
$2 \ell$ components, $\ell \in {\mathbb N}$, come from Lax equations with 
Lax operators of the form
\begin{equation}
L^{\rm ext} = \partial^N + \sum^{N-2}_{i = 1} u_i \partial^{N-2-i} 
+ \sum^\ell_{s=1} \varphi_s \partial^{-1} \varphi_{\ell +s} \label{18.23}
\end{equation}
(see \S~3.5 in [16]).  If $L$ is self-adjoint then we can take 
\begin{equation}
L^{\rm ext} = L + \sum^\ell_{s=1} \xi_s \partial^{-1} \xi_s, 
\label{18.24}
\end{equation}
where the $\xi$'s are odd.  If $L$ is skew-adjoint, we can take 
\begin{equation}
L^{\rm ext} = L+ \sum^\ell_{s=1} \varphi_s \partial^{-1} \varphi_s, 
\label{18.25}
\end{equation}
where the $\varphi$'s are even.  The simplest case of the latter 
sort is
\begin{gather}
L = \partial^3 + u \partial + \partial u, \label{18.26}\\
L^{\rm ext} = \partial^3 + u \partial + \partial u + \varphi 
\partial^{-1} \varphi. \label{18.27}
\end{gather}

\section{Composition of Hamiltonian extensions}

Suppose we have a Hamiltonian matrix $B = B\{q\} = (B_{nm})$ in 
$q$-variables, and suppose we have two Hamiltonian extensions of 
$B$, $B^1$ and $B^2$, of the form
\begin{gather}
B^1 = \bordermatrix{& q_m & \varphi_\beta \cr
q_n& B_{nm} & {\mathcal O}_{\beta n}^\dagger \cr
\varphi_\alpha& {\mathcal O}_{\alpha m} & 0 \cr}, \label{19.1}\\
B^2 = \bordermatrix{& q_m & \psi_\nu \cr
q_n&B_{nm} & - \bar {\mathcal O}^\dagger_{\nu n} \cr
\psi_\mu & \bar {\mathcal O}_{\mu m} & 0 \cr}, \label{19.2}
\end{gather}
where the operators ${\mathcal O}_{\alpha m}$ depend only on $\{\varphi\}$, 
and operators $\bar {\mathcal O}_{\mu m}$ depend only on $\{\psi\}$.  

\setcounter{proposition}{2}
\setcounter{equation}{3}

\begin{proposition}\label{proposition:19.3}
 If $B^1$ and $B^2$ are Hamiltonian then 
so is their direct sum
\begin{equation}
{\mathcal B} = \bordermatrix{& q_m & \varphi_\beta & \psi_\nu \cr
q_n & B_{nm} & - {\mathcal O}^\dagger_{\beta n} & - \bar {\mathcal O}^\dagger_{\nu n} \cr
\varphi_\alpha & {\mathcal O}_{\alpha m}& 0 & 0 \cr
\psi_\mu&\bar {\mathcal O}_{\mu m} & 0 & 0 \cr} \label{19.4}
\end{equation}
\end{proposition}

\begin{proof} By the main result of the Hamiltonian formalism 
[14, p.~49], to show that a matrix is Hamiltonian we need only verify 
the Jacobi identity for arbitrary {\it linear} Hamiltonians:
\begin{equation}
\{H, \{F, G\}\} + \{F, \{G, H\}\} + \{G, \{H, F\}\} \sim 0, \quad
\forall \; H, G, F. \label{19.5}
\end{equation}

1)~If all Hamiltonians $H$, $F$, $G$, are $q$-independent, $\{H, \{F, 
G\}\} = 0$, $\forall \; H, F, G$;

2)~If two out of three Hamiltonians $H$, $F$, $G$ are $q$-independent,  
$\{H, \{F, G\}\} = 0$, $\forall \; H, F, G$;

3)~If all Hamiltonians $H$, $F$, $G$ are $\varphi$- and $\psi$-independent, 
then (\ref{19.5}) is the criterion for the extendee matrix $B$ to be 
Hamiltonian; 

4)~It leaves us with the case when two out of three Hamiltonians 
are $\varphi$- and $\psi$-in\-de\-pen\-dent, and the third one is $q$-independent.  
If that third Hamiltonian is $\psi$-independent, of the form
\begin{subequations}\label{19.6}
\begin{equation}
{\pbf{U}}^t {\pbf{\varphi}}, \label{19.6a}
\end{equation}
then (\ref{19.5}) is the condition for the extended matrix $B^1$ to be 
Hamiltonian; if that third Hamiltonian is $\varphi$-independent, 
of the form
\begin{equation}
{\pbf{V}}^t {\pbf{\psi}}, \label{19.6b}
\end{equation}
\end{subequations}
then (\ref{19.5}) is the condition for the extended matrix $B^2$ to be 
Hamiltonian; in general, any linear $q$-independent Hamiltonian is 
a sum of two terms, of the type (\ref{19.6}), and the LHS of (\ref{19.5}) is 
trilinear in $H$, $F$, $G$. 
\end{proof}

\subsection*{Acknowledgements}

This paper is based on the material I lectured about in the fall 2000 
at the University of Tennessee Space Institute.  I'm grateful to 
Lesong Wang for useful discussions.  I also thank George Wilson for 
the suggestion mentioned in Section~17, and the anonymous referees for their 
comments.

\label{kupersh-lastpage}

\end{document}